\documentclass[a4paper]{article}
\usepackage{RRA4}
\RRNo{6229}
\usepackage{hyperref}
\usepackage{macroRap}
\RRdate{Juin 2007}

\RRauthor{
Pierre Deransart\thanks{INRIA Rocquencourt, Pierre.Deransart@inria.fr}%
  \and
Mireille Ducass\'e\thanks{INSA-Rennes, Mireille.ducasse@irisa.fr}%
  \and
G\'erard Ferrand\thanks{LIFO-Orl\'eans, Gerard.Ferrand@lifo.univ-orleans.fr}

}
\authorhead{Deransart \& al}
\RRtitle{Une s\'emantique observationnelle du mod\`ele des bo\^ites pour la r\'esolution de programmes logiques (version \'etendue)}
\RRetitle{An Observational Semantics of the Logic Program Resolution Boxe's Model (Extended Version)}
\titlehead{Le mod\`ele des \bts}
\RRresume{Dans ce rapport on \'etudie une pr\'esentation originale du mod\`ele
  des bo\^ites de Byrd bas\'ee sur la notion de s\'emantique
  observationnelle. Cette approche permet de rendre compte de la
  s\'emantique des traceurs Prolog ind\'ependamment de toute
  implantation particuli\`ere.

  Le sch\'ema explicatif obtenu est une pr\'esentation formelle
  \'epur\'ee d'une trace consid\'er\'ee en g\'en\'eral comme plut\^ot
  obscure et difficile \`a utiliser. Il peut constituer une approche
  simple et p\'edagogique tant pour l'enseignement (par sa forme
  \'epur\'ee) que pour les implantations de traceurs Prolog dont il
  constitue une forme de sp\'ecification.

   Notre approche met en \'evidence les qualit\'es du mod\`ele des
  bo\^ites qui en ont fait son succ\`es, mais aussi ses
  inconv\'enients et ses limites.

 Ceci, en fait, n'est qu'un exemple pour illustrer une
  probl\'ematique g\'en\'erale relative aux traceurs et aux processus
  observants qui ne connaissent du processus observ\'e que sa
  trace. La question est alors de pouvoir reconstituer par l'analyse
  de la trace l'essentiel du processus observ\'e, et si possible, sans
  perte d'information.
}
\RRabstract{This report specifies an observational semantics and gives an
  original presentation of the Byrd's box model. The approach accounts
  for the semantics of Prolog tracers independently of a particular
  implementation.

  Traces are, in general, considered as rather obscure and difficult
  to use. The proposed formal presentation of a trace constitutes a
  simple and pedagogical approach for teaching Prolog or for
  implementing Prolog tracers. It constitutes a form of declarative
  specification for the tracers.

  Our approach highlights qualities of the box model which made its
 success, but also its drawbacks and limits.

 As a matter of fact, the presented semantics is only one example to
  illustrate general problems relating to tracers and observing
  processes.  Observing processes know, from observed processes, only
  their traces. The issue is then to be able to reconstitute by the
  sole analysis of the trace the main part of the observed process, and if
  possible, without any loss of information.
}
\RRmotcle{trace, traceur, pilote de tracer, analyseur, outils d'analyse, analyse de programme, analyse dynamique, manageur, s\'emantique observationnelle, trace ad\'equate, d\'eveloppement de logiciel, d\'eboggage, environnement de programmation, mod\`ele de Byrd, mod\`ele des bo\^{\i}tes, Prolog, programmation en logique, validation de trace}
\RRkeyword{trace, tracer, tracer driver, analyser, analyser manager, program analysis, dynamic analysis, observational semantics, trace adequation, software engineering, debugging, programming environment, Byrd's trace, Prolog, Logic Programming, Prolog debugging, box model, trace validation}
\RRprojets{Contraintes}
\RRtheme{\THCog} 
\URRocq 
\begin{document}
\makeRR   
\section{Introduction}
\label{intro}

Ce rapport pr\'esente un mod\`ele de trace Prolog (souvent appel\'e ``mod\`ele des bo\^ites de Byrd'') d'une mani\`ere originale, bas\'ee sur la notion de s\'emantique observationnelle (SO). Cette s\'emantique a \'et\'e introduite dans \cite{ercimlnai}  afin de rendre compte de la s\'emantique de traceurs ind\'ependamment de la s\'emantique du processus trac\'e.

\vspace{2mm}
Ce n'est pas l'objet de ce rapport d'\'etudier cette s\'emantique.
Notre objectif est de l'illustrer ici avec un exemple simple mais non trivial. Le r\'esultat est une s\'emantique originale de la trace Prolog telle qu'usuellement implant\'ee, sans tenir compte d'aucune implantation particuli\`ere ni d\'ecrire la totalit\'e du processus de r\'esolution.
Une telle s\'emantique constitue \'egalement une forme de sp\'ecification formelle de traceur Prolog et permet d'en comprendre facilement quelques propri\'et\'es essentielles.

\vspace{2mm}
``Comprendre une trace'' c'est d'une certaine mani\`ere tenter de retrouver le fonctionnement du processus trac\'e \`a partir d'un \'etat initial connu et d'une suite d'\'ev\'enements de trace. Cette d\'emarche  suppose une connaissance suffisante mais non n\'ecessairement compl\`ete du mod\`ele de fonctionnement du processus, et de savoir relier les \'ev\'enements de trace \`a ce mod\`ele.
C'est cette d\'emarche que nous voulons capturer avec la notion de s\'emantique observationnelle, de sch\'ema de trace et de sch\'ema de reconstruction du mod\`ele original \`a partir de la trace, c'est \`a dire d'``ad\'equation'' de la trace au mod\`ele observ\'e. 

\vspace{2mm} Le ``mod\`ele des bo\^{\i}tes'' a \'et\'e introduit pour
la premi\`ere fois par Lawrence Byrd en 1980 \cite{byrd80} dans le but
d'aider les utilisateurs du ``nouveau'' langage Prolog (il fait alors
r\'ef\'erence aux implantations d'Edinburgh \cite{proedinb} et de
Marseille\cite{promars}) \`a ma\^{\i}triser la lecture
op\'erationnelle du d\'eroulement du programme.  D\`es les d\'ebuts,
en effet, les utilisateurs se sont plaints des difficult\'es de
compr\'ehension du contr\^ole li\'es au non d\'eterminisme des
solutions. M\^eme si, par la suite, d'autres mod\`eles ont \'et\'e
adopt\'es avec des strat\'egies bien plus complexes\footnote{Le
  mod\`ele de Byrd se limite \`a la strat\'egie standard de
  parcours/construction d'arbre.},
les quatre ``ports'' introduits par Byrd ({\bf Call}, {\bf Exit}, {\bf Redo} et {\bf Fail}), associ\'es aux quatre coins d'une bo\^{\i}te et manipulables dans une sorte d'alg\`ebre de poup\'ees russes, sont rest\'es c\'el\`ebres et se retrouvent dans toutes les traces des syst\`emes Prolog existants. 

Le mod\`ele des bo\^{\i}tes de Byrd fascine par sa simplicit\'e apparente.
Toujours et souvent cit\'e mais rarement bien expliqu\'e, le mod\`ele des bo\^ites garde l'aura des premiers essais r\'eussis. 
C'est sans doute pour cela qu'il reste l'objet de publications ponctuelles mais r\'eguli\`eres depuis 1980, comme \cite{boiz84} (1984), \cite{ToBe93} (1993), \cite{jahier00} (2000), \cite{kulas03} (2003).
Pour autant, il 
reste souvent difficile \`a ``expliquer'' parce que ses diverses d\'efinitions sont soit trop informelles, soit noy\'ees dans une formalisation compl\`ete de la s\'emantique de Prolog.




\vspace{2mm}
Dans ce rapport nous proposons une description formelle d'une variante du mod\`ele initial d\'efini informellement par Byrd en 1980. L'originalit\'e de cette description r\'eside dans le fait que, bien que contenant les ingr\'edients du mod\`ele original et limit\'ee 
aux \'el\'ements de contr\^ole dont elle veut rendre compte (c'est \`a dire, sans, ou en tous cas le moins possible, faire r\'ef\'erence aux m\'ecanismes de choix de clauses et d'unification propres \`a la r\'esolution Prolog), elle est formellement compl\`ete.

Apr\`es une introduction aux traces et leur s\'emantique
observationnelle (sections~\ref{introtrace} et \ref{semobs}), nous
pr\'esentons la SO qui sp\'ecifie le mod\`ele des \bts
(section~\ref{soprologpur}) et l'extraction de la trace
(section~\ref{generation}). Enfin nous donnons le mod\`ele de
reconstruction (section~\ref{reconstruction})~; la preuve
d'ad\'equation est en annexe,
\'etablissant ainsi une grille de lecture possible de la trace,
bas\'ee sur la SO.

Notre approche met en \'evidence les qualit\'es du mod\`ele des bo\^ites qui en ont fait son succ\`es, mais aussi
ses d\'efauts principaux (section~\ref{commdisc}). Elle montre aussi
l'int\'er\^et 
de l'approche observationnelle.


\section{Introduction aux traces}
\label{introtrace}

Nous donnons ici un aper\c{c}u rapide du contexte de cette \'etude. Pour plus de d\'etails sur les motivations on pourra se reporter \`a \cite{pierrewlpe06} et \cite{ercimlnai} .

D'une mani\`ere g\'en\'erale, on veut s'int\'eresser \`a l'observation de processus dynamiques \`a partir des traces qu'ils laissent ou qu'on leur fait produire\footnote{Il faut bien distinguer ce qui rel\`eve de ce que nous appelons ici ``trace'' et ce qui rel\`eve d'outils l'analyse de processus (``monitoring'', pr\'esentation particuli\`ere de la trace ou ``jolies'' impressions, visualisation, analyse de performance, d\'ebogage, \ldots) qui tous d'une mani\`ere ou d'une autre, directement ou indirectement, de l'int\'erieur ou ind\'ependemment, en mode synchrone ou asynchrone, utilisent ce que nous appelons une ``trace virtuelle''. Cette \'etude n'est pas concern\'ee par la nature ni la forme de ces processus observants.}.

On peut toujours consid\'erer qu'entre un observateur et un ph\'enom\`ene observ\'e il y a un objet que nous appellerons {\em trace}. La trace est l'empreinte reconnaissable laiss\'ee par un processus et donc ``lisible'' par d'autres processus. Le ph\'enom\`ene observ\'e sera consid\'er\'e ici comme un processus ferm\'e (ceci concernant toutes les donn\'es et fonctions qu'il manipule) dont on ne conna\^{\i}t que la trace. La trace est une suite d'\'ev\'enements repr\'esentant l'\'evolution d'un \'etat 
qui contient tout ce que l'on peut ou veut conna\^{\i}tre de ce processus. Celle-ci peut \^etre formalis\'ee par un mod\`ele de transition d'\'etats, c'est \`a dire  par 
un domaine d'\'etats et une fonction de transition formalisant le
passage d'un \'etat \`a un autre. Cette s\'emantique sera appel\'ee
{\em s\'emantique observationnelle} (SO) car elle repr\'esente ce que
l'on est susceptible de conna\^{\i}tre ou de d\'ecrire du processus,
vu de l'``ext\'erieur''.

La SO se caract\'erise par le fait que chaque transition donne lieu \`a un \'ev\'enement de trace. Si une trace peut \^etre infinie, les diff\'erents types d'actions (ou ensemble d'actions) du processus observ\'e r\'ealisant les transitions sont suppos\'es en nombre fini. 
A chaque type d'action correspond une seule transition. On consid\'erera ici que la SO est sp\'ecifi\'ee par un ensemble fini de r\`egles de transition nomm\'ees, not\'e $R$.

Pour formaliser cette approche on introduit la notion de {\em trace int\'egrale virtuelle}.

%

\begin{definition} [Trace int\'egrale virtuelle]

Une trace int\'egrale virtuelle est une suite d'\'ev\'enements de trace qui sont de la forme {\bf $e_t: (t, a_t, S_{t+1}),\ t \ge 0$} o\`u:
\begin{itemize}
  \item $e_t$: est l'{\bf identificateur} unique  de l'\'ev\'enement.
  \item $t$: est le {\bf chrono}, temps de la trace. C'est un entier incr\'ement\'e d'une unit\'e \`a chaque \'ev\'enement.
  \item $S_{t+1} = p_{1, t+1}, ..., p_{n, t+1}$: $S_{t+1}$ est l'\'etat courant suivant l'\ev de trace au moment $t$ et les $p_{i, t+1}$ sont des valeurs des {\bf param\`etres} $p_i$ de l'\'etat obtenu, une fois l'action r\'ealis\'ee.
  \item $a_t$: un identificateur d'{\bf action} caract\'erisant le type des actions r\'ealis\'ees pour effectuer la transition de l'\'etat $S_t$ \`a $S_{t+1}$. 
\end{itemize}
Une trace est produite \`a partir d'un \'etat initial not\'e $S_0$ et peut \^etre infinie. Une suite finie d'\'ev\'enements de trace $e_t e_{t-1} \ldots e_0 $ de taille $t+1, t \geq 0$ sera d\'enot\'ee ${e_t}^+$ (${e_t}^*$ si la suite vide est incluse). La suite vide sera d\'enot\'ee $\epsilon$. Une portion finie non vide de trace sera d\'enot\'ee $<S_0, e^+_t>$. 
Une trace a au moins un \'ev\'enement vide.
\end{definition}

La trace int\'egrale virtuelle repr\'esente ce que l'on souhaite ou ce qu'il est possible d'observer d'un processus donn\'e. Comme l'\'etat courant (virtuel) du processus est int\'egralement d\'ecrit dans cette trace, on ne peut esp\'erer ni la produire ni la communiquer efficacement. En pratique on effectuera une sorte de ``compression'', et on s'assurera que le processus observant puisse la ``d\'ecompresser''. La trace effectivement diffus\'ee sera extraite de la trace virtuelle et communiqu\'ee sous forme de {\em trace actuelle}.

\begin{definition} [Trace actuelle, sch\'ema de trace]
 
Une trace actuelle est une  suite d'\'ev\'enements de trace de la forme {\bf $e_t: (t, a_t, A_t),\ t \geq 0$}, d\'eriv\'es de la transition $<S_t, S_{t+1}>$ par la fonction ${\cal E}$, dite {\em fonction d'extraction}, telle que $e_t = {\cal E}(S_t, S_{t+1})$. 

Si $A_t = S_{t+1}$, la trace actuelle est la trace virtuelle int\'egrale.

\noindent
$A_t$ d\'enote une suite finie de valeurs d'attributs. 

La fonction d'extraction est une famille de fonctions d\'efinies pour
chaque r\`egle de transition $r$ de la SO. 
\\Soit: ${\cal E} = \{{\cal
  E}_r | r \in R \}$ telle que $\forall <r,S,S'> \in SO, {\cal
  E}_r(S,S') = e_r$, o\`u $e_r$ d\'enote les calculs des valeurs de
chaque attribut.

\noindent
La description de la famille de fonctions ${\cal E}_r$, avec les calculs d'attributs, constitue un {\em sch\'ema de trace}.
\end{definition}


La trace actuelle est la trace \'emise par le traceur du processus observ\'e.
 
La trace int\'egrale virtuelle est un cas particulier de trace
actuelle o\`u les attributs d\'ecrivent compl\`etement les \'etats obtenus par la
suite des transitions.

\vspace{3mm}
La question se pose maintenant de l'utilit\'e d'une trace, c'est \ag dire la possibilit\'e de reconstruire une suite d'\'etats, \'eventuellement partielle, \ag partir d'une trace produite, sans le recours direct \`a la SO, mais qui corresponde, pas \ag pas, aux transitions de la SO qui ont produit cette trace. C'est ce que tente de capturer la notion d'ad\'equation.

La notion d'ad\'equation est relative \ag des \'etats limit\'es \ag un sous-ensemble des param\`etres. On note $S{/Q}$ la restriction d'un \'etat quelconque $S$ aux param\`etres $Q$. $Q$ sera appel\'e {\em \'etat actuel courant} et $S/Q$ l'\'etat virtuel restreint aux param\`etres de $Q$, ou, s'il n'y a pas d'ambiguit\'e, {\em \'etat virtuel courant restreint}.

On supposera donc que la SO est d\'ecrite par un ensemble fini de
r\`egles qui constituent un ``mod\`ele de trace'', tel que chaque
r\`egle donne lieu \`a la production d'un \ev de trace. La fonction
d'extraction est donc constitu\'ee d'autant de composants qu'il y a de
\rgs et d\'enot\'es ${\cal E}_r$ pour chaque \rg $r$. De m\^eme on
utilisera une {\em fonction de reconstruction} ${\cal C}_r$ d\'ecrite
par autant de composants qu'il y a de \rgs et d\'enot\'es ${\cal C}_r$
pour chaque \rg $r$. La description de ${\cal C}$ constitue un {\em
  sch\'ema de reconstruction}.

\begin{definition} [Trace ad\'equate]

  Etant donn\'es un \'etat actuel $Q$ restriction de $S$ \ag un sous
  ensemble de ses param\`etres, une SO d\'efinie sur $S$ par un
  ensemble fini de transitions $R$ et une trace actuelle $T_w = <Q_0,
  w^*_t>$ telle que $Q_0 = {S_0}{/Q}$ \\ $T_w $ est {\em ad\'equate} pour $Q$
  par rapport \ag la trace virtuelle int\'egrale $T_v = <S_0, v^*_t>, \forall t \geq 0$
  s'il existe
une fonction ${\cal F}$ telle que
\begin{quote}
$\forall t \geq 0, \ {\cal F}(w^*_t,Q_0) = Q_t\ $ et

$\forall i \in [0 .. t-1], \ Q_i = S_i/Q \ \wedge 
\exists r \in R, \ \ \, $tel que$\,\  w_i = {\cal E}_r(S_i,S_{i+1})$.
\end{quote}

\end{definition}

L'ad\'equation stipule qu'\ag toute suite d'\'etats, engendr\'ee par une trace actuelle, il correspond une suite de transitions de la SO qui a engendr\'e cette trace et dont la suite des \'etats restreints est la m\^eme.

\vspace{2mm} L'ad\'equation pour un sous-\'etat $Q$ donne \`a une
trace une s\'emantique propre: la lecture de la trace peut \^etre
comprise comme l'\'evolution d'un \'etat restreint. De plus, cette
suite d'\'etats est exactement la suite des \'etats restreints
correspondants observables sur le processus observ\'e. L'ad\'equation
assure que toute l'information possible est bien dans la trace
actuelle modulo le fait que seule une partie de ce qui est observable
est communiqu\'e dans la trace.

\vspace{1mm} Si $Q = S$ et $T_w$ est une trace ad\'equate, alors $T_w$
est une trace int\'egrale.  

Une telle trace actuelle, ad\'equate et
int\'egrale est la garantie que l'observateur est capable de
reconstituer toute l'\'evolution observable du processus et donc de
reconstituer l'int\'egralit\'e des objets observ\'es et leur
\'evolution \ag partir de la trace observ\'ee. Si l'on consid\`ere que
${\cal E}$ est une forme de fonction de compression, ${\cal C}$ peut
\^etre vue, dans ce cas, comme une fonction de d\'ecompression sans
perte de donn\'ees\footnote{On n'utilise pas ici les termes
  ``compression/d\'ecompression'' afin de se d\'emarquer du cas o\`u
  le flot de trace (par exemple cod\'e en XML) peut lui-m\^eme
  faire l'objet d'une ``compression/d\'ecompression'' num\'erique. Ces
  termes sont donc r\'eserv\'es pour ce cas.}.

\vspace{2mm}


En r\'ealit\'e, et c'est le but de la trace virtuelle, un observateur
ne sera int\'eress\'e qu'\`a une partie de la trace virtuelle, c'est
\`a dire qu'\ag un sous-ensemble $Q$ de ses param\`etres. Par contre
il est essentiel que la trace actuelle soit ad\'equate par
rapport \`a $Q$, garantissant ainsi la transmission et compr\'ehension
compl\`etes des \'etats partiels que l'on peut retrouver alors par la
seule lecture de la trace.

On propose maintenant une  condition suffisante pour prouver l'ad\'equation d'une trace
actuelle qui utilisent des couples d'\'ev\'enements de trace. En effet, un
\ev de trace actuelle, produit par une transition $<r,S,S'>$ en
appliquant une \rg $r$, peut ne pas comporter suffisamment d'attributs
pour restituer les param\`etres souhait\'es de l'\'etat $S'$ (ceux qui
se trouvent dans l'\ev correspondant de la trace virtuelle
int\'egrale). Il est alors tr\`es probable que l'information manquante soit contenue dans l'\ev de trace suivant.
Il est donc parfois n\'ecessaire de recourir \ag deux\footnote{Cela signifie qu'en fait
la reconstruction repose sur toute la trace connue, plus un \ev ``en avant''.
D'un point de vue th\'eorique on peut vraisemblablement se limiter \ag un \'ev\`enement ``en avant'', mais
en pratique il doit \^etre possible de trouver des situations o\`u plus d'\'ev\'enements 
peuvent s'av\'erer n\'ecessaires. Ceci
n'est pas un probl\`eme dans la mesure o\`u les traces sont en fait non born\'ees.}
\evs de trace pour pouvoir reconstruire les param\`etres souhait\'es
de l'\'etat courant obtenu. 
La fonction de reconstruction sera donc d\'ecrite par une
famille de {\em fonctions locales de reconstruction} d'\'etats
restreints $Q$, ${\cal C} = \{{\cal C}_r | r \in R \}$ telle que
$\forall <r,S,S'> \ \in SO, \ Q' = {\cal C}_r(e,e',Q)$. La description
des fonctions locales de reconstruction constitue un {\em sch\'ema de
  reconstruction}.

Par ailleurs il est \ag noter que l'identification de la \rg qui a produit
l'\ev de trace a pu \^etre perdue au cours de l'extraction de l'\'ev\'enement.
Il est donc n\'ecessaire de pouvoir associer \ag un \ev de
trace la transition, donc la \rg de la SO, qui l'a produit. Pour ce
faire on utilise \'egalement une famille de conditions $Cond_r(e,e')$
qui, \'etant donn\'e un couple d'\evs de trace, identifient sans
ambiguit\'e la \rg $r$ utilis\'ee donc la transition qui a produit le
premier \'ev\'enement $e$.

\begin{proposition} [Condition d'ad\'equation]
\label{theoadeq}

Etant donn\'es une SO d\'efinie avec un ensemble de \rgs $R$, un sch\'ema de trace ${\cal E}$ et un sch\'ema de reconstruction  ${\cal C}$ pour un sous-ensemble de param\`etres $Q$. Si les deux propri\'et\'es suivantes sont satisfaites pour chaque \rg $r \in R$:
\begin{quote}
$\forall\, e,\,e',\,r', S,\,S',\,S'',$

${\cal E}_r(S,S') = e\  \wedge \ {\cal E}_{r'}(S',S'') = e'$

(1) seule  $Cond_r(e,e')$ est vraie, i.e. $Cond_r(e,e') \bigwedge_{s \not = r} \neg Cond_{s}(e,e')$.

(2) $\ {\cal C}_r(e,e',S/Q) = S'/Q$.
\end{quote}

\vspace{2mm}
alors toute trace actuelle $T_w = <Q_0, w^+_t>$ , d\'efinie par le sch\'ema de trace ${\cal E}$ et telle que $Q_0 = {S_0}{/Q}$, est {\em ad\'equate} pour $Q$ par rapport \ag la trace int\'egrale virtuelle $T_v = <S_0, v^+_t>$.

%
%
\end{proposition}

Noter que la condition~1 est trivialement satisfaite s'il y a autant de types d\'ev\'enement de trace distincts que de transitions possibles dans la SO.
Dans ce cas, la condition d'ad\'equation se r\'eduit \ag la condition~2.

Il y a deux mani\`eres d'interpr\'eter cette proposition. Elle montre que l'ad\'equation est d'abord une propri\'et\'e de correction de trace. Dans ce sens elle signifie que si l'on suit pas \`a pas l'extraction d'une trace \`a partir de l'\'etat initial $S_0$ durant $t+1$ \'etapes, l'\'etat obtenu reconstruit \`a partir de la portion de trace $T_w = <S_0/Q, w^*_t>$ est le m\^eme que celui obtenu par application des \rgs de la SO, restreint \`a $Q$, soit $S_{t+1}/Q$. Ceci correspond \ag la deuxi\`eme condition.

Mais il y a une deuxi\`eme lecture li\'ee \ag la question de compr\'ehension du processus observ\'e \ag travers la trace et qui correspond aux deux conditions prises ensembles. Connaissant l'\'etat initial et la suite des \'ev\'enements de trace, la proposition assure que l'on peut en d\'eduire une suite d'\'etats d\'eriv\'es par transitions dans la SO, suite identique dans le sens o\`u elle engendre la m\^eme trace. 

Ceci met en \'evidence la distinction qu'il y a lieu de faire entre
``correction'' de la trace (condition~2) et la capacit\'e de
compr\'ehension du processus observ\'e \ag travers la trace. La
seconde lecture est aussi li\'ee \ag la capacit\'e d'associer une
transition ou \rg \ag un pas de trace (condition~1) et ainsi de
retrouver des informations sur les \'etats virtuels au del\ag de leur
restriction.  L'ad\'equation assure que la trace extraite repr\'esente
bien l'\'evolution possible d'un sous-\'etat virtuel, mais aussi, que
si on conna\ii t la SO, on peut \'egalement
appr\'ehender le fonctionnement du processus observ\'e.

\section{S\'emantique Observationnelle et fonctions associ\'ees}
\label{semobs}

La S\'emantique Observationnelle (SO) 
se distingue d'une s\'emantique op\'era\-tion\-nelle par le fait que
son objet est avant tout la description d'un flot de donn\'ees
\'eventuellement infini sans faire explicitement r\'ef\'erence \`a un
processus particulier ni \ag une s\'emantique concr\`ete particuli\`ere.


La S\'emantique Observationnelle (SO) rend compte de toutes les traces virtuelle possibles, c'est \`a dire de toutes les  suites d'\'etats d\'ecrits par un ensemble fini de param\`etres, et d\'efinies par un \'etat initial, une fonction de transition d'\'etats, et telles qu'\`a chaque transition un \'el\'ement de trace puisse \^etre produit. Une SO est donc d\'efinie par un domaine d'\'etats et une fonction de transition d'\'etats.



Dans la SO, la fonction de transition est d\'ecrite par un ensemble
fini de r\`egles nomm\'ees. L'application d'une r\`egle produit un
\'ev\'enement de trace. Une r\`egle a quatre composants.

\begin{itemize}
\item Un identificateur de r\`egle (nom).
\item Un num\'erateur comportant des conditions sur l'\'etat ant\'erieur et des calculs pr\'eliminaires.
\item Un d\'enominateur comportant la description de l'\'etat obtenu (ce qui reste invariant peut \^etre omis), par les calculs des nouvelles valeurs des param\`etres.
\item Des conditions externes (entre accolades) ou propri\'et\'es portant sur des \'el\'ements non d\'ecrits par des param\`etres, mais intervenant dans le choix des r\`egles ou les valeurs des param\`etres.
\end{itemize}

\vspace{2mm} 

Noter que la distinction entre les \'el\'ements figurant
au num\'erateur et dans les accolades est arbitraire. Toutefois, toute
expression contenant des \'el\'ements externes sera dans les
accolades.

Chaque \rg de transition de la SO sera pr\'esent\'ee formellement par un triplet\footnote{Pour la raison indiqu\'ee plus haut on omettra ici la partie externe.} $<nom, S,S'>$ o\`u par abus de notation on d\'enotera $S$ les conditions portant sur un \'etat courant $S_t$ et auquel la transition peut alors s'appliquer, et par $S'$ l'\'etat r\'esultant de la transition (dont l'instance est alors $S_{t+1}$), mais simplement d\'ecrit ici par les calculs des nouvelles valeurs de param\`etres. On y ajoutera \'egalement d'\'eventuels facteurs externes\footnote{Ces facteurs sont dits ``externes'' du point de vue de la SO. Ils ne le sont pas du point de vue du processus observ\'e. Il ne s'agit donc pas d'interaction.} d\'ecrits par des {\em conditions externes}.
Une \rg sera donc pr\'esent\'ee de la mani\`ere suivante.

\vspace{1mm}
\noindent
\reglecontrole{Nom}
{Conditions \,caracterisant\,l'etat\,courant}
{Calcul\,des\,nouveaux\,parametres}
{\{ Conditions\, externes\}}

\vspace{2mm}
Pour d\'ecrire la SO, on utilisera deux types de fonctions: celles relatives aux objets d\'ecrits et leur \'evolution dans la trace virtuelle et celles relative \`a des \'ev\'enements ou objets non d\'ecrits dans cette trace, mais susceptibles de se produire dans les processus observ\'es et d'y \^etre interpr\'et\'ees. Les fonctions de la premi\`ere cat\'egorie sont dites ``utilitaires'', celles de la seconde ``externes''. Elles concernent des param\`etres non pris en compte dans la trace virtuelle. Enfin on distinguera \'egalement les fonctions exclusivement utilis\'ees pour le calcul des attributs lors de l'extraction de la trace, dites ``auxiliaires d'extraction'' et celles exclusivement utilis\'ees pour la reconstruction dites ``auxiliaires de reconstruction''.


\vspace{2mm}
La fonction d'extraction ${\cal E}$ sera d\'ecrite par le m\^eme type
de r\`egles, mais leur d\'enominateur comportera exclusivement
l'\'ev\'enement de trace actuelle correspondant, c'est \`a dire le
port et les attributs.
Il y a un seul \'ev\'enement de trace par \rg. L'ensemble des règles qui
d\'ecrivent la fonction d'extraction constitue un {\em sch\'ema de
  trace}. Chaque \rg du sch\'ema de trace a la forme suivante.

\vspace{1mm}
\noindent
\def\et{,\ \ }
\reglecontrole{Nom}
{Calcul  \,\, des \,\, attributs}
{< Evenement \,\, de \,\, trace >}
{\{ Cond. \,\, externes\}}

\vspace{2mm}
La description de la reconstruction utilisera une fonction locale de reconstruction ${\cal C} = \{{\cal C}_r | r \in R  \}$. Elle sera d\'ecrite avec le m\^eme type de r\`egles.

\vspace{1mm}
\noindent
\reglecontrole{Nom}
{Conditions \,\, d'identification}
{Calculs \,\, de \, reconstruction}
{\{ 
Evenements \,\,de\,\, trace \}}


\noindent
La trace est cette fois consid\'er\'ee comme une information
``externe'' et se situe en position de composant externe (dans les
accolades, o\`u il y a au plus deux \evs de trace).
Le num\'erateur de la r\`egle contient la condition permettant
d'identifier la \rg de la SO qui s'applique (condition de
``compr\'ehensibilit\'e'').
Le d\'enominateur contient les calculs de reconstruction (calcul des
param\`etres de l'\'etat virtuel restreint \`a partir de la trace).
L'ensemble des \rgs de reconstruction constitue un {\em sch\'ema de
  reconstruction}.

\vspace{1mm}
Noter que les trois ensembles de \rgs (SO, sch\'emas de trace et de reconstruction) 
sont en bijection deux \ag deux.


\section{Une s\'emantique Observationnelle du mod\`ele des bo\^{\i}tes}
\label{soprologpur}

\begin{figure}[t]
\begin{center}
\includegraphics[width=0.75\linewidth]{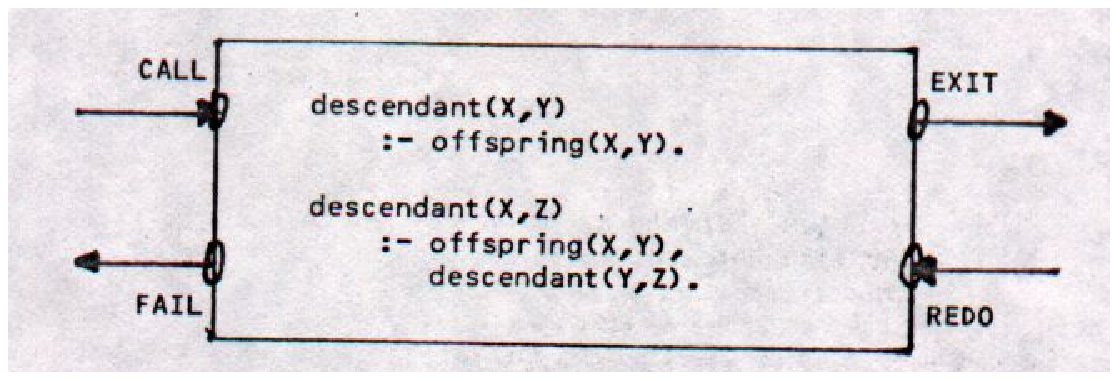}
\end{center}
\caption[Mod\`ele de bo\^{\i}te tel que dessin\'e par Byrd]{Mod\`ele de bo\^{\i}te tel que dessin\'e par Byrd \cite{byrd80}}
\label{fig:boite}
\end{figure}

Dans ses articles \cite{byrd80,byrd80ter}, Byrd illustre son mod\`ele
\`a l'aide de deux sch\'emas: une bo\^{\i}te avec les quatre fameux
ports (voir figure~\ref{fig:boite}) et un ``arbre et/ou'', structure
d\'ej\`a tr\`es r\'epandue \`a cette \'epoque qui combine les
repr\'esentations d'arbre de preuve et d'arbre de recherche. Il
n'utilise ni la notion d'arbre de preuve partiel, ni celle d'arbre de
recherche (arbre SLD), encore peu connus,
le rapport de Clark \cite{clark79} venant \ag peine de para\ii tre.

Byrd fustige n\'eanmoins les implanteurs qui, lors du retour arri\`ere (ce qui se traduit dans la trace par un \'ev\'enement de port {\bf Redo}) vont directement au point de reprise et n'expriment pas dans la trace tout le cheminement inverse. Byrd estime que ceci est de nature \`a perdre l'utilisateur et qu'il est pr\'ef\'erable de d\'efaire pas \`a pas ce qui a \'et\'e explicitement fait lors des recours successifs aux clauses pour r\'esoudre des buts.

M\^eme si nous voulons 
rester le plus proche possible de ce mod\`ele, nous ne suivrons
cependant pas ce point de vue et adopterons celui des implanteurs,
plus r\'epandu actuellement, et qui nous semble tout aussi facile \`a
comprendre \`a partir du moment o\`u tout ce qui est utile est
formalis\'e. En effet, le mod\`ele des bo\^{\i}tes oblige \`a suivre
les appels de clauses \`a travers un syst\`eme de bo\^{\i}tes
encastr\'ees. Il est alors facile de comprendre qu'\`a partir du moment
o\`u chaque bo\^{\i}te a un identificateur unique, l'acc\`es \`a un
point de choix profond\'ement enfoui dans les profondeurs de
l'empilement peut se faire aussi clairement en sautant directement sur
la bonne bo\^{\i}te qu'en descendant l'escalier r\'esultant de
l'empilement ou en faisant strictement le chemin inverse. On \'evitera
ainsi de d\'etailler explicitement la mani\`ere d'acc\'eder \`a la
bonne \bt.

M\^eme si, au final, nous ne d\'ecrivons pas exactement le mod\`ele initialement d\'efini par Byrd, nous estimons que nous en gardons les \'el\'ements historiquement essentiels, \ag savoir le parcours construction d'arbre et les \bts dans lesquelles les clauses, ou un sous-ensemble, sont stock\'ees. L'approche formalis\'ee ici sera qualifi\'ee de {\em \md des \bts simplifi\'e}.

\vspace{2mm}
L'empilement des bo\^{\i}tes et son \'evolution seront donc d\'ecrits par un parcours construction d'arbre dont chaque n\oe ud correspond \`a une bo\^{\i}te. La strat\'egie de parcours correspond \`a la strat\'egie de Prolog standard (ISO Prolog \cite{alipie96}), celle d'un parcours construction descendant gauche droite. Chaque n\oe ud nouveau, ou bo\^{\i}te, re\c{c}oit un num\'ero qui est incr\'ement\'e de 1 \`a chaque cr\'eation.

Chaque n\oe ud est \'etiquet\'e avec une pr\'edication et un paquet de clauses. 
Chaque bo\^{\i}te est donc la racine d'un sous-arbre qui se d\'eploie \`a la mani\`ere d'un ``treemap'', r\'ealisant ainsi un jeu de bo\^ites encastr\'ees.

\vspace{2mm}
Dans la mesure du possible nous utilisons le vocabulaire ISO-Prolog \cite{alipie96}.

\vspace{2mm}
{\underline{\bf Param\`etres de la trace virtuelle}}

\noindent
L'\'etat courant comporte 9 param\`etres:

$\ \ \ \ \ \ \ \ \ \{ T , u , n, num, pred , claus , first, ct , flr\}$. 

\begin{enumerate}
\item {\bf $T$}: $T$ est un arbre \'etiquet\'e avec un num\'ero de cr\'eation, une pr\'edication et un sous-ensemble de clauses du programme $P$. Il est d\'ecrit ici par ses fonctions de construction-reconstruction et parcours (cf plus bas) et \'etiquetage. 
Aucune repr\'esentation particuli\`ere n'est requise. Nous utiliserons cependant dans les exemples une notation ``\`a la Dewey''. Chaque n\oe ud est repr\'esent\'e 
par une suite de nombres entiers 
 et d\'enot\'es 
$\epsilon$, $1$, $11$, $12$, $112$, $\dots$. L'ordre lexicographique est le suivant: $u,v,w$ sont des mots, $ui < uiv (v \not = \epsilon$), et $uiv < ujw \ $si$ \ i < j$, $\epsilon$ est le mot vide.

\item {\bf $u \in T$}:  $u$ est le n\oe ud courant dans $T$ (bo\ii te visit\'ee).

\item {\bf $n \in {\cal N}$}:  $n$ est un entier positif associ\'e \`a chaque n\oe ud dans $T$ par la fonction $num$ (ci-dessous). 
C'est le num\'ero du dernier n\oe ud cr\'e\'e. 

\item {\bf $num: T \rightarrow {\cal N}$}. Abbrev.~: {\bf $nu$}.
 $nu(v)$ est le num\'ero (entier positif) associ\'e au n\oe ud $v$ dans $T$. 

\item {\bf $pred: T \rightarrow {\cal H}$}. Abbrev.: {\bf $pd$}.
 $pd(v)$ est la pr\'edication associ\'ee au n\oe ud $v$ dans $T$. C'est un \el de l'ensemble d'atomes non clos ${\cal H}$ (base de Herbrand non close).

\item {\bf $claus: T \rightarrow 2^P$}. Abbrev.~: {\bf $cl$}.
  $cl(v)$ est une liste de clauses de $P$ (m\^eme ordre que dans $P$) contribuant \ag la d\'efinition du pr\'edicat  $pred(v)$ associ\'ee au n\oe ud $v$ dans $T$. 
$[]$ est la liste vide. Selon les clauses de $cl(v)$, on peut obtenir diff\'erents mod\`eles. On ne met ici dans la bo\ii te $v$ que les clauses dont la t\^ete est unifiable avec la pr\'edication $pred(v)$. 
Si la bo\ii te est vide, la predication $pred(v)$ ne peut \^etre r\'esolue et le n\oe ud sera en \'echec (cf. {\em failure}). Cette liste de clauses est d\'efinie par un ordre externe lorsque la predication est appel\'ee 
(voir $claus\_pred\_init$ dans les fonctions externes) et mise \ag jour chaque fois que le n\oe ud est visit\'e (voir $update\_claus\_and\_pred$ dans les fonctions utilitaires). 

\item {\bf $first: T \rightarrow Bool$}. Abbrev.~: $fst$.
 $fst(v)$ est vrai ssi $v$ est un n\oe ud de $T$ qui n'a pas encore \'et\'e  visit\'e (c'est une feuille). 

\item {\bf $ct \in Bool$}:  $ct$ est l'indicateur de construction achev\'ee (compl\`ete) de $T$: $true$ ssi le n\oe ud courant est redevenu $\epsilon$ (retour \ag la racine) lors d'une remont\'ee dans l'arbre (en succ\`es ou \'echec).

\item {\bf $flr \in Bool$}:  $flr$ est l'indicateur d'\'etat d'\'echec du sous-arbre ($true$ si en \'echec, $false$ sinon, ce qui n'est pas synonyme de succ\`es).

\end{enumerate}


\vspace{2mm}
{\bf Etat initial $S_0$}:

Pour des raisons d'espace typographique, on utilisera parfois $T$ ($F$) pour $true$ (resp. $false$).

\vspace{1mm}
\noindent $\{ \{ \epsilon \} , \epsilon , 1, \{ (\epsilon, 1)\}, \{ (\epsilon, goal)\}, \{ (\epsilon, list\_of\_goal\_claus)\}, \{(\epsilon,T)\}, F, F \}$


\vspace{1mm}
Le mod\`ele est bas\'e sur un parcours de construction d'arbres de preuve partiels, construits puis reconstruits apr\`es des retours arri\`eres. Les \nds ne sont construits que juste avant d'\^etre visit\'es pour la premi\`ere fois.
La relation avec le mod\`ele des bo\^ites de Byrd est fond\'ee sur l'id\'ee que chaque n\oe ud est une \bt qui contient les clauses susceptibles de donner des d\'eveloppements alternatifs. Si la bo\^ite est vide au moment de sa cr\'eation, le n\oe ud sera en \'echec. Chaque visite d'un n\oe ud (bo\^ite) donne lieu \`a un \'ev\'enement de trace.

\newpage
{\underline{\bf Fonctions utilitaires}} (manipulation des objets d\'ecrits): 

\begin{itemize}
\item {\bf $parent: T \rightarrow T$}. Abbrev.: {\bf $pt$}.
$pt(v)$ est l'anc\^etre direct de $v$ dans $T$. 
Pour simplifier le mod\`ele, on suppose que $pt(\epsilon) = \epsilon$. 

\item {\bf $leaf: T \rightarrow Bool$ }. Abbrev.: {\bf $lf$}.
$lf(v)$ est vraie ssi $v$ est une feuille dans $T$. 

\item {\bf $may\_have\_new\_brother: T \rightarrow Bool$}. Abbrev.: {\bf $mhnb$ }.
$mhnb(v)$ est vrai ssi 
$pred(v)$ n'est pas la derni\`ere pr\'edication dans le corps de la clause courante, elle-m\^eme la premi\`ere clause dans la bo\ii te du n\oe ud parent 
de $v$ dans $T$. La racine ($\epsilon)$ n'a pas de fr\`ere. 

\item {\bf $create\_child: T \rightarrow T$ }. Abbrev.: {\bf $crc$}.
$crc(v)$ est le nouvel enfant de $v$ dans $T$. 

\item {\bf $create\_new\_brother: T \rightarrow T$ }. Abbrev.: {\bf $crnb$}.
$crnb(v)$ est le nouveau fr\`ere de $v$ dans $T$. Defini si $v$ diff\'erent de $\epsilon$. 

\item {\bf $has\_a\_choice\_point: T \rightarrow Bool$ }. Abbrev.: {\bf $hcp$}.
$hcp(v)$ est vrai ssi il existe un point de choix $w$ dans le sous-arbre de racine $v$ dans $T$ 
($claus(w)$ contient au moins une clause). 

\item {\bf $greatest\_choice\_point: T \rightarrow T$ }.  Abbrev.: {\bf $gcp$}.
$w= gcp(v)$ est le plus grand point de choix dans le sous-arbre de racine $v$ (dans $T$,
$claus(w)$ contient au moins une clause) selon l'ordre lexicographique des n\oe uds dans $T$ . 

\item {\bf $fact: T \rightarrow Bool$ }.  Abbrev.: {\bf $ft$ }.
$ft(v)$ est vrai ssi la premi\`ere clause dans $claus(v)$ est un fait.


\item {\bf $update\_number: F,T \rightarrow F $}.  Abbrev.: {\bf $upn$}.
$upn(nu,v)$ met \ag jour la fonction $num$ en supprimant toutes les r\'ef\'erences aux n\oe uds d\'econstruits de $T$ jusqu'au n\oe ud $v$ (conserv\'e),

\item {\bf $update\_claus\_and\_pred: F,T,{\cal H} \rightarrow F $}. Abbrev.: {\bf $upcp$}.
 ($F$ ensemble de fonctions): $upcp(claus, v)$, $upcp(pred,v)$ (2 arguments) ou $upcp(pred, v,p)$ 
(3 arguments): met \ag jour les fonctions $claus$ et $pred $ en supprimant toutes les r\'ef\'erences aux n\oe uds d\'econstruits de $T$ jusqu'au n\oe ud $v$ (conserv\'e),
 et mettant \'egalement \ag jour, si cela est requis par la fonction externe $pred\_update$, la valeur de $pred(v)$ avec la paire $(v,p)$ 
 ainsi que les valeurs de la fonction $claus$ au n\oe ud $v$ en
 enlevant la derni\`ere clause utilis\'ee.
\end{itemize}

\vspace{2mm}
{\underline{\bf Fonctions externes}}:

Elles correspondent aux actions non d\'ecrites dans la trace virtuelle mais qui l'influencent effectivement, en particulier tous les aspects de la r\'esolution li\'es \`a l'unification et qui sont omis dans cette SO.

\begin{itemize}
\item {\bf $success: T \rightarrow Bool$}: Abbrev.: {\bf $scs$}. $scs(v)$ est vrai ssi 
$v$ est une feuille et la pr\'edication courante a \'et\'e unifi\'ee avec succ\`es avec la t\^ete de la clause utilis\'ee dans cette bo\ii te. 

\item {\bf $failure: T \rightarrow Bool$}. Abbrev.: {\bf$flr$}.
$flr(v)$ est vrai ssi $v$ est une feuille et aucune clause du programme ne s'unifie avec la pr\'edication courante (dans ce mod\`ele la bo\ii te ne contient alors aucune clause). 

\item {\bf $claus\_pred\_init: T \rightarrow (pred, list\_of\_clauses)$}. Abbrev.: {\bf $cpini$}.
$(c,p) = cpini(v)$ met \`a jour 1- la fonction $claus$ avec la paire $(v, c)$ o\`u  $c$ est la liste des clauses dont la t\^ete est unifiable avec la pr\'edication  $pred(v)$ et qui sont donc 
utilisables pour essayer diff\'erentes alternatives pour la r\'esolution  (si la liste est vide il n'y a pas de solution), et 2- la fonction $pred$ avec la paire $(v, p)$ o\`u  $p$ est la pr\'edication \ag associer au n\oe ud $v$.  On notera $c\_cpini(v)$ et $p\_cpini(v)$ les arguments respectifs (clauses et pr\'edication) r\'esultants de $cpini(v)$.

\item {\bf $pred\_update: T \rightarrow {\cal H}$}. Abbrev.: {\bf $pud$ }.
$pud(v)$ est la nouvelle valeur de la pr\'edication attach\'ee au n\oe ud $v$ de $T$, suite \ag une unification r\'eussie.
\end{itemize}

\begin{figure*}[ht]\small
\noindent
\reglecontrole{\callun{}}
{fst(u) \wedge lf(u) \wedge \neg ct \wedge ft(u) }
{cl' \gets upcp(cl,u) \et fst'(u) \gets F \et flr' \gets F }
{\{ \}}
\saut
\reglecontroledeux{\calldeux{}}
{fst(u) \wedge lf(u) \wedge \neg ct \wedge \neg ft(u) \et v \gets crc(u)}
{T' \gets T \cup  \{v\} \et u' \gets v \et n'\gets n+1 \et nu' \gets nu \cup \{(v,n')\} \et pd' \gets pd \cup \{(v,p)\} \et }
{ cl' \gets upcp(cl,u) \cup \{(v,c)\} \et fst'(u) \gets F \et fst' \gets fst' \cup \{(v,T)\} \et flr' \gets F}
{\{ \\ \phantom{xxxxxxxxxxxxxxxxxxxxxxxxxx} scs(u) \et (c,p) = cpini(v)\}}
 \saut
\reglecontrole{\exitun{}} 
{\neg fst(u) \wedge \neg mhnb(u) \wedge \neg ct \wedge \neg flr \et v \gets pt(u)}
{u' \gets v \et pd' \gets upcp(pd,u,p) \et (u = \epsilon) \Rightarrow (ct' \gets T)}
{\{ scs(u) \et \\ \phantom{xxxxxxxxxxxxxxxxxxxxxxxxxxxxxxxxxxxxxxxxxxxxxxxxxxxxxx} p=pud(u)\}}
 \saut
\reglecontroledeux{\exitdeux{}}
{ \neg fst(u) \wedge mhnb(u) \wedge \neg ct \wedge \neg flr \et v \gets crnb(u)}
{T' \gets T \cup  \{v\} \et u' \gets v \et n'= n+1 \et nu' \gets nu \cup \{(v,n')\} \et}
{ pd' \gets upcp(pd,u,p') \cup \{(v,p)\} \et cl' \gets cl \cup \{(v,c)\} \et  fst' \gets fst \cup \{(v,T)\}}
{ \{ \\
\phantom{xxxxxxxxxxxxxxxxxxxxxxxxxxxxxxxxx} scs(u) \et p' = pud(u) \et (c,p) = cpini(v) \}}
 \saut
\reglecontrole{\faildeux{}}
{\neg fst(u) \wedge \neg ct \wedge \neg hcp(u) \et  v \gets pt(u)}
{u' \gets v \et (u = \epsilon) \Rightarrow (ct' \gets T) \et flr' \gets T}
{\{flr(u) \ \vee \  flr\}}
\saut
\reglecontroledeux{\redoun{}}
{ v \gets gcp(u) \et \neg fst(u) \wedge hcp(u) \wedge ft(v) \wedge (flr \ \vee \ ct)}
{T' \gets T - \{y | y > v \}\et u' \gets v \et cl' \gets upcp(cl,v) \et}
{ ct \Rightarrow (ct' \gets F) \et flr' \gets F}
{\{ \}}
 \saut
\reglecontroledeuxshort{\redodeuxshort{}}
{ v \gets gcp(u) \et \neg fst(u) \wedge hcp(u) \wedge (flr \ \vee \ ct) \wedge \neg ft(v) \et w \gets crc(v)}
{ T' \gets T - \{y | y > v \} \cup \{w\}\et u' \gets w  \et n'= n+1 \et nu' \gets upn(nu,v) \cup \{(w,n')\} \et flr' \gets F \et }
{pd' \gets upcp(pd,v) \cup \{(w,p)\}  \et cl' \gets upcp(cl,v) \cup \{(w,c)\} \et fst' \gets fst \cup \{(w,T)\} \et ct' \Rightarrow (ct \gets F)}
{\{ \\
\phantom{xxxxxxxxxxxxxxxxxxxxxxxxxxxxxxxxxxxxxxx}scs(v) \et (c,p) = cpini(w)\}}
\caption{Semantique Observationnelle de la r\'esolution Prolog (trace int\'egrale virtuelle)}
\label{sovirttracefig}
\end{figure*}

Noter que $\forall u, flr(u) \Rightarrow flr = true$ (voir r\`egle \faildeux{}) 

\vspace{3mm} La SO est d\'ecrite par les \rgs de la
figure~\ref{sovirttracefig}. Chaque r\`egle est comment\'ee dans ce
qui suit.

\begin{itemize}
\item \callun{}: Le n\oe ud courant est une feuille et la pr\'edication appel\'ee doit \^etre r\'esolue par un fait. Ce n\oe ud restera donc une feuille. Le point de choix est mis \`a jour (une clause de moins dans la bo\^{\i}te).

\item \calldeux{}: Le n\oe ud courant est une feuille mais la
  pr\'edication associ\'ee est r\'esolvable avec une clause dont la
  t\^ete a \'et\'e unifi\'ee avec succ\eg s et dont le corps n'est pas
  vide. Ce n\oe ud va \^etre d\'evelopp\'e. Un nouveau n\oe ud est
  cr\'e\'e dont la bo\^ite $v$ est remplie avec les clauses utiles
  (susceptibles de r\'eussir) et une pr\'edication appelante est
  associ\'ee. Le point de choix est mis \`a jour.

\item \exitun{}: sortie en succ\`es de la derni\`ere pr\'edication d'un corps de clause.  $pred(u)$ est mis \`a jour (ce n'est pas n\'ecessairement le m\^eme que lors de l'appel). Remont\'ee en succ\`es dans l'arbre sans cr\'eation de nouvelle branche.


\item \exitdeux{}: sortie en succ\`es avec cr\'eation d'une nouvelle branche ``s\oe ur'' (nouvelle feuille $v$, cas du traitement d'une clause avec plus d'une pr\'edication dans le corps). La bo\^ite $v$ est remplie avec les clauses utiles (susceptibles de r\'eussir) et une pr\'edication appelante est associ\'ee.



\item \faildeux{}: remont\'ee dans l'arbre en \'echec tant qu'il n'y a pas de point de choix dans le sous-arbre. 

\item \redoun{}: reprise suite \`a succ\`es ou \'echec, s'il y a un point de choix dans le sous-arbre ouvrant une possibilit\'e de solution ou de nouvelle solution si on est \`a la racine.
Comme discut\'e au d\'ebut de cette section,  
dans ce mod\`ele, on ne refait pas tous les ``redo'' en suivant le chemin jusqu'au point de reprise, comme dans le mod\`ele original de Byrd.
\item \redodeux{}: reprise suite \`a succ\`es ou \'echec, s'il y a un point de choix dans le sous-arbre ouvrant une possibilit\'e de solution ou une nouvelle solution si on est \`a la racine. 
Comme pr\'ec\'edemment, mais avec cr\'eation d'un descendant comme dans le cas de \calldeux{}.
\end{itemize}


\vspace{2mm}
A l'\'etat initial $S_0$, seule une des r\`egles \callun{} ou \calldeux{} s'applique.
Quel que soit l'\'etat, une seule \rg peut s'appliquer
tant qu'un arbre complet n'a pas \'et\'e construit. Aucune \rg ne s'applique si l'arbre construit est complet et qu'il n'y a plus de point de choix.
Pour les r\`egles \callun{} et \calldeux{} le port associ\'e est {\bf Call}, pour les r\`egles \exitun{} et \exitun{}, {\bf Exit}, pour \faildeux{},  {\bf Fail}, et pour \redoun{} \redodeux{}, le port est {\bf Redo}.

\section{Extraction de la trace actuelle}
\label{generation}

Chaque application d'une r\`egle de la SO donne lieu \`a l'extraction d'un \'ev\'enement de trace dont le chrono est incr\'ement\'e d'une unit\'e \`a chaque fois. Pour l'extraction on a besoin d'une fonction auxiliaire.

\vspace{2mm}
{\underline{\bf Fonction auxiliaire d'extraction}}.
\begin{itemize}
\item {\bf $lpath: T \rightarrow {\cal N}$ }. Abbrev.: {\bf $lp$ }. 
Byrd l'appelle la profondeur de r\'ecursion.
 $lp(v)$  est le nombre de n\oe uds sur le chemin de la racine au n\oe ud $v$. C'est donc la longueur du chemin de la racine au n\oe ud $+ 1$. $lp(\epsilon) = 1$. 
\end{itemize}

\vspace{2mm}
Comme la trace de Byrd (voir Annexe A), dans sa forme originale\footnote{Une description d\'etaill\'ee de la trace de Byrd originale est donn\'ee dans les annexes A et D. La trace actuelle originale correspondante ne dit rien sur l'\'evolution des clauses elles-m\^emes dans les bo\ii tes. Pour cette raison les informations sur les clauses sont omises dans l'\'etat actuel courant.}, se contente de d\'ecrire le parcours construction d'arbre dont les \nds sont \'etiquet\'es avec des pr\'edications et de donner, en cas de succ\`es, le squelette final complet d\'ecor\'e avec les \'etiquettes finales correctes, on obtient ainsi au final les instances de clauses utilis\'ees, sans n\'ecessairement savoir quelle clause a effectivement \'et\'e utilis\'ee \ag un \nd donn\'e.

Pour rendre compte des \'el\'ements propres \ag la trace de Byrd seulement (\'evolution de l'arbre ou des \bts et \'etiquettes) 4 param\`etres sont suffisants.
On prendra donc comme \'etat virtuel restreint les param\`etres suivants:
\begin{quote}
$Q = \{ T , u , num, pred \}$.
\end{quote}
Noter que l'on aurait pu ajouter les param\`etres $ct$ et $flr$. Mais
cela ne parait pas n\'ecessaire a priori car $ct$ est vrai (sauf au
premier \'ev\'enement de trace) ssi le premier ou le deuxi\`eme
attribut est $1$ (en fait ils le sont enesembles); et $flr$ devient
faux (\'echec) pour tout \ev de trace de port {\bf Fail}. En
particulier si on est \`a la racine, on sait alors si on est en
\'echec (\ev de port {\bf Fail} \ag la racine) ou en succ\`es (\ev de
port {\bf Exit} \ag la racine). On sait alors si on a \ag faire \ag un
arbre en \'echec ou un arbre de preuve complet (succ\`es).

\vspace{2mm}
L'\'etat initial $S_0/Q$ est donc: (voir l'\'etat complet \ag la section pr\'ec\'edente)
\begin{quote}
\noindent $\{ \{ \epsilon \} , \epsilon , \{ (\epsilon, 1)\}, \{ (\epsilon, goal)\}\}$
\end{quote}


\vspace{2mm}
La trace actuelle a 3 attributs et chaque \ev a la forme
\begin{verbatim}
         t    r    l    port    p 
\end{verbatim}
o\`u
\begin{itemize}
\item \verb.t. est le chrono.
\item \verb.r. est le num\'ero de cr\'eation du \nd  $u$ concern\'e par l'\'ev\'enement de trace, soit $nu(u)$.
\item \verb.l. est la profondeur dans l'arbre $T$ du \nd concern\'e, soit $lp(u)$.
\item \verb.port. est l'identificateur d'action ayant produit l'\'ev\'enement de trace ({\bf Call, Exit, Fail} ou {\bf Redo}).
\item \verb.p. est la pr\'edication associ\'ee au \nd concern\'e, soit $pd(u)$.
\end{itemize}

\vspace{1mm}

L'exemple 1 ci-dessous pr\'esente un programme et la trace extraite correspondant au but \verb,:-goal., ($u$ \nd courant)
\begin{verbatim}
c1: goal:-p(X),eq(X,b).
c2: p(a).
c3: p(b).
c4: eq(X,X).

:- goal. 
                    
chrono nu(u) lp(u)  port    pd(u)    Etat virtuel atteint

  1     1     1     Call    goal              S2
  2     2     2     Call    p(X)              S3
  3     2     2     Exit    p(a)              S4
  4     3     2     Call    eq(a,b)           S5
  5     3     2     Fail    eq(a,b)           S6
  6     2     2     Redo    p(a)              S7
  7     2     2     Exit    p(b)              S8
  8     4     2     Call    eq(b,b)           S9
  9     4     2     Exit    eq(b,b)           S10
 10     1     1     Exit    goal              S11
\end{verbatim}
L'exemple est d\'etaill\'e dans l'annexe B.

\vspace{2mm}
Le sch\'ema de trace est d\'ecrit \`a la figure\rf{tracegenfig}.
\begin{figure*}[ht]\small
\noindent
\reglecontrole{\callun{}}
{}
{<nu(u) \ \ lp(u)\ \ {\bf Call} \ \ pd(u)>}
{\{ \}}
\saut
\reglecontrole{\calldeux{}}
{}
{<nu(u) \ \ lp(u)\ \ {\bf Call} \ \ pd(u)> }
{\{\}}
 \saut
\reglecontrole{\exitun{}} 
{}
{<nu(u) \ \ lp(u)\ \ {\bf Exit} \ \ p>}
{\{p=pud(u)\}}
 \saut
\reglecontrole{\exitdeux{}}
{}
{<nu(u) \ \ lp(u)\ \ {\bf Exit} \ \ p> }
{ \{ p = pud(u) \}}
 \saut
\reglecontrole{\faildeux{}}
{}
{<nu(u) \ \ lp(u)\ \ {\bf Fail} \ \ pd(u)>}
{\{ \}}
\saut
\reglecontrole{\redoun{}}
{ v \gets gcp(u) }
{<nu(v) \ \ lp(v)\ \ {\bf Redo} \ \ pd(v)> }
{\{ \}}
 \saut
\reglecontrole{\redodeux{}}
{ v \gets gcp(u) }
{ <nu(v) \ \ lp(v)\ \ {\bf Redo} \ \ pd(v) >}
{\{ \}}
\caption{Sch\'ema de trace (fonction d'extraction de la Trace)}
\label{tracegenfig}
\end{figure*}

\vspace{2mm}
Afin de faciliter la lecture, toutes les informations non n\'ecessaires \`a l'extraction sont omises. 
En fait, un \'ev\'enement de trace est extrait lors de chaque transition de la SO, donc chaque \rg peut se lire aussi avec l'ensemble des param\`etres de l'\'etat virtuel. Ainsi par exemple pour la r\`egle \faildeux{}, 
le description compl\`ete de l'extraction ${\cal E}_{\faildeux{}}$ est:


\vspace{1mm}
\noindent
\reglecontroledeux{\faildeux{}}
{\neg fst(u) \wedge \neg ct \wedge \neg hcp(u) \et  v \gets pt(u)}
{u' \gets v \et (u = \epsilon) \Rightarrow (ct' \gets true) \et flr' \gets true}
{<nu(u) \ \ lp(u)\ \ {\bf Fail} \ \ pd(u)>}
{\{flr(u) \ \vee \  flr\}}

\vspace{1mm} On peut y observer clairement la remont\'ee ``directe''
dans l'arbre, suite \ag un \'echec (extraction d'\'ev\'enement de
port {\bf Fail}), jusqu'\ag ce qu'un point de choix puisse se trouver
dans le sous-arbre, ou jusqu\`a la racine de l'arbre sinon.

\section{Reconstruction d'une trace virtuelle restreinte}
\label{reconstruction}

On d\'ecrit maintenant la fonction de reconstruction ${\cal C}$ de la trace virtuelle restreinte, \`a partir d'un \'etat actuel initial et de la trace actuelle, ainsi que l'ad\'equation de la trace actuelle pour cet \'etat relativement \`a la trace virtuelle.

\vspace{1mm}
{\underline{\bf Fonction auxiliaire de reconstruction}}:

Pour reconstruire l'\'etat courant partiel, une fonction auxiliaire seulement est n\'ecessaire, \`a savoir la fonction inverse de $num$, not\'ee $node$. 
\begin{itemize}
\item {\bf $node: {\cal N} \rightarrow T$}. Abbrev.: {\bf $nd$}.
 Fonction inverse de $num$. $v= nd(n)$ est le n\oe ud de $T$ dont le rang de cr\'eation est $n$ (tel que $nu(v) = n$). Par d\'efinition $nd(nu(v))=v$ et $nu(nd(n))=n$.

\end{itemize}

\vspace{2mm}
Le sch\'ema de reconstruction est donn\'e dans la figure~\ref{traceregenefig} par la famille $\{{\cal C}_r | r \in R \}$.

Chaque \rg comporte en num\'erateur la condition d'identification de la \rg \ag partir de la trace, au d\'enominateur les calculs du nouvel \'etat virtuel restreint \ag partir des \evs de trace qui figurent entre accolades et, \'eventuellement, des param\`etres de l'\'etat virtuel restreint courant. Elles traduisent que

Si $Cond_r(e_t,e'_{t+1}) \ $ alors ${\cal C}_r(e_t,e'_{t+1},Q_t) = Q_{t+1}$.

Ces \rgs permettent en particulier de reconstruire pas \ag pas un arbre (ou de mani\`ere \'equivalente les bo\^ites encastr\'ees), son parcours construction re-construction, ainsi que les fonctions $num$ et $pred$. L'\'etat virtuel restreint comporte donc 4 param\`etres, \ag savoir $Q = S/Q = \{ T, u, num, pred \}$

\begin{figure*}[ht]\small
\noindent
\reglecontrole{\callun{}}
{r' = r}
{}
{\{< r \ \ l \ \ {\bf Call} \ \ p > \pv < r' >\}}
\saut
\reglecontrole{\calldeux{}}
{r' > r}
{u' \gets crc(nd(r)) \et T' \gets T \cup  \{u'\} \et nu'(u') \gets r' \et pd'(u') \gets p' }
{\{\\ \phantom{xxxxxxxxxxxxxxxxxxxxxxxxxxxxxxxxxxxxxxxxxxx} 
< r \ \ l \ \ {\bf Call} \ \ p > \pv <r' p'>\}}
 \saut
\reglecontrole{\exitun{}} 
{r' < r \vee u = \epsilon}
{u' \gets pt(u) \et pd'(u) \gets p}
{\{ < r \ \ l \ \ {\bf Exit} \ \ p > \pv < r' > \}}
 \saut
\reglecontroleshort{\exitdeuxshort{}}
{  r' > r \wedge u \not = \epsilon}
{u' \gets crnb(u) \et T' \gets T \cup  \{u'\} \et nu'(u') \gets r' \et pd'(u) \gets p \et pd'(u') \gets p'}
{\{ \\ \phantom{xxxxxxxxxxxxxxxxxxxxxxxxxxxxxxxxxxxx}
< r \ \ l \ \ {\bf Exit} \ \ p > \pv <r' \ \ p'> \}}
 \saut
\reglecontrole{\faildeux{}} 
{\,\,\,\,\,\,\,\,\,\,\,\,}
{u' \gets pt(u)}
{\{< r \ \ l \ \ {\bf Fail} \ \ p > \}}
 \saut
\reglecontrole{\redoun{}}
{r' = r}
{u' \gets nd(r) \et T' \gets T - \{y | y > u'\} }
{\{ < r \ \ l \ \ {\bf Redo} \ \ p > \pv < r' >\}}
 \saut
\reglecontroledeuxshort{\redodeuxshort{}}
{r' > r }
{v \gets nd(r) \et T' \gets T - \{y | y > v\}\cup \{u'\} \et u' \gets crc(v) \et}
{ nu' \gets upn(nu,v)\cup\{(u',r')\} \et pd' \gets upcp(pd,v)\cup\{(u',p')\}}
{\{ 
\\ \phantom{xxxxxxxxxxxxxxxxxxxxxxxxxxxxxxxxxxxx}
 < r \ \ l \ \ {\bf Redo} \ \ p > \pv < r' \ \ p' >\}}
 \saut
\caption{Reconstruction de la trace virtuelle restreinte (\md des \bts simplifi\'e) \ag partir de la trace actuelle}
\label{traceregenefig}
\end{figure*}

\vspace{2mm}
Cette trace exige de lire deux \'ev\'enements de trace successifs pour pouvoir \^etre comprise.

\vspace{2mm}
Il faut aussi remarquer qu'\`a partir du moment o\`u la trace est ad\'equate, et que l'on peut reconstituer ainsi le ``fonctionnement'' de la SO \ag partir de la trace, on peut rendre explicite une telle lecture de la trace en incluant dans les \rgs de reconstruction tous les param\`etres de la trace virtuelle.
A titre d'exemple, voici ce que donne la \rg de reconstruction  \calldeux{} avec tous les param\`etres.

\vspace{1mm}
\noindent
\reglecontrolecinqshort{\calldeuxshort{}}
{r' > r}
{v \gets crc(u) \et fst(u) \wedge lf(u) \wedge \neg ft(u) \wedge \neg ct}
{T' \gets T \cup  \{v\} \et u' \gets v \et n'= n+1 \et nu' \gets nu \cup \{(v,n')\} \et}
{ pd' \gets pd \cup \{(v,p')\} \et fst(u) \gets false \et fst' \gets fst \cup \{(v,true)\} \et flr' \gets false}
{\{ \\ \phantom{xxxxxxxxxxxxxxxxxxxxxxxxxxxxxxxxxxxx} < r \ \ l \ \ {\bf Call} \ \ p > \pv < r'\ \  p' > \}}

\vspace{1mm} Cette r\`egle indique que si apr\`es un \'ev\'enement de
port {\bf Call}, les num\'eros de bo\^{\i}tes croissent avec
l'\'ev\'enement de trace suivant ($r' > r$), alors c'est la \rg
\calldeux{} qui s'applique. Les conditions s'appliquant \ag l'\'etat
courant (sous-d\'enominateur du num\'erateur) sont alors
v\'erifi\'ees, un n\oe ud $v$ a \'et\'e cr\'e\'e, descendant du n\oe
ud courant $u$, et \'etiquet\'e avec la pr\'edication donn\'ee dans
l'\ev de trace suivant $p'$.
On sait \'egalement que l'arbre courant $T$  n'est pas complet et qu'il n'est pas en \'echec. Noter \'egalement que les conditions de la \rg utilis\'ee (sous-d\'enominateur du num\'erateur) sont toujours v\'erifi\'ees.

\vspace{1mm}
Sur l'exemple 1 de la section pr\'ec\'edente, cette \rg est utilis\'ee pour passer des \'etats $S_1$ \ag  $S_2$ (voir annexe B pour les d\'etails). Elle donne une lecture de la transition $S_1$ \ag $S_2$ avec les \'ev\'enements de trace de chrono 1 et 2.

\vspace{1mm}
\noindent
\reglecontrolecinq{\calldeux{}}
{2 > 1}
{1 = crc(\epsilon) \et fst(\epsilon) \wedge lf(\epsilon) \wedge \neg ft(\epsilon) \wedge \neg ct}
{T' = \{\epsilon,1\} \et u' = 1 \et n'= 2 \et nu' = \{(\epsilon,1),(1,2)\} \et}
{ pd' = \{(\epsilon,goal),(1,p(X))\} \et fst'=\{(\epsilon,false),(1,true)\} \et flr'= false}
{\{ \\ \phantom{xxxxxxxxxxxxxxxxxxxxxxxxxxxxxxx} < 1 \,\, 1 \,\, {\bf Call} \,\, goal > \pv <2 \,\, p(X)> \}}

\begin{figure}[h]
\begin{center}
\includegraphics[width=0.40\linewidth]{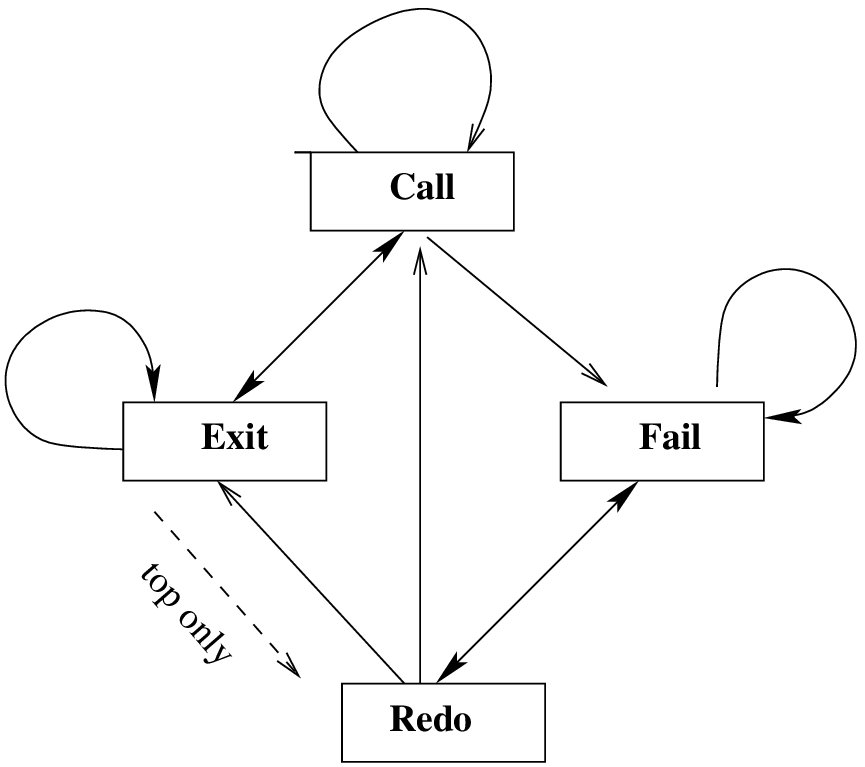}
\end{center}
\caption[Alg\`ebre des ports dans une trace de Byrd]{Alg\`ebre des ports dans une trace de Byrd (mod\`ele simplifi\'e)}
\label{figadequat00}
\end{figure}

\vspace{1mm}
La preuve compl\`ete de l'ad\'equation du sch\'ema de reconstruction pour $Q$ relativement \ag la SO est donn\'ee dans l'annexe C. 
Elle comporte trois parties: 
lemmes \'etablissant quelques propri\'et\'es g\'en\'erales de la SO (encha\^inement des \rgs et des ports, voir figure~\ref{figadequat00}); 
v\'erification de l'exclusivit\'e des conditions associ\'ees \ag chaque \rg du sch\'ema de reconstruction; enfin, pour chaque \rg de $R$, 
v\'erification que le sous-\'etat reconstruit est bien le m\^eme que l'\'etat virtuel restreint \ag $Q$ correspondant. 

\vspace{2mm}
A titre d'exemple les \'etapes de preuve sont illustr\'ees ci-dessous pour la \rg \calldeux{}, dont la figure~\ref{extranscall}
montre l'\'etat virtuel r\'esultant. L'\'etat virtuel restreint r\'esultant est alors:

\vspace{2mm}

\begin{figure*}[ht]\small
\noindent
\reglecontroledeux{\calldeux{}}
{fst(u) \wedge lf(u) \wedge \neg ct \wedge \neg ft(u)}
{T'= T \cup  \{u'\} \et u'= crc(u) \et n'= n+1 \et nu'(u')= n'  \et pd'(u')= p\_cpini(u') \et }
{ cl'(u)= upcp(u) \et cl'(u')= c\_cpini(u') \et fst'(u)=false \et fst'(u') = true\} \et flr'= false}
{\{ 
scs(u) \}}
\caption{R\`egle de transition \calldeux{} et \'etat $S'$ obtenu}
\label{extranscall}
\end{figure*}

\noindent
$S'/Q = \{T \cup  \{u'\} , u' = crc(u), nu'(u')= n' , pd'(u')= p\_cpini(u') \}$

\vspace{2mm}
L'\'ev\'enements de trace extrait, conform\'ement au sch\'ema de trace pour cette \rg, est:

\vspace{1mm}
${\cal E}_{\calldeux{}}(S,S') = < nu(u) \ \ lp(u) \ \ Call \ \ pd(u)>$

\vspace{1mm}
Il peut \^etre suivi d'un \ev de trace $e'$ qui contient les deux attributs suivants (on ne pr\'ecise pas les ports possibles, mais le diagramme de la figure~\ref{figadequat} montre que seuls des \evs avec des ports {\bf Exit} ou {\bf Call} sont possibles): $nd(u')= n'$ et $pd'(u')= p\_cpini(u')$, soit:

\vspace{1mm}
${\cal E}_{s}(S',S'') = < n' ... p\_cpini(u')>$ avec $u'= crc(u)$.

\vspace{2mm} 
On utilise alors la \rg correspondante du sch\'ema de
reconstruction, instanci\'ee avec les \evs de trace $e$ et $e'$.

\vspace{2mm}
\reglecontroledeuxshortter{\calldeuxshortter{}}
{Cond_{\calldeux{}}(e,e')}
{u'= crc(nd(nu(u))) \et T'= T\cup\{u'\} \et nu'(u')= n'\et }
{pd'(u')= p\_cpini(u')}
{\{ 
e \pv e'\}}
\saut

On v\'erifie que la condition discriminant la \rg \calldeux{} est bien v\'erifi\'ee:

\noindent
$Cond_{\calldeux{}}(e,e') = (nu'(crc(u)) > nu(u))$, soit $n' > n$. En effet tout nouveau n\oe ud cr\'e\'e l'est avec un num\'ero sup\'erieur \ag tous ceux d\'ej\`a existants.

\vspace{2mm}
Enfin l'\'etat $Q'$ reconstruit \ag l'aide de la \rg du sch\'ema de reconstruction est bien identique \ag l'\'etat virtuel restreint \ag $Q$.

\vspace{1mm}
\noindent
$Q' = \{T \cup \{u'\}, u'= crc(u), nu'(u') = n', pd'(u')= p\_cpini(u') \} = S'/Q$

\vspace{2mm}
Il r\'esulte de l'ad\'equation du sch\'ema de reconstruction que la lecture de la trace peut se faire en utilisant une partie quelconque de l'\'etat courant virtuel (une fois identifi\'ee la \rg qui s'applique), ce qui en simplifie consid\'erablement la compr\'ehension. 

Ainsi on peut ``voir'' sur les r\`egles de la
figure~\ref{traceregenefig} (\'evolution de l'\'etat restreint $Q$)
comment l'arbre de preuve partiel \'evolue, ou comment se fait le
parcours dans les bo\^{\i}tes. Par exemple, il est assez clair qu'une
succession de {\bf Exit} jusqu'\`a la racine de l'arbre (bo\^{\i}te $r
= 1$) va permettre d'obtenir un arbre de preuve complet dont toutes
les pr\'edications associ\'ees aux n\oe uds auront \'et\'e mises \`a
jour conform\'ement \`a la s\'emantique attendue de la r\'esolution
(non d\'ecrite ici); c'est \`a dire que l'on aura obtenu une preuve du
but associ\'e \`a la racine en utilisant toutes les instances de
clauses correspondant aux pr\'edications associ\'ees \`a chaque n\oe
ud et leurs descendants\footnote{Dans ce mod\`ele, si des clauses
  diff\'erentes peuvent avoir des instances identiques, on ne saura
  pas n\'ecessairement quelle clause a \'et\'e effectivement
  utilis\'ee.} et dont les pr\'edications correspondantes sont
associ\'ees aux \ev de trace de port {\em exit}. 

Ainsi l'examen exclusif de tous les \evs de trace de port {\em exit}
permet de reconstituer les arbres de preuve obtenus, partiels ou
complets.

\section{Conclusion sur le mod\`ele des bo\^{\i}tes}
\label{commdisc}

Nos premi\`eres observations porteront sur la compr\'ehension de la trace que donnent les r\`egles de la figure~\ref{traceregenefig}. Celles-ci peuvent se comprendre en effet sans avoir recours \`a la SO compl\`ete, mais en se limitant \ag un \'etat restreint (d\'enot\'e $Q$). Tout ce qui est n\'ecessaire y est formalis\'e, le recours \`a la SO n'\'etant utile que pour aller plus avant dans la compr\'ehension. Les r\`egles en donnent le squelette dynamique (parcours construction re-construction d'arbre) et leurs conditions optionnelles associ\'ees (toujours valides pour la reconstruction d'une trace actuelle produite avec la SO) donnent l'interpr\'etation imm\'ediate des attributs de la trace.

Cette approche met aussi imm\'ediatement en \'evidence les difficult\'es d'interpr\'etation d'un tel mod\`ele. Nous en retiendrons deux. 
En premier lieu on observera que s'il est normal que
l'interpr\'etation de la trace n\'ecessite d'appr\'ehender l'ensemble
de la trace depuis le d\'ebut (pour avoir une id\'ee de l'\'etat de la
r\'esolution), il l'est moins que la lecture d'un \ev ``en avant''
soit n\'ecessaire, ce qui est en soi un facteur de difficult\'e. Ceci
pourrait \^etre \'evit\'e si une information sur la clause utilis\'ee
figurait dans un attribut\footnote{Dans les conditions en effet, les
  facteurs discriminant les \rgs utilis\'ees portent sur la nature des
  clauses.} (par exemple la clause choisie avant un \ev de port {\bf
  Call}). La repr\'esentation avec des \bts avait essentiellement pour
objectif de ``contenir'' les clauses potentiellement utiles. On ne les
retrouve plus dans la trace, ce qui retire au mod\`ele une grande
partie de son int\'er\^et, en le limitant de fait \`a la seule
description d'un parcours d'arbre.

En deuxi\`eme remarque on observera a contrario que la trace contient
un attribut inutile. La profondeur (attribut {\tt l}) ne contribue
finalement pas \`a la compr\'ehension de la trace et la surcharge
inutilement. En fait la profondeur pourrait contribuer \`a la
compr\'ehension de l'arbre de preuve partiel en la combinant avec un
codage ad\'equat des \nds. Ce choix est fait par exemple dans la trace
de \gprolog\ \cite{gnuprolog} o\`u les \nds sont cod\'es, non par leur
ordre de cr\'eation, mais par leur rang dans l'arbre. La combinaison
des deux attributs permet alors un rep\'erage direct dans l'arbre $T$
du \nd courant. Ce choix constitue bien une am\'elioration de la trace
originale\footnote{Beaucoup de travaux introduisent des visualisations
  de la trace avec indentation et utilisent l'attribut $lpath$ pour ce
  faire. Cela montre que cet attribut a une utilit\'e pratique, mais
  il n'est pas utile \ag la reconstruction.}.

\vspace{3mm}
Les quelques articles cit\'es dans l'introduction traduisent la recherche permanente d'am\'eliorations de la compr\'ehension du contr\^ole et aussi de l'unification. Ainsi \cite{boiz84} (1984) \cite{NumFuji85} (1985) proposent des am\'eliorations de la trace de Byrd avec un nombre d'\ev plus r\'eduit, apportant ainsi une vision plus synth\'etique de l'arbre parcouru, et ils proposent \'egalement de nouveaux ports concernant l'unification et le choix des clauses. \cite{ToBe93} (1993) introduit explicitement une alg\`ebre de \bts avec graphiques \`a l'appui, mais ce mod\`ele qui veut saisir tous les aspects de la r\'esolution reste assez complexe. \cite{jahier00} (2000) propose une s\'emantique de trace fond\'ee sur une s\'emantique d\'enotationnelle de Prolog. L'inconv\'enient principal est que la compr\'ehension de la trace passe par une bonne compr\'ehension d'un mod\`ele complet de Prolog, synth\'etique mais n\'ecessitant une certaine familiarit\'e avec les continuations. L'article \cite{kulas03} (2003) rel\`eve d'une d\'emarche analogue, mais celle-ci s'appuie directement sur les 
ports dont les encha\ii nements possibles constituent son squelette. Le r\'esultat est \'egalement que la compr\'ehension de la trace passe par l'assimilation d'une s\'emantique relativement complexe de Prolog qui s'apparente plus \`a une s\'emantique bas\'ee sur les ``magic sets'' qu'\`a une explication directe de la trace.

\vspace{2mm} Ces \'etudes montrent que l'on a beaucoup cherch\'e \`a
am\'eliorer les moyens de comprendre la r\'esolution. Au fil du temps
les travaux se sont concentr\'es sur des m\'ethodes d'analyse et de
visualisation de plus en plus complexes (par exemple \cite{Opium:JLP}
pour l'analyse  des traces Prolog) pour des formes de
r\'esolution elles aussi de plus en plus complexes comme la
r\'esolution de CSP \cite{oadymppac}. Il n'en reste pas moins
cependant que la trace de Byrd reste la base des traceurs pour les
syst\`emes de r\'esolution et ses fameux ports inspirent encore, de
temps en temps, les chercheurs.

Dans cet exemple on a trait\'e une instance particuli\`ere du \md des
bo\ii tes. Il serait int\'eressant, et ce sera notre prochaine
\'etape, d'obtenir un \md plus g\'en\'erique susceptible d'engendrer
potentiellement diverses implantations connues de ce mod\`ele. Cela est
possible avec l'approche pr\'esent\'ee ici. Une premi\`ere description
de diff\'erents mod\`eles est faite dans l'annexe D, avec une SO
propos\'ee \ag l'annexe E.

\section{Conclusion g\'en\'erale}
\label{conclusion}

\vspace{2mm} Le point essentiel de ce rapport est l'illustration d'une
approche originale pour donner une s\'emantique \`a des traces
d'ex\'ecution.
L'exemple utilis\'e ici a essentiellement un caract\`ere anecdotique,
m\^eme si, in fine, le r\'esultat est sans doute une formalisation
compl\`ete parmi les plus simples (car restreinte aux seuls
\'el\'ements n\'ecessaires \`a sa compr\'ehension) que l'on ait pu
formuler juqu'\`a pr\'esent d'un mod\`ele des bo\^ites de Byrd.

La notion de trace virtuelle a pour but de capturer l'id\'ee du ``bon'' niveau d'observation d'un processus physique. M\^eme pour un programme, le bon niveau d'observation n'est pas \'evident. Quels sont les \'el\'ements significatifs ou utiles \`a observer? Toute ex\'ecution d'un programme met en \oe uvre une s\'erie de couches de logiciels jusque dans les composants mat\'eriels. Certaines erreurs peuvent m\^eme provenir d'interf\'erences de particules \'energ\'etiques avec des composants \'electroniques. Le ``bon'' niveau d'observation ne peut donc \^etre d\'efini de mani\`ere absolue. Toute trace virtuelle ne peut \^etre dite int\'egrale que si l'on se fixe une limite a priori quant \`a la granularit\'e du ph\'enom\`ene observ\'e (mais penser que l'on puisse atteindre un niveau ``ultime'' de description rel\`everait d'une approche excessivement r\'eductionniste). Dans le cas d'un langage de programmation,
le niveau d'observation sera usuellement d\'efini par le langage
lui-m\^eme, ne serait-ce que pour des raisons \'evidentes de capacit\'e
de compr\'ehension (celui que l'auteur du programme est seul \`a
m\^eme d'appr\'ehender).

Le point important ici est que le niveau d'observabilit\'e est en fait arbitraire et qu'en aucun cas le niveau choisi ne peut \^etre consid\'er\'e comme ultime. Il est donc normal que pour un niveau d'observation donn\'e, on soit oblig\'e de tenir compte dans la description, aussi pr\'ecise soit-elle, d'\'el\'ements externes \`a celle-ci. C'est pourquoi la SO constitue un mod\`ele \`a la fois ind\'ependant d'un processus particulier observ\'e (c'est en ce sens qu'elle est ``g\'en\'erique''), mais \'egalement comportant des r\'ef\'erences \ag des aspects non formellement d\'ecrits 
associables aux processus que l'on souhaite observer.

\bibliographystyle{plain}

\bibliography{toutenun,pierre}

\appendix

\clearpage

\section{ANNEXE: Le \md de Byrd}
\label{appendixB}

\begin{figure}[h]
\begin{center}
\includegraphics[width=0.8\linewidth]{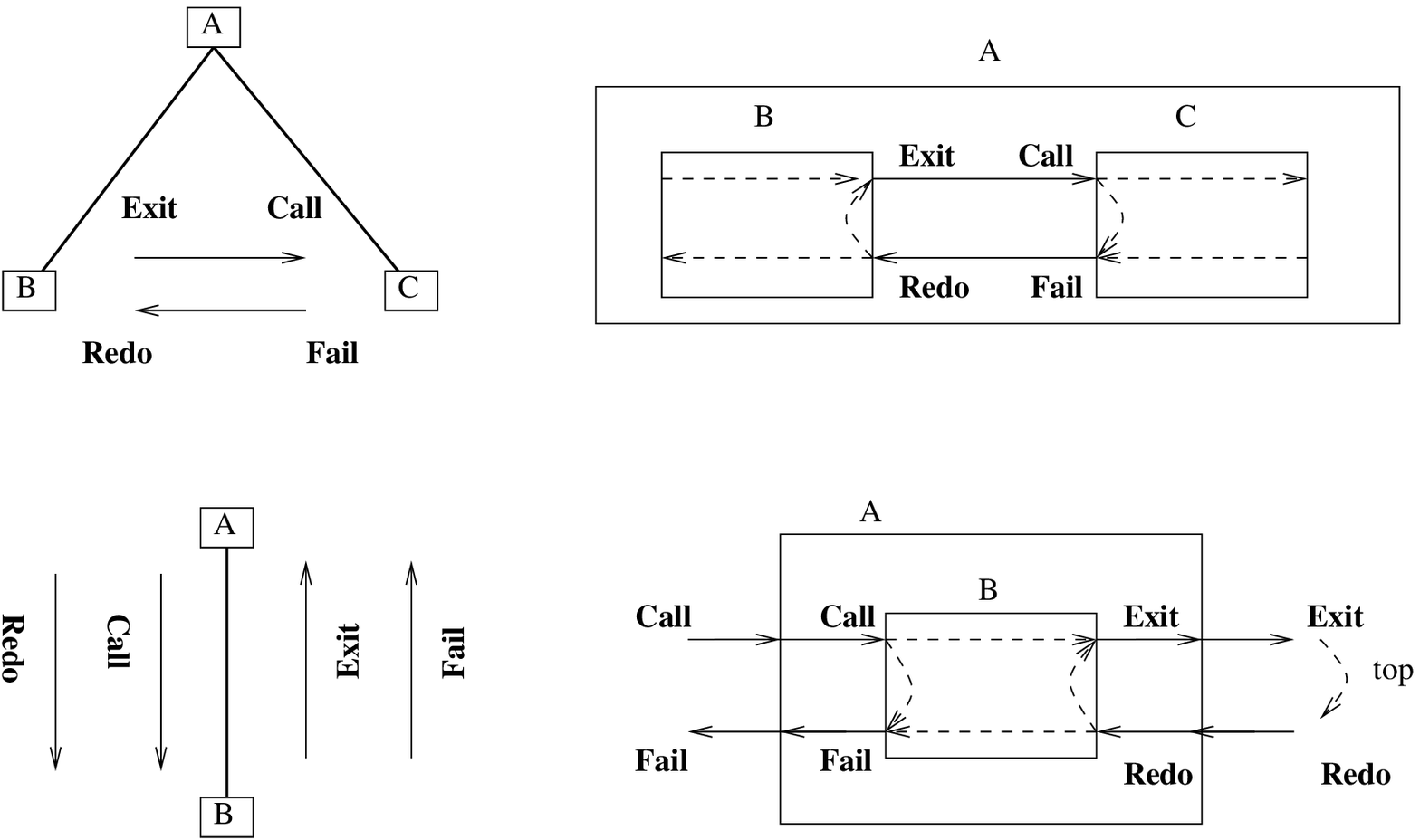}
\end{center}
\caption[Correspondance des repr\'esentations arbres /\bts: combinaisons possibles des ports (\bts adjacentes ou ench\^ass\'ees)]{Correspondance des repr\'esentations arbres /\bts: combinaisons possibles des ports (\bts adjacentes ou ench\^ass\'ees)}
\label{fig:arbbox}
\end{figure}

On d\'ecrit ici le \md de Byrd avec ses repr\'esentations possibles en utilisant des \bts ou des arbres.
La figure~\ref{fig:arbbox} montre la mani\`ere dont les \bts se combinent donnant une sorte d'alg\`ebre des ports. On indique \'egalement la correspondance graphique avec la repr\'esentation sous forme d'arbre de la combinaison des bo\ii tes.

Il y a deux points de vue possibles: avec \bts ou avec arbres. Les \bts en effet sont encastrables les unes dans les autres et peuvent donc \^etre repr\'esent\'ees \ag la mani\`ere d'un treemap\footnote{Un treemap est une mani\`ere de repr\'esenter un arbre, au d\'epart repr\'esent\'e en deux dimensions verticales (mani\`ere adopt\'ee ici), dans un plan horizontal. La racine est un grand rectangle. Celui-ci est subdivis\'e en autant de parties qu'il y a d'enfants. Chaque enfant est \`a son tour subdivis\'e \ldots\ etc. Afin de pouvoir mettre en \'evidence les communications (ports) entre descendants de m\^eme niveau, la partition d'une ``\bt'' est en fait une juxtaposition de sous-\bts mises dans le m\^eme ordre que les \nds correspondants de l'arbre. Il en est de m\^eme pour \bts encastr\'ees dont aucun contour n'est jointif afin de mettre en \'evidence les sous-bo\ii tes et les ports liant les descendants.}. Les deux repr\'esentations sont isomorphes. La trace de Byrd consiste \`a tracer syst\'ematiquement les passages d'un port \`a l'autre des \bts, tout en respectant un ordre de visite fix\'e a priori et en refaisant dans l'ordre inverse tout le parcours d\'ej\`a effectu\'e afin de trouver d'autres solutions possibles. Cet ordre est repr\'esent\'e par des fl\`eches dont la succession est limit\'ee \`a certaines combinaisons ainsi qu'il est indiqu\'e sur la figure.

Comme la repr\'esentation par arbres ou par \bts est isomorphe, les parcours possibles sont \'egalement isomorphes. Du point de vue des \bts cela signifie que l'on ne peut sortir d'une \bt qu'en y \'etant rentr\'e d'abord et r\'eciproquement et toute travers\'ee gauche droite compl\`ete d'une \bt doit (trajet aller) \^etre suivie d'un trajet retour. Seuls les parcours externes sont trac\'es. L'effet recherch\'e par Byrd \'etait une vision globale imm\'ediate du non d\'eterminisme des solution en utilisant une repr\'esentation syst\'ematique structur\'ee avec une \'evolution d\'eterministe.

Du point de vue des arbres, chaque \bt est repr\'esent\'ee par un n\oe ud; les parcours entre \bts de m\^eme niveau correspondent aux parcours entre \nds voisins de m\^eme niveau dans l'arbre; et les parcours entre descendants correspondent \ag des parcours entre \bts encastr\'ees. Le parcours contraint des \bts correspond dans l'arbre \ag des parcours de visite descendant gauche-droite mais avec la particularit\'e qu'on ne peut remonter dans l'arbre, suite \`a un \'echec, que par la premi\`ere branche visit\'ee (i.e. la branche la plus \ag gauche).

\begin{figure}[h]
\begin{center}
\includegraphics[width=0.40\linewidth]{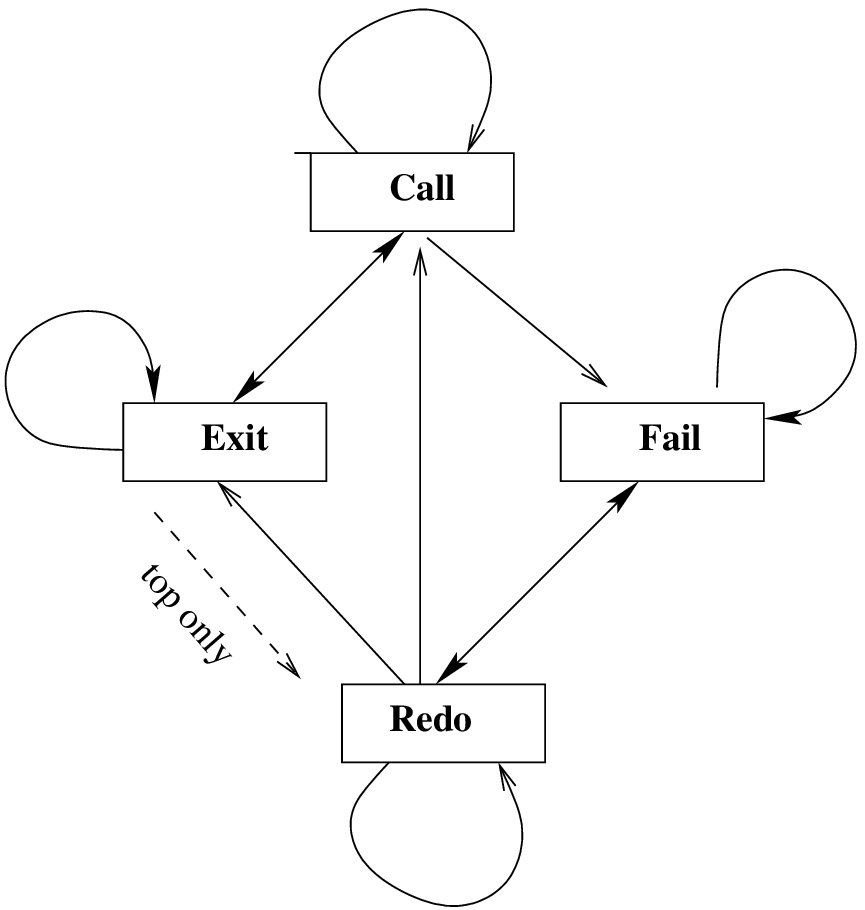}
\end{center}
\caption[Alg\`ebre des ports dans une trace de Byrd]{Alg\`ebre des ports dans une trace de Byrd}
\label{figadequat}
\end{figure}

Il r\'esulte de cette repr\'esentation que les ports ne peuvent se suivre que selon certaines combinaisons.
La figure~\ref{figadequat} montre les combinaisons de ports possibles qui r\'esultent de cette alg\`ebre.

La trace de Byrd originale peut \^etre produite en instrumentant le programme avec le m\'eta-interpr\`ete suivant propos\'e par Byrd \cite{byrd80}. Il est int\'eressant d'observer que malgr\'e la simplicit\'e du m\'eta-programme qui instrumente tout programme afin de le tracer, sa compr\'ehension\footnote{Le m\'eta-interpr\`ete utilise les pr\'edicats syst\`eme suivants: {\em call}, ex\'ecution d'une pr\'edication, {\em fail}, \'echec provoqu\'e (suivi d'actions de retour arri\`ere), {\em \pv}, d\'enotant la disjonction. Le pr\'edicat {\em display} est ici une \'evocation d'\'ecriture de trace sur une sortie standard. Seuls les ports sont trac\'es ici. La production de la trace actuelle compl\`ete de la section~\ref{generation} est simple \ag programmer, mais d\'etruit la clart\'e de cette pr\'esentation du fait de la n\'ecessit\'e d'instrumenter, dans le m\'eta-programme, tous les programmes avec un argument suppl\'ementaire correspondant \ag la profondeur.} n\'ecessite une excellente connaissance de la s\'emantique op\'erationnelle des interpr\`etes Prolog. La s\'emantique d'une telle implantation du traceur est donc loin d'\^etre \'evidente.

\begin{verbatim}
:- trace(goal).

goal:- trace(p(X)), trace(eq(X,b)).
p(a).
p(b).
eq(X,X).

trace(Pred) :-  display('Call',Pred),
               (call(Pred) ; (display('Fail',Pred), fail)),
               (display('Exit',Pred) ; (display('Redo',Pred), fail)).
\end{verbatim}


\clearpage

\section{ANNEXE: Exemple simple}
\label{appendixA}

On illustre ici la SO de la section~\ref{semobs}, l'extraction et la reconstruction de la trace avec un exemple simple de programme. La trace obtenue est la m\^eme pour les diff\'erents mod\`eles de traces consid\'er\'es ici (voir Annexes suivantes D et E).

\vspace{1mm}
{\bf Exemple 1}

\begin{verbatim}
Programme:

c1: goal:-p(X),eq(X,b).
c2: p(a).
c3: p(b).
c4: eq(X,X).

:- goal. 
                    
chrono nu(u) lp(u)  port    pd(u)    Etat virtuel atteint

  1     1     1     Call    goal              S2
  2     2     2     Call    p(X)              S3
  3     2     2     Exit    p(a)              S4
  4     3     2     Call    eq(a,b)           S5
  5     3     2     Fail    eq(a,b)           S6
  6     2     2     Redo    p(a)              S7
  7     2     2     Exit    p(b)              S8
  8     4     2     Call    eq(b,b)           S9
  9     4     2     Exit    eq(b,b)           S10
 10     1     1     Exit    goal              S11
\end{verbatim}

\vspace{1mm}
La trace est celle d\'efinie dans la section~\ref{generation}, figure~\ref{tracegenfig}.

Elle a la forme suivante (le ``\nd concern\'e'' correspond au \nd courant vu du point de vue de la trace).
\begin{verbatim}
         t    r    l   port  p
\end{verbatim}
o\`u
\begin{itemize}
\item \verb.t. est le chrono.
\item \verb.r. est le num\'ero de cr\'eation du \nd  $u$ concern\'e par l'\'ev\'enement de trace, soit $nu(u)$ ($v$ au lieu de $u$ pour les \rg de port {\bf Redo}).
\item \verb.l. est la profondeur dans l'arbre $T$ du \nd concern\'e, soit $lp(u)$ ($lp(\epsilon) = 1$).
\item \verb.port. est l'identificateur d'action ayant produit l'\'ev\eg nement de trace ({\bf Call, Exit, Fail} ou {\bf Redo}).
\item \verb.p. est la pr\'edication associ\'ee au \nd concern\'e, soit $pd(u)$ [ou $pud(u)$ (\'ev\'enements de port {\bf Exit})].
\end{itemize}
L'identificateur d'\'ev\'enement de trace est omis ici, le chrono en tenant lieu.

\vspace{2mm}
L'\'etat courant a la forme: $\{ T , u , n, nu, pd , cl , fst, ct , flr\}$.

\vspace{1mm}
Les param\`etres sont:

\noindent
\begin{enumerate}
\item {\bf $T$}: $T$ est l'arbre courant.
\item {\bf $u \in T$}:  $u$ est le n\oe ud courant dans $T$ (\bt visit\'ee).
\item {\bf $n \in {\cal N}$}:  $n$ est le num\'ero de cr\'eation du dernier \nd cr\'e\'e.
\item {\bf $nu: T \rightarrow {\cal N}$}:  $nu(u)$ est le num\'ero de cr\'eation de $u$ dans $T$.
\item {\bf $pd: T \rightarrow {\cal H}$}:  $pd(u)$ est la pr\'edication associ\'ee au \nd $u$.
\item {\bf $cl: T \rightarrow 2^P$}:  $cl(u)$ est une liste de clauses du programme trac\'e d\'efinissant le pr\'edicat de $pred(u)$.
\item {\bf $fst: T \rightarrow Bool$}:  $fst(u)$ est vrai ssi $u$ n'a pas encore \'et\'e  visit\'e.
\item {\bf $ct \in Bool$}:  $ct$ vrai ssi le sous-arbre de racine $u$ a \'et\'e compl\`etement visit\'e.
\item {\bf $flr \in Bool$}:  $flr$ est vrai si le sous-arbre de racine $u$ est en \'echec.
\end{enumerate}

On ommet $S_0$ et la transition $S_0$ vers $S_1$ (action ``top level'') et l'\'etat ``initial'' consid\'er\'e est donc l\'etat $S_1$.

La trace actuelle est ad\'equate pour l'\'etat $Q = \{ T , u , nu, pd \}$. On a besoin de la fonction $nu$, son inverse $nd$ \'etant utilis\'e pour la reconstruction.

\vspace{2mm}
L'exemple est donn\'e int\'egralement: pour chaque transition d\'ecrite (de $t=1$ \ag $10$) \`a partir de l'\'etat $S_t$, on donne: la \rg de transition, l'\'etat $S_{t+1}$ obtenu, l'\'ev\'enement de trace extrait et l'\'etat $Q_{t+1}$ reconstruit, dont l'identit\'e avec la partie $Q$ de $S_{t+1}$ peut \^etre v\'erifi\'ee.

\vspace{3mm}
{\bf Etat ``initial''}:

\noindent {\bf $S_1$}: $\{ \{ \epsilon \} , \epsilon , 1 , \{(\epsilon, 1)\} , \{(\epsilon, goal)\} , \{(\epsilon, [c1])\} , \{(\epsilon, true)\}, false, false\}$

\vspace{2mm}
Sa restriction \ag $Q$ est:

\noindent {\bf $S_1/Q$}: $\{ \{ \epsilon \} , \epsilon , \{(\epsilon, 1)\} , \{(\epsilon, goal)\}\}$

\vspace{2mm}
\underline{\bf De $S_1$ \ag $S_2$}

Seule la r\`egle (1) \calldeux{} s'applique (premi\`ere visite du \nd courant). On vient de ``rentrer'' dans la \bt \verb.goal. et la premi\`ere clause but ({\verb.c1.}) n'est pas un fait. Un descendant est construit dont l'\'etiquette $p(X)$ n'apparaitra que dans l'\ev de trace suivant.

\vspace{2mm}
\noindent
(1) \reglecontroledeux{\calldeux{}}
{fst(\epsilon) \wedge leaf(\epsilon) \wedge \neg ct \wedge \neg fact(c1) \et v \gets 1}
{T' = \{\epsilon, 1\} \et u' = 1 \et n'= 2 \et nu' = \{(\epsilon, 1),(1,2)\} \et pd' = \{(\epsilon, goal)(1,p(X))\} \et }
{ cl' = \{(\epsilon, [])(1,[c2,c3])\} \et fst' = \{(\epsilon, false),(1,true)\} \et flr' = false}
{\{ \\ \phantom{xxxxxxxxxxxxxx} scs(\epsilon) \et ([c2,c3],p(X)) = cpini(1)\}}

\vspace{2mm}
\noindent
{\bf $S_2$}: $\{ \{ \epsilon, 1 \} , 1 , 2 , \{(\epsilon, 1),(1, 2)\} , \{(\epsilon, goal),(1,p(X))\},$

\indent
$\ \ \ \ \ \ \ \ \{ (\epsilon, []),(1 ,[c2,c3])\}, \{(\epsilon, false),(1,true)\}, false, false\}$

\vspace{1mm}
L'\'ev\'enement de trace extrait est (r\`egle d'extraction \calldeux{} de la section~\ref{generation}):

\verb.< 1    1   1  Call  goal >. (suivi de \verb.< 2   2   2  Call  p(X) >.).

\vspace{1mm}
L'\'etat reconstruit est (r\`egle de reconstruction \calldeux{} de la section~\ref{reconstruction}):

\noindent
{\bf $Q_2 = S_2/Q$}: $\{ \{ \epsilon, 1 \} , 1 , \{(\epsilon, 1),(1, 2)\} , \{ (\epsilon, goal),(1,p(X))\}\}$

\vspace{1mm}
Pour la g\'en\'eration de $Q_2$, on utilise les informations $<r'\ \  p'>$ du second \ev de trace.

\vspace{2mm}
\underline{\bf De $S_2$ \ag $S_3$}

Seule la r\`egle (2) \callun{} s'applique (la premi\`ere clause utilis\'ee ({\verb.c2.}) est un fait), on entre dans la \bt \verb.p..

\vspace{2mm}
\noindent
(2) \reglecontrole{\callun{}}
{first(1) \wedge leaf(1) \wedge \neg ct \wedge fact(c2)}
{cl' = \{ (\epsilon, []), (1 ,[c3]) \} \et fst' = \{(\epsilon, false),(1,false)\}}
{\{\}}

\vspace{2mm}
\noindent
{\bf $S_3$}: $\{ \{ \epsilon, 1 \} , 1 , 2 , \{ (\epsilon, 1),(1, 2)\} , \{ (\epsilon, goal),(1,p(X))\},$

\indent
$\ \ \ \ \ \ \ \ \{ (\epsilon, []),(1,[c3]) \}, \{(\epsilon, false),(1,false)\}, false, false\}$

\vspace{1mm}
L'\'ev\'enement de trace extrait est (r\`egle d'extraction \callun{} de la section~\ref{generation}):

\verb.< 2     2    2  Call  p(X) >. 

\vspace{1mm}
L'\'etat reconstruit est (r\`egle de reconstruction \callun{} de la section~\ref{reconstruction}):

\noindent
{\bf $Q_3 = S_3/Q$}: $\{ \{ \epsilon, 1 \} , 1 , \{(\epsilon, 1),(1, 2)\} , \{ (\epsilon, goal),(1,p(X))\}\}$

\vspace{1mm}
mais l'\ev de trace laisse l'\'etat courant actuel invariant: $Q_3 = Q_2$.

\vspace{2mm}
\underline{\bf De $S_3$ \ag $S_4$}

La r\`egle (3) \exitdeux{} s'applique car la pr\'edication appel\'ee $p(X)$ est un succ\`es et devient $p(a)$, et peut avoir un fr\`ere (existence d'une deuxi\`eme pr\'edication dans le corps de la clause \verb.c1. utilis\'ee). Un \nd est cr\'e\'e dont l'\'etiquette est $eq(a,b)$.

\vspace{2mm}
\noindent
(3) \reglecontroledeuxshort{\exitdeuxshort{}}
{ \neg fst(1) \wedge mhnb(1) = true \wedge \neg ct \wedge \neg flr \et v \gets 2}
{T'= \{ \epsilon, 1,2\}\et u'= 2\et n'= 3\et nu'= \{(\epsilon, 1),(1, 2),(2,3)\}\et pd'= \{(\epsilon, goal),(1,p(a)),(2,eq(a,b))\}\et }
{ cl'= \{(\epsilon, []),(1,[c3]),(2,[c4])\} \et  fst'= \{(\epsilon, false),(1,false),(2,true)\}}
{ \{ \\
\phantom{xxxxxxxxxxxxxxxxxxxxxx} scs(1) \et p' = p(a) \et ([c4], eq(a,b)) = cpini(2) \}}

\vspace{2mm}
\noindent
{\bf $S_4$}: $\{ \{ \epsilon,1,2 \} , 2, 3 , \{(\epsilon,1),(1, 2),(2,3)\} , \{ (\epsilon, goal),(1,p(a)),(2,eq(a,b))\},$

\indent
$\ \ \ \ \ \ \ \ \{ (\epsilon, []),(1,[c3]),(2,[c4])\}, \{(\epsilon,false),(1,false),(2,true)\}, false, false\}$

\vspace{1mm}
L'\'ev\'enement de trace extrait est (r\`egle d'extraction \exitdeux{} de la section~\ref{generation}):

\verb.< 3   2   2  Exit  p(a) >. (suivi de \verb.< 4   3   2  Call  eq(a,b) >.).

\vspace{1mm}
L'\'etat actuel reconstruit est (r\`egle de reconstruction \exitdeux{} de la section~\ref{reconstruction}):

\noindent
{\bf $Q_4 = S_4/Q$}: $\{ \{ \epsilon, 1,2 \} , 2 , \{(\epsilon, 1),(1, 2),(2,3)\} , \{ (\epsilon, goal),(1,p(a)),(2,eq(a,b))\}\}$

\vspace{1mm}
mais sa reconstruction n\'ecessite de conna\ii tre l'\ev de trace suivant.

\vspace{2mm}
\underline{\bf De $S_4$ \ag $S_5$}

La r\`egle (4) \callun{} est la seule qui s'applique (premi\`ere visite du n\oe ud).

\vspace{2mm}
\noindent
(4) \reglecontroleshort{\callunshort{}}
{first(3) \wedge leaf(3) \wedge \neg ct \wedge fact(3)}
{cl' = \{(\epsilon, []) (1 ,[c3]),(2,[])\}, fst' = \{(\epsilon,false),(1,false),(2,false)\}}
{\{ \}}

\vspace{2mm}
\noindent
{\bf $S_5$}: $\{ \{ \epsilon,1,2\} , 2 , 3 , \{(\epsilon,1),(1, 2),(2,3)\} , \{ (\epsilon, goal),(1,p(a)),(2,eq(a,b))\},$

\indent
$\ \ \ \ \ \ \ \ \{ (\epsilon, []),(1,[c3]),(2,[])\}, \{(\epsilon,false),(1,false),(2,false)\}, false, false\}$

\vspace{1mm}
L'\'ev\'enement de trace extrait est (r\`egle d'extraction \callun{} de la section~\ref{generation}):

\verb.< 4   3   2  Call  eq(a,b) >. (suivi de \verb.< 5   3   2  Fail  eq(a,b) >.)

\vspace{1mm}
L'\'etat actuel reconstruit est (r\`egle de reconstruction \callun{} de la section~\ref{reconstruction}) est invariant ($Q_5 = Q_4$):

\noindent
{\bf $Q_5 = S_5/Q$}: $\{ \{ \epsilon,1,2 \} , 2 , \{(\epsilon, 1),(1, 2),(2,3)\} , \{ (\epsilon, goal),(1,p(a)),(2,eq(a,b))\} \}$


\vspace{2mm}
\underline{\bf De $S_5$ \ag $S_6$}

L'unification de $eq(a,b)$ et $eq(X,X)$ (clause \verb.c4.) \'echoue, le \nd est donc en \'echec ($flr(2)=true$) et seule la \rg (5) \failun{} s'applique (en effet ce n'est ni une premi\`ere visite, ni un succ\`es, ni un \'echec en cours ($flr = false$), ni une visite de la racine de l'arbre ($ct = false$)).

\vspace{2mm}
\noindent
(5) \reglecontroleshort{\faildeuxshort{}}
{\neg first(2) \wedge \neg ct }
{ u' = \epsilon \et ct' = true \et flr' = true}
{\{flr(2)(=true) \vee flr(=false)\}}

\vspace{2mm}
\noindent
{\bf $S_6$}: $\{ \{ \epsilon,1,2\} , \epsilon , 3 , \{(\epsilon,1),(1, 2),(2,3)\} , \{(\epsilon, goal),(1,p(a)),(2,eq(a,b)\},$

\indent
$\ \ \ \ \ \ \ \ \{ (\epsilon, []),(1,[c3]),(2,[])\}, \{(\epsilon,false),(1,false),(2,false)\}, true, true\}$

\vspace{1mm}
Le \nd courant devient la racine, l'arbre complet en cours est donc un arbre d'\'echec et la raison de l'\'echec est la pr\'edication $eq(a,b)$.

L'\'ev\'enement de trace extrait est (r\`egle d'extraction \faildeux{} de la section~\ref{generation}):

\verb.< 5     3    2  Fail  eq(a,b) >.

\vspace{1mm}
L'\'etat reconstruit (r\`egle de reconstruction \faildeux{} de la section~\ref{reconstruction}) est: (on aurait pu en fait ajouter le param\`etre $flr$ dans l'\'etat restreint pour rendre compte de l'\'etat d'\'echec de l'arbre; ici seul l'\ev de trace porte cette information).

\noindent
{\bf $Q_6 = S_6/Q$}: $\{ \{ \epsilon,1,2 \} , \epsilon , \{(\epsilon, 1),(1, 2),(2,3)\} , \{ (\epsilon, goal),(1,p(a)),(2,eq(a,b))\}$


\vspace{2mm}
\underline{\bf De $S_6$ \ag $S_7$}

Du fait que le \nd courant est la racine, seul un retour arri\`ere peut intervenir. Seul le \nd $1$ a une clause dans sa \bt (clause \verb.c3.), et la clause est un fait. Seule la r\`egle (6) \redoun{} s'applique (pas de cr\'eation de nouveau \nd) et le \nd $2$ est supprim\'e). Le \nd courant devient $1$.

\vspace{2mm}
\noindent
(6) \reglecontroledeuxshort{\redounshort{}}
{  v \gets 1 \et \neg fst(\epsilon) \wedge hcp(\epsilon) \wedge ft(1) \wedge (flr (=T)\ \vee \ ct (=T)) }
{T' = \{\epsilon, 1\} \et u' = 1 \et pd' = \{(\epsilon, goal),(1,p(a))\} \et cl' = \{(\epsilon, []), (1 ,[]))\} \et }
{ ct = false \et flr' = false }
{\{ \}}

\vspace{2mm}
\noindent
{\bf $S_7$}: $\{ \{ \epsilon,1\} , 1 , 3 , \{(\epsilon,1),(1, 2)\} , \{(\epsilon,goal),(1,p(a))\},$

\indent
$\ \ \ \ \ \ \ \ \{(\epsilon,[]),(1,[])\}, \{(\epsilon,false),(1,false)\}, false, false\}$

\vspace{1mm}
L'\'ev\'enement de trace extrait est (r\`egle d'extraction \redoun{} de la section~\ref{generation}):

\verb.< 6     2    2  Redo  p(a) >. (suivi de \verb.< 7     2    2  Exit  p(b) >.)

\vspace{1mm}
L'\'etat reconstruit (r\`egle de reconstruction \redoun{} de la section~\ref{reconstruction}) est: 

\noindent
{\bf $Q_7 = S_7/Q$}: $\{ \{ \epsilon,1\} ,1 , \{(\epsilon, 1),(1, 2)\} , \{ (\epsilon, goal),(1,p(a))\}$


\vspace{2mm}
\underline{\bf De $S_7$ \ag $S_8$}

L'unification r\'eussit \ag nouveau, toujours avec la m\^eme clause utilis\'ee \ag la racine. Le fr\`ere du \, \nd courant (\nd $2$) est donc recr\'e\'e, avec num\'ero de cr\'eation $4$. C'est donc la r\`egle (7) \exitdeux{} qui s'applique. La pr\'edication appel\'ee (toujours ``faussement'' not\'ee $p(a)$) est un succ\`es et devient $p(b)$. Elle peut avoir un fr\`ere (existence d'une deuxi\`eme pr\'edication dans le corps de la clause \verb.c1. utilis\'ee). Le \nd re-cr\'e\'e a pour \'etiquette $eq(b,b)$ et a une clause dans sa \bt (\verb.c4.).

\vspace{2mm}
\noindent
(7) \reglecontroledeuxshort{\exitdeuxshort{}}
{ \neg first(1) \wedge mhnb(1) \wedge \neg ct \wedge \neg flr \et v \gets 2}
{u' = 2 \et n'= 4 \et nu' = \{(\epsilon,1),(1, 2),(2,4)\} \et pd' =\{(\epsilon,goal),(1,p(b)),(2,eq(b,b))\} \et}
{ T' = \{ \epsilon,1,2\} \et  cl' = \{(\epsilon,[]),(1,[]),(2,[c4])\} \et fst' = \{(\epsilon,false),(1,false),(2,true)\}}
{ \{ \\
\phantom{xxxxxxxxxxxxxxxx} scs(1) \et p(b) = pud(1) \et (c4,eq(b,b)) = cpini(2)\}}

\vspace{2mm}
\noindent
{\bf $S_8$}: $\{ \{ \epsilon,1,2\} , 2 , 4 , \{(\epsilon,1),(1, 2),(2,4)\} , \{(\epsilon,goal),(1,p(b)),(2,eq(b,b))\},$

\indent
$\ \ \ \ \ \ \ \ \{(\epsilon,[]),(1,[]),(2,[c4])\}, \{(\epsilon,false),(1,false),(2,true)\}, false, false\}$

\vspace{1mm}
L'\'ev\'enement de trace extrait est (r\`egle d'extraction \exitdeux{} de la section~\ref{generation}):

\verb.< 7   2   2  Exit  p(b) >. (suivi de \verb.< 8   4   2  Call  eq(b,b) >.)

\vspace{1mm}
L'\'etat reconstruit (r\`egle de reconstruction \exitdeux{} de la section~\ref{reconstruction}) est:

\noindent
{\bf $Q_8 = S_8/Q$}: $\{ \{ \epsilon,1,2 \} ,2 , \{(\epsilon, 1),(1, 2),(2,4)\} , \{ (\epsilon, goal),(1,p(b)),(2,eq(b,b))\}$

\vspace{1mm}
mais sa reconstruction n\'ecessite de conna\ii tre l'\ev de trace suivant (reconnaissance du fait que le prochain \nd visit\'e sera un nouveau n\oe ud).

\vspace{2mm}
\underline{\bf De $S_8$ \ag $S_9$}

Le \nd courant n'ayant pas encore \'et\'e visit\'e et la clause utilis\'ee \verb.c4. \'etant un fait, seule  la r\`egle (8) \callun{} s'applique .

\vspace{2mm}
\noindent
(8) \reglecontrole{\callun{}}
{first(2) \wedge leaf(2) \wedge \neg ct \wedge fact(2) }
{cl' = \{(\epsilon,[]),(1,[]),(2,[])\} \et fst'(2) = false \et flr'= false }
{\{ \}}

\vspace{2mm}
\noindent
{\bf $S_9$}: $\{ \{ \epsilon,1,2\} , 2 , 4 , \{(\epsilon,1),(1, 2),(2,4)\} , \{(\epsilon,goal),(1,p(b)),(2,eq(b,b))\},$

\indent
$\ \ \ \ \ \ \ \ \{(\epsilon,[]),(1,[]),(2,[])\}, \{(\epsilon,false),(1,false),(2,false)\}, false, false\}$

\vspace{1mm}
L'\'ev\'enement de trace extrait est (r\`egle d'extraction \callun{} de la section~\ref{generation}):

\verb.< 8     4    2  Call  eq(b,b) >.

\vspace{1mm}
L'\'etat reconstruit (r\`egle de reconstruction \callun{} de la section~\ref{reconstruction}) est:

\noindent
{\bf $Q_9 = S_9/Q$}: $\{ \{ \epsilon,1,2 \} ,2 , \{(\epsilon, 1),(1, 2),(2,4)\} , \{ (\epsilon, goal),(1,p(b)),(2,eq(b,b))\}$

\vspace{1mm}
mais l'\ev de trace laisse l'\'etat actuel courant invariant $Q_9 = Q_8$.

\vspace{2mm}
\underline{\bf De $S_9$ \ag $S_{10}$}

L'unification de $eq(b,b)$ et $eq(X,X)$ (clause \verb.c4. du \nd num\'ero $4$) r\'eussit, il n'y a pas de fr\`ere potentiel et on n'est pas dans un \'etat d'\'echec ($flr = false$). Seule la r\`egle (9) \exitun{} s'applique donc. $pred(2) = eq(b,b)$ reste invariant (fait clos).

\vspace{2mm}
\noindent
(9) \reglecontrole{\exitun{}} 
{\neg first(2) \wedge \neg mhnb(2) \wedge \neg ct \wedge \neg flr \et v \gets \epsilon}
{u' = \epsilon \et pd' = pd }
{\{ scs(2) \et \\ 
   \phantom{xxxxxxxxxxxxxxxxxxxxxxxxxxxxxxxxxxxxxxxxxxxxxx} eq(b,b)=pud(2)\}}

\vspace{2mm}
\noindent
{\bf $S_{10}$}: $\{ \{ \epsilon,1,2\} , \epsilon , 4 , \{(\epsilon,1),(1, 2),(2,4)\} , \{(\epsilon,goal),(1,p(b)),(2,eq(b,b))\},$

\indent
$\ \ \ \ \ \ \ \ \{(\epsilon,[]),(1,[]),(2,[])\}, \{(\epsilon,false),(1,false),(2,false)\}, false, false\}$

\vspace{1mm}
L'\'ev\'enement de trace extrait est (r\`egle d'extraction \exitun{} de la section~\ref{generation}):

\verb.< 9   4   2  Exit  eq(b,b) >. (suivi de \verb.< 10   1   1  Exit  goal >.)

\vspace{1mm}
L'\'etat reconstruit (r\`egle de reconstruction \exitun{} de la section~\ref{reconstruction}) est:

\noindent
{\bf $Q_{10} = S_{10}/Q$}: $\{ \{ \epsilon,1,2 \} ,\epsilon , \{(\epsilon, 1),(1, 2),(2,4)\} , \{ (\epsilon, goal),(1,p(b)),(2,eq(b,b))\}$


\vspace{2mm}
\underline{\bf De $S_{10}$ \ag $S_{11}$}

Finalement, la r\`egle (10) \exitun{} s'applique \ag nouveau (pas de fr\`ere possible). On obtient alors un arbre de preuve complet. La pr\'edication de la racine $goal$ est invariante.

\vspace{2mm}
\noindent
(10) \reglecontrole{\exitun{}} 
{\neg first(\epsilon) \wedge \neg mhnb(\epsilon) \wedge \neg ct \wedge \neg flr \et v \gets \epsilon}
{u' = \epsilon \et pd' \gets upcp(pd,u,p) \et ct' = true}
{\{ scs(\epsilon) \et \\
   \phantom{xxxxxxxxxxxxxxxxxxxxxxxxxxxxxxxxxxxxxxxxxxxxxxxx} goal = pud(\epsilon)\}}

\vspace{2mm}
\noindent
{\bf $S_{11}$}: $\{ \{ \epsilon,1,2\} , \epsilon , 4 , \{(\epsilon,1),(1, 2),(2,4)\} , \{\bar{(\epsilon,goal)},(1,p(b)),(2,eq(b,b))\},$

\indent
$\ \ \ \ \ \ \ \ \{(\epsilon,[]),(1,[]),(2,[])\}, \{(\epsilon,false),(1,false),(2,false)\}, true, false\}$

\vspace{1mm}
L'\'ev\'enement de trace extrait est (r\`egle d'extraction \exitun{} de la section~\ref{generation}):

\verb.< 10    1    1  Exit  goal >. 

(suivi d'aucun autre car la racine de l'arbre est \ag nouveau atteinte et il n'y a plus de point de choix dans l'arbre).

\vspace{1mm}
L'\'etat reconstruit (r\`egle de reconstruction \exitun{} de la section~\ref{reconstruction}) est:

\noindent
{\bf $Q_{11} = S_{11}/Q$}: $\{ \{ \epsilon,1,2 \} ,\epsilon , \{(\epsilon, 1),(1, 2),(2,4)\} , \{ (\epsilon, goal),(1,p(b)),(2,eq(b,b))\}$

\vspace{1mm}
mais l'\ev de trace n'engendre aucune modification de l'\'etat actuel courant $Q_{11} = Q_{10}$.

\begin{figure}[h]
\begin{center}
\includegraphics[width=0.7\linewidth]{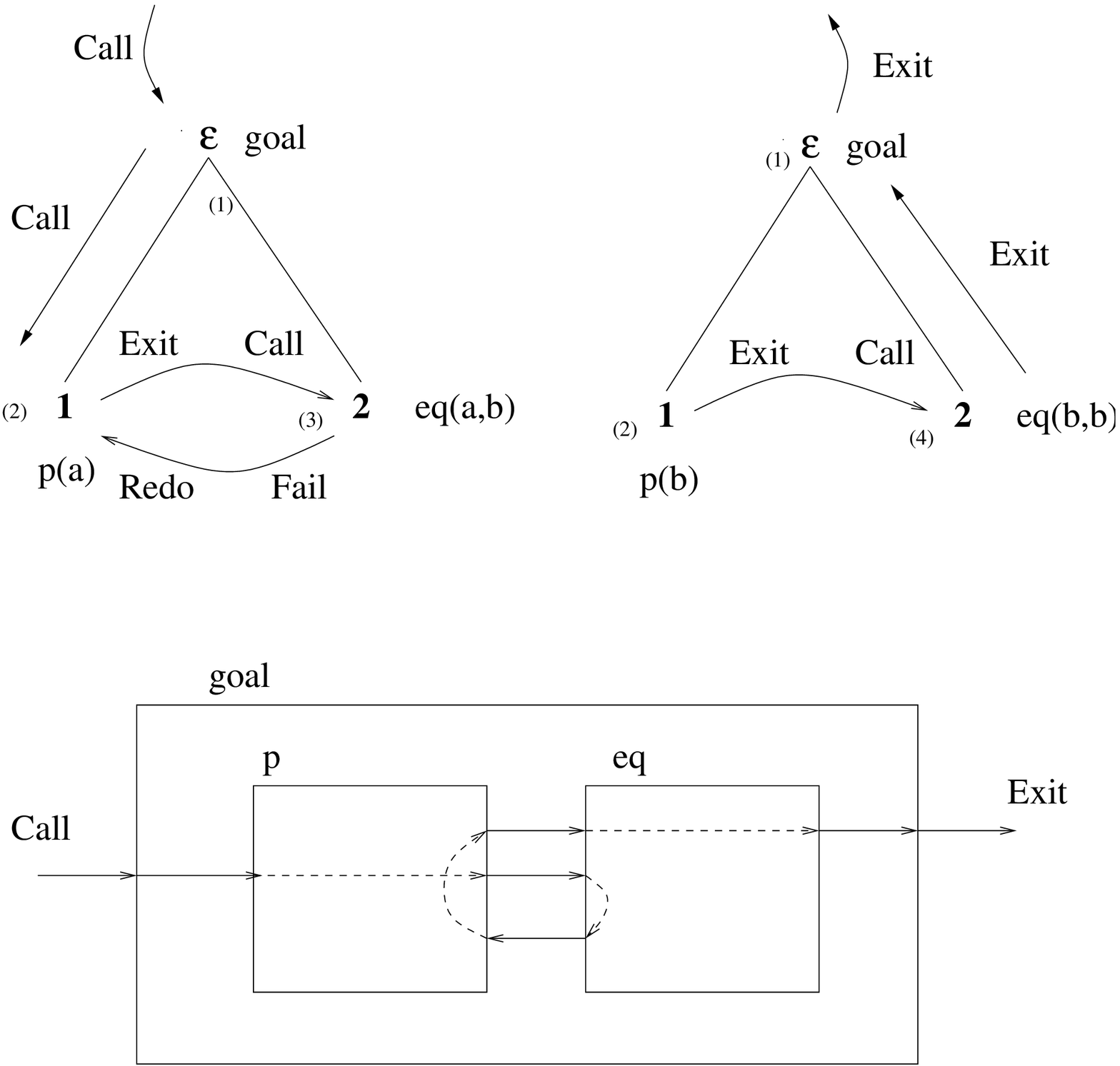}
\end{center}
\caption[Arbres construits et parcours dans les \bts]{Arbre de preuve obtenu et parcours dans les \bts}
\label{fig:adpres}
\end{figure}

\vspace{2mm}
Comme l'arbre $T$ est complet ($ct = true$)
et qu'il n'y a plus de point de choix dans l'arbre ($hcp(\epsilon) = false$), aucune r\eg gle de retour arri\`ere ne peut s'appliquer. On a obtenu une solution, et aucune r\`egle ne s'applique plus. La trace (virtuelle) se termine donc et on a obtenu un arbre de preuve complet. La figure~\ref{fig:adpres} illustre les arbres (partiel et complet) construits, ainsi que les encha\^{\i}nements des ports dans la trace de cet exemple.

\clearpage

\section{ANNEXE: Ad\'equation de la trace actuelle (Byrd simplifi\'e)}
\label{appendixC}

{\bf [Rappel: Condition d'ad\'equation]}

Etant donn\'es une SO d\'efinie avec un ensemble de \rgs $R$, un sch\'ema de trace ${\cal E}$ et un sch\'ema de reconstruction  ${\cal C}$ pour un sous-ensemble de param\`etres $Q$. Si les deux propri\'et\'es suivantes sont satisfaites pour chaque \rg $r \in R$:
\begin{quote}
$\forall\, e,\,e',\,r', S,\,S',\,S'',$

${\cal E}_r(S,S') = e\  \wedge \ {\cal E}_{r'}(S',S'') = e'$

(1) seule  $Cond_r(e,e')$ est vraie, i.e. $Cond_r(e,e') \bigwedge_{s \not = r} \neg Cond_{s}(e,e')$.

(2) $\ {\cal C}_r(e,e',S/Q) = S'/Q$.
\end{quote}
alors la trace actuelle $T_w = <S_0/Q, w^*_t>$, d\'efinie par le sch\'ema de trace ${\cal E}$, est ad\'equate pour $Q$ par rapport \ag la trace virtuelle int\'egrale $T_v = <S_0, v^*_t>$.

\vspace{2mm}
La preuve se fait en plusieurs \'etapes
\begin{itemize}
\item Etude de quelques propri\'et\'e dynamiques de la SO concernant l'enchainement de \rgs et des ports. Ces lemmes sont illustr\'es par les sch\'emas de la figures \ref{figadequat2}.
\item V\'erification de la condition 1: exclusivit\'e des conditions d'identification des \rgs dans le sch\'ema de reconstruction), cons\'equence directe des lemmes pr\'ec\'edents.
\item Enfin v\'erification de la condition 2: correction du sch\'ema de reconstruction pour le sous-ensemble de param\`etres $\{ T, u, num, pred \}$.
\end{itemize}

\begin{proof} [Preuve des lemmes]
On aura besoin des lemmes suivants \'etablis sur la trace virtuelle, illustr\'es par la  figure~\ref{figadequat2}, qui montrent les encha\ii nements possibles des \rgs et des ports pour toute trace. Le point d'entr\'ee oblig\'e dans la figure (\ev faisant suite \`a un \ev initial non trait\'e ici) est l'\'etat dit ``top'' ou un \ev de port {\bf Call}.

\begin{figure}[h]
\begin{center}
\includegraphics[width=\linewidth]{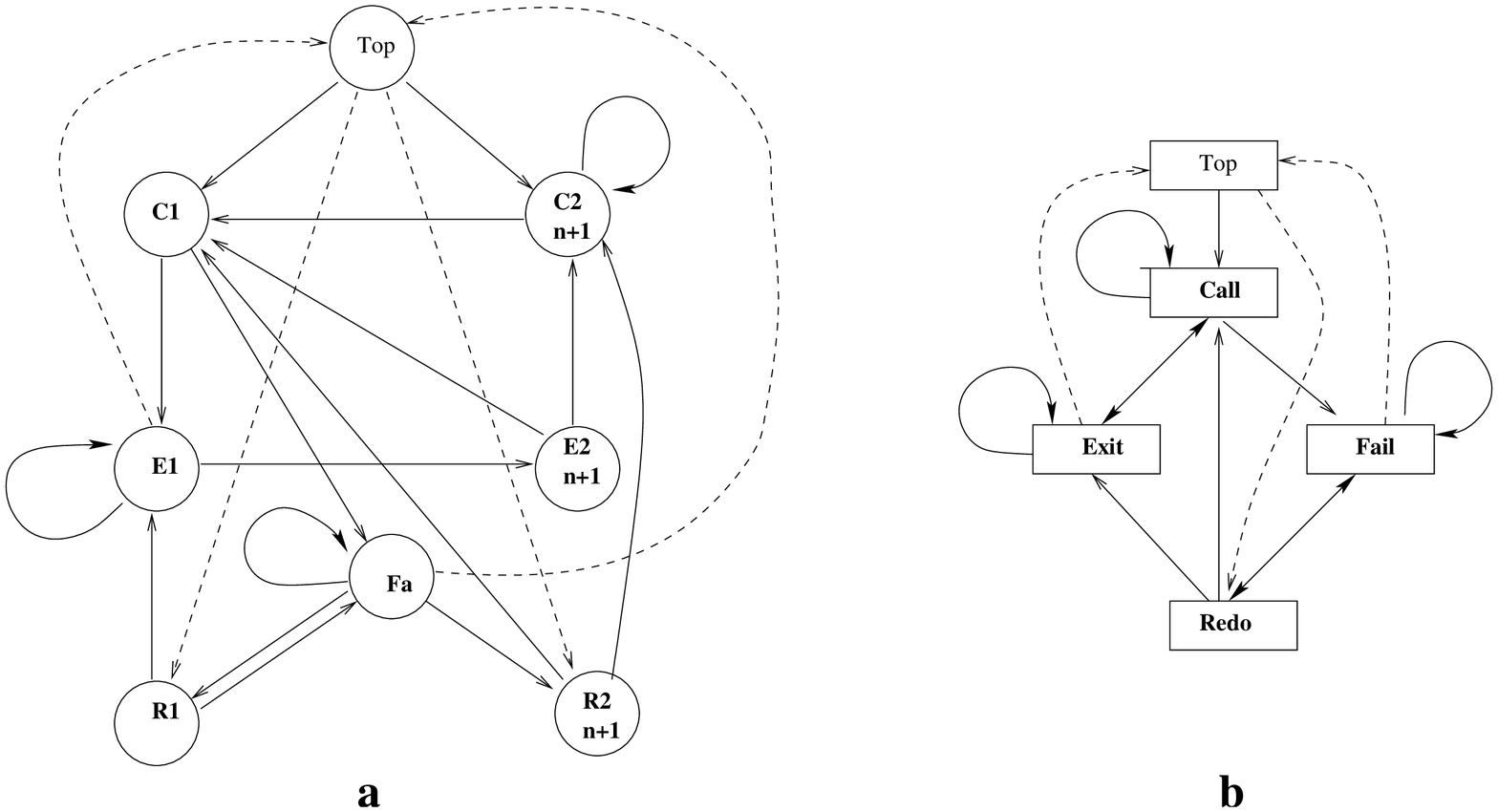}
\end{center}
\caption{Encha\ii nement des \rgs dans la SO ({\bf a}) et des ports dans toute trace ({\bf b}) (le diagramme {\bf b} est une abstraction du diagramme {\bf a})}
\label{figadequat2}
\end{figure}

La figure~\ref{figadequat2}~{\bf a} illustre l'encha\ii nement des \rgs dans la SO. Elle inclue dans les \'etats {\bf c2}, {\bf e2} et {\bf r2} l'incr\'ement du compteur de \nds cr\'e\'es ($n$), montrant ainsi que seule la travers\'ee de ces \nds fait cro\ii tre celui-ci. Cette propri\'et\'e est utilis\'e explicitement pour discriminer les \rgs dans la deuxi\`eme partie de la d\'emonstration.

\vspace{2mm}
Pour l'encha\ii nement des ports dans une trace actuelle (figure~\ref{figadequat2}~{\bf b}),
les fl\`eches sont pour la plupart \'evidentes. On observera ici simplement que la possibilit\'e d'avoir un {\bf Redo} apr\`es un {\bf Exit} est li\'ee \ag la condition $ct$.
La possibilit\'e d'avoir des {\bf Call} successifs est li\'ee \`a l'usage exclusif de la \rg \calldeux{}. L'absence de transition de {\bf Fail} vers {\bf Call} est li\'e \ag la condition $fst$ qui l'emp\^eche. L'absence de transition {\bf Call} vers {\bf Redo} est due \ag la condition $flr \vee ct$.  Enfin les {\bf Redo} ne peuvent boucler sur eux-m\^emes.
\end{proof}

\begin{proof} [Condition 1]

L'ensemble des conditions exclusives telles que d\'ecrites dans le sch\'ema de reconstruction (figure~\ref{traceregenefig}) est rappel\'e ici (figure~\ref{condfig}). Seuls les \'el\'ements de trace utiles sont indiqu\'es.

\begin{figure*}[ht]\small
\noindent
\reglecontrole{\callun{}}
{r' = r}
{}
{\{< r \ \ l \ \ {\bf Call} \ \ p > \pv < r' >\}}
\saut
\reglecontrole{\calldeux{}}
{r' > r}
{}
{\{
< r \ \ l \ \ {\bf Call} \ \ p > \pv <r'\ \ p'>\}}
 \saut
\reglecontrole{\exitun{}} 
{r' < r \vee nd(r) = \epsilon}
{}
{\{ < r \ \ l \ \ {\bf Exit} \ \ p > \pv < r' > \}}
 \saut
\reglecontrole{\exitdeux{}}
{  r' > r \wedge nd(r) \not = \epsilon}
{}
{\{ 
< r \ \ l \ \ {\bf Exit} \ \ p > \pv <r' \ \ p'> \}}
 \saut
\reglecontrole{\faildeux{}} 
{\,\,\,\,\,\,\,\,\,\,\,\,}
{}
{\{< r \ \ l \ \ {\bf Fail} \ \ p > \}}
 \saut
\reglecontrole{\redoun{}}
{r' = r}
{}
{\{ < r \ \ l \ \ {\bf Redo} \ \ p > \pv < r' >\}}
 \saut
\reglecontrole{\redodeux{}}
{r' > r }
{}
{\{ 
 < r \ \ l \ \ {\bf Redo} \ \ p > \pv < r' \ \ p' >\}}
 \saut
\caption{Conditions exclusives du sch\'ema de reconstruction}
\label{condfig}
\end{figure*}
La discrimination des \rgs se fait en premier \ag partir du port qui est toujours pr\'esent dans les \evs de trace (ceci est suffisant pour les \evs de port {\bf Fail}), puis sur l'accroissement ou non du compteur de \nds, avec une particularit\'e pour les \evs de port {\bf Exit} o\`u la condition est un peu plus complexe.

Noter que l'on utilise ici une information qui d\'ecoule de la propri\'et\'e de correction: le num\'ero de cr\'eation du \nd courant $nd(u)$ est toujours \'egal au premier attribut de l'\ev de trace associ\'e $r$, soit $nd(u) = r$ pour tous les \evs de port diff\'erent de {\bf Redo}. Dans le cas en effet de la \rg \redodeux{}, le \nd trac\'e ($nd(v)$) n'appara\ii t ni dans $S$ ni dans $S'$; dans le cas de la \rg \redoun{}, il appara\ii t seulement dans l'\'etat $S'$.
\end{proof}

\vspace{2mm}
\begin{proof} [Condition 2: correction]

Pour chaque \rg $r \in R$ de la SO  (il y en a 7), on indique les \'el\'ements suivants.
\begin{itemize}
\item Description de la transition correspondant \ag la \rg $r$ en mettant en \'evidence ce qui rel\`eve de l'\'etat ``avant'' ($S$) et ``apr\`es'' ($S'$) (tous les param\`etres de cet \'etat sont prim\'es).
\item On en d\'eduit imm\'ediatement l'\'etat virtuel $S'/Q$  restreint \ag $Q$.
\item On regarde alors l'\'ev\'enements de trace extrait $e = {\cal E}_r(S,S')$ et les suivants possibles $e' = {\cal E}_{r'}(S',S'')$, en se limitant aux attributs utilis\'es dans la \rg correspondante du sch\'ema de reconstruction (l'\ev correspondant \ag la \rg $r'$ n'est pas d\'ecrit compl\`etement, laissant la possibilit\'e d'avoir en fait diff\'erents \evs et/ou \rgs $r'$). En fait on se limite aux propri\'et\'es communes aux attributs des \'ev\'enements suivants possibles.
\item On donne la \rg correspondante du sch\'ema de reconstruction ${\cal C}_r$ d\'ecrivant la transition $<S/Q,Q'>$.
\item On v\'erifie que la condition $Cond_r(e,e')$ est bien v\'erifi\'ee.
\item On donne la description de l'\'etat  $Q'$ r\'esultants de la \rg de reconstruction ($Q' = {\cal C}_r(e,e',S/Q)$).
\item Enfin et si cela n'est pas \'evident par construction, on montre l'\'egalit\'e de $S'/Q$ et $Q'$.
\end{itemize}

\vspace{2mm}
\noindent \underline{\bf \Rg \callun{}}

\vspace{1mm}
\noindent
\reglecontroledeux{\callun{}}
{fst(u) \wedge lf(u) \wedge \neg ct \wedge ft(u) }
{T'=T \et u'=u \et nu' = nu \et pd'=pd \et cl'= upcp(cl,u) \et }
{fst'(u)= F \et flr'= F }
{\{ \}}

\vspace{2mm}
\noindent
$S' = S$

\vspace{2mm}
\noindent
${\cal E}_{\callun{}}(S,S') = < nu(u) \ \ lp(u) \ \ Call \ \ pd(u)>$

\noindent
${\cal E}_{r'}(S',S'') = < nu'(u') ...>$

\vspace{2mm}
\noindent
${\cal C}_{\callun{}}$ \ \ \reglecontrole{\callun{}}
{Cond_{\callun{}}(e,e')}
{}
{\{< ... >\}}

(aucun \'el\'ement utile)

\vspace{2mm}
\noindent
$Cond_{\callun{}}(e,e') = (nu(u) = nu'(u'))$  (vraie, car $u' = u$ et $nu' = nu$)

\vspace{2mm}
La restriction \ag $Q$ de $S$ ou $S'$ est invariante. De plus la \rg de reconstruction ne modifie aucun des param\`etres de de $S/Q$. Il en r\'sulte que $Q' = Q = S'/Q$.

\vspace{2mm}
\noindent \underline{\bf \Rg \calldeux{}}

\vspace{1mm}
\noindent
\reglecontroledeuxshort{\calldeuxshort{}}
{fst(u) \wedge lf(u) \wedge \neg ct \wedge \neg ft(u) \et v = crc(u)}
{T'= T \cup  \{v\} \et u'= crc(u) \et n'= n+1 \et nu'(u')= n'  \et pd'(u')= p\_cpini(u') \et }
{ cl'(u)= upcp(u) \et cl'(u')= c\_cpini(u') \et fst'(u)=F \et fst'(u') = T\} \et flr'= F}
{\{ \\ \phantom{xxxxxxxxxxxxxxxxxxxxxxxxxxxxxxxxxxxxxxxxx} scs(u) \et (c,p) = cpini(v)\}}

\vspace{2mm}
\noindent
$S'/Q = \{T \cup  \{u'\} , u' = crc(u), nu'(u')= n' , pd'(u')= p\_cpini(u') \}$

\vspace{2mm}
${\cal E}_{\calldeux{}}(S,S') = < nu(u) \ \ lp(u) \ \ Call \ \ pd(u)>$

${\cal E}_{r'}(S',S'') = < nu'(u') ... pd'(u')>$

\vspace{2mm}
\noindent
\reglecontroledeuxshort{\calldeuxshort{}}
{Cond_{\calldeux{}}(e,e')}
{u'= crc((nd(nu(u))) \et T' = T \cup \{u'\} \et nu'(u')= nu'(u') \et }
{pd'(u')= pd'(u')= p\_cpini(u')}
{\{ 
e \ \ e'\}}

\vspace{2mm}
\noindent
$Cond_{\calldeux{}}(e,e') = (nu'(crc(u)) > nu(u))$ en effet tout nouveau \nd cr\'e\'e l'est avec un num\'ero sup\'erieur \ag tous ceux d\'ej\`a existants ($n' > n$).

\vspace{2mm}
\noindent
$Q' = \{T \cup \{u'\}, u'= crc(u), nu'(u') = n', pd'(u')= p\_cpini(u') \} = S'/Q$

\vspace{2mm}
\noindent \underline{\bf \Rg \exitun{}}

\vspace{1mm}
\noindent
\reglecontroleshort{\exitunshort{}}
{\neg fst(u) \wedge \neg mhnb(u) \wedge \neg ct \wedge \neg flr}
{T' =T \et u'= pt(u) \et nu' = nu \et pd'(u)= pud(u) \et (u = \epsilon) \Rightarrow (ct'= T)}
{\{ \\ \phantom{xxxxxxxxxxxxxxxxxxxxxxxxxxxxxxxxxxxxxxxxxxx} scs(u) \et p=pud(u)\}}

\vspace{2mm}
\noindent
$S'/Q = \{T  , u'= pt(u), nu'= nu , pd'(u)= pud(u)\}$

\vspace{2mm}
${\cal E}_{\exitun{}}(S,S') = < nu(u) \ \ lp(u) \ \ Exit \ \ pud(u)>$

${\cal E}_{r'}(S',S'') = < nu'(pt(u)) ... >$

\vspace{2mm}
\noindent
\reglecontrole{\exitun{}} 
{Cond_{\exitun{}}(e,e')}
{u' = pt(u) \et pd'(u) = pud(u)}
{\{ < nu(u) \ \ lp(u) \ \ Exit \ \ pud(u)> \pv \\
\phantom{xxxxxxxxxxxxxxxxxxxxxxxxxxxxxxxxxxxxxxxxxxx} < nu'(pt(u)) ...> \}}

\vspace{2mm}
\noindent
$Cond_{\exitun{}}(e,e') = (nu'(pt(u)) < nu(u) \vee u = \epsilon)$ est v\'erifi\'ee; en effet tout \nd inf\'erieur dans l'ordre lexicographique a \'et\'e cr\'e\'e avec un num\'ero d'ordre inf\'erieur.

\vspace{2mm}
\noindent
$Q' = \{T, u'= pt(u), nu'= nu, pd'(u')= pud(u) \} = S'/Q$.

\vspace{2mm}
\noindent \underline{\bf \Rg \exitdeux{}}

\vspace{1mm}
\noindent
\reglecontroledeuxshort{\exitdeuxshort{}}
{ \neg fst(u) \wedge mhnb(u) \wedge \neg ct \wedge \neg flr}
{u'= crnb(u) \et T'= T \cup  \{u'\} \et n'= n+1 \et nu'(u')= n' \et}
{ pd'(u)= pud(u) \et pd'(u')= p  \et cl'(u')= c \et  fst'(u')= true }
{ \{ scs(u) \et \\
\phantom{xxxxxxxxxxxxxxxxxxxxxxxxxxxxxxxxxxxxxxxxxx} (c,p) = cpini(u') \}}

\vspace{2mm}
\noindent
$S'/Q = \{T \cup \{u'\} , u'= crnb(u), nu'= nu \cup \{(u',n')\}, pd'= upcp(pd,u,pud(u)) \cup \{(u', p\_cpini(u'))\}\}$

\vspace{2mm}
${\cal E}_{\exitdeux{}}(S,S') = < nu(u) \ \ lp(u) \ \ Exit \ \ pud(u)>$

${\cal E}_{r'}(S',S'') = < nu'(crnb(u))= n'\ \ p\_cpini(u')>$

\vspace{2mm}
\noindent
\reglecontroledeuxshort{\exitdeuxshort{}}
{Cond_{\exitdeuxshort{}}(e,e')}
{u'= crnb(u) \et T'= T \cup \{u'\} \et nu'(u') = n' \et }
{pd'(u)= pud(u) \et pd'(u')= p\_cpini(u')}
{\{ < nu(u) \ \ lp(u) \ \ {\bf Exit} \ \ pud(u) > \pv \\ 
\phantom{xxxxxxxxxxxxxxxxxxxxxxxxxxxxxxxxxxxxxxxxx} < n' \ \ p\_cpini(u') > \}}

\vspace{2mm}
\noindent
$Cond_{\exitdeux{}}(e,e') = (n'= n+1) > nu(u) \wedge u \not = \epsilon$ est v\'erifi\'ee; en effet tout \nd inf\'erieur dans l'ordre lexicographique a \'et\'e cr\'e\'e avec un num\'ero d'ordre inf\'erieur, soit $n \geq nu(u) $. De plus, comme $u'$ est fr\`ere de $u$, $u$ ne peut \^etre racine.

\vspace{2mm}
\noindent
$Q' = \{T'= T \cup \{u'\}, u'= crnb(u), nu'= nu \cup \{(u',n')\},  pd'= upcp(pd,u,pud(u)) \cup \{(u', p\_cpini(u'))\} \} = S'/Q$.

\vspace{2mm}
\noindent \underline{\bf \Rg \faildeux{}}

\vspace{1mm}
\noindent
\reglecontrole{\faildeux{}}
{\neg fst(u) \wedge \neg ct \wedge \neg hcp(u) \et  v \gets pt(u)}
{u'= pt(u) \et (u = \epsilon) \Rightarrow (ct' \gets T) \et flr' \gets T}
{\{flr(u) \ \vee \  flr\}}
\saut

\vspace{2mm}
\noindent
$S'/Q = \{T  , u'= pt(u), nu , pd\}$

\vspace{2mm}
${\cal E}_{\faildeux{}}(S,S') = < nu(u) \ \ lp(u) \ \ Fail \ \ pd(u)>$

${\cal E}_{r'}(S',S'') = < nu'(pt(u)) ... >$

\vspace{2mm}
\noindent
\reglecontrole{\faildeux{}} 
{Cond_{\faildeux{}}(e,e')}
{u'= pt(u)}
{\{< nu(u) \ \ lp(u) \ \ {\bf Fail} \ \ pd(u) > \}}
 \saut

\vspace{2mm}
\noindent
Il n'y a pas de condition autre que le nom du port.

\vspace{2mm}
\noindent
$Q' = \{T, u'= pt(u), nu, pd \} = S'/Q$.

\vspace{2mm}
\noindent \underline{\bf \Rg \redoun{}}

\vspace{1mm}
\noindent
\reglecontroledeux{\redoun{}}
{ v = gcp(u) \et \neg fst(u) \wedge hcp(u) \wedge ft(v) \wedge (flr \ \vee \ ct)}
{T'= T - \{y | y > v\} \et u'= gcp(u) \et cl'= upcp(cl,v) \et}
{ ct \Rightarrow (ct' \gets F) \et flr' \gets F}
{\{ \}}

\vspace{2mm}
\noindent
$S'/Q = \{T - \{y | y > u'\}, u'= gcp(u), nu, pd \}$

\vspace{2mm}
${\cal E}_{\redoun{}}(S,S') = < nu(v) \ \ lp(v) \ \ Redo \ \ pd(v)>$ avec $v= gcp(u)$.

${\cal E}_{r'}(S',S'') = < nu'(v) ... >$ avec $nu'= nu$.

\vspace{1mm}
L'\ev de trace produit a pour premier attribut $nu'(v)$ qui est \'egal \ag $nu(u)$ car les \evs suivants sont diff\'erents de {\bf Redo}.

\vspace{2mm}
\noindent
\reglecontrole{\redoun{}}
{ Cond_{\redoun{}}(e,e')}
{u'= nd(nu(gcp(u)))= gcp(u)= v \et T'= T - \{y | y > v\} }
{\{ \\ \phantom{xxxxxxxxxxxxxxxxxxxxxxx}
< nu(v) \ \ lp(v) \ \ {\bf Redo} \ \ pd(v) > \pv < nu'(v) >\}}

\vspace{2mm}
\noindent
$Cond_{\redoun{}}(e,e') = (nu(v) = nu'(v))$ est v\'erifi\'ee (cf ci-dessus).

\vspace{2mm}
\noindent
$Q' = \{T - \{y | y > v\}, u'= gcp(u), nu, pd\} = S'/Q$.

\vspace{2mm}
\noindent \underline{\bf \Rg \redodeux{}}

\vspace{1mm}
\noindent
\reglecontroledeuxshort{\redodeuxshort{}}
{ v= gcp(u) \et \neg fst(u) \wedge hcp(u) \wedge \neg ft(v) \wedge (flr \vee ct) \et w= crc(v)}
{ T'= T - \{y | y > v \} \cup \{w\} \et u'= w  \et n'= n+1 \et nu'= upn(nu,v) \cup \{(w,n')\} \et flr'= F \et }
{pd'= upcp(pd,v) \cup \{(w,p)\} \et cl'= upcp(cl,v) \cup \{(w,c)\} \et fst'= fst \cup \{(w,T)\} \et ct' \Rightarrow (ct \gets F)}
{\{ \\
\phantom{xxxxxxxxxxxxxxxxxxxxxxxxxxxxxxxxxxxx} scs(v) \et (c,p) = cpini(w)\}}

\vspace{2mm}
\noindent
$S'/Q = \{T- \{y | y > v\} \cup \{w\}  , u'= w, nu'= upn(nu,v) \cup \{(u',n')\} , pd'= upcp(pd,v) \cup \{(u',p\_cpini(w))\} \}$  avec $v= gcp(u)$ et $w= crc(v)$.

\vspace{2mm}
${\cal E}_{\redodeux{}}(S,S') = < nu(v) \ \ lp(v) \ \ Redo \ \ pd(v)>$

${\cal E}_{r'}(S',S'') = < nu'(w)=n' \ \  pd'(w)= p\_cpini(w) >$

\vspace{2mm}
\noindent
\reglecontroledeuxshort{\redodeuxshort{}}
{Cond_{\redodeux{}}(e,e')}
{v= nd(nu(v))= gcp(u) \et u'= crc(v) \et T'= T - \{y | y > v\}\cup \{u'\} \et}
{ nu'= upn(nu,v)\cup\{(u',n')\} \et pd'= upcp(pd,v)\cup\{(u',p\_cpini(u'))\}}
{\{ 
\\ \phantom{xxx}
 < nu(v) \ \ lp(v) \ \ {\bf Redo} \ \ pd(v) > \pv < nu'(w)=n+1 \ \ pd'(w)=p\_cpini(w) >\}}

\vspace{2mm}
\noindent
$Cond_{\redodeux{}}(e,e') = nu'(w) > nu(v)$ est v\'erifi\'ee;  en effet tout nouveau \nd cr\'e\'e l'est avec un num\'ero sup\'erieur \ag tous ceux d\'ej\`a existants. Le \nd $nd(u)$ peut avoir \'et\'e supprim\'e, mais tout $y < w$ dans l'ordre lexicographique des \nds est tel que $n+1 > nu(y)$.

\vspace{2mm}
\noindent
$Q' = \{T - \{y | y > v\}\cup \{u'\}, u'= crc(v), nu'= upn(nu,v)\cup\{(u',n')\}, pd'= upcp(pd,v)\cup\{(u',p\_cpini(u'))\}\} = S'/Q$ (avec $v= gcp(u)$).

\end{proof}

\clearpage

\section{ANNEXE: Analyse des diff\'erents mod\`eles de traces}
\label{appendixD}

On compare ici le mod\`ele pr\'esent\'e ici, dit ``Byrd simplifi\'e'' (m1), celui de GNU Prolog (m2), et le mod\`ele de Byrd avec l'implantation par m\'eta-interpr\`ete (m3).

Sur le m\^eme exemple (Annexe B) la trace produite par la SO pr\'esent\'ee ici (comme celle de Byrd) et celle de GNU Prolog 1.2.16 copyright (C) 1999-2002 Daniel Diaz \cite{gnuprolog} sont les m\^emes avec une seule diff\'erence sur les \'ev\'enements \verb.8. et \verb.9.. Cela r\'esulte du fait que dans GNU \verb.r. est le rang du \nd courant atteint dans l'arbre de preuve (selon l'ordre lexicographique). Le chrono est ajout\'e. Seuls les \evs 8 et 9 diff\`erent donc (voir la trace compl\`ete plus bas).

\begin{verbatim}
GNU     
8     3    2    Call:   eq(b,b) ? 
9     3    2    Exit:   eq(b,b) ?        

Byrd
8     4    2    Call    eq(b,b)
9     4    2    Exit    eq(b,b)
\end{verbatim}

On observera dans la suite que, si l'on se limite \`a la suite des ports, les mod\`eles sont inclus les uns dans les autres ( \verb.m1. $\subseteq$ \verb.m2. $\subseteq$ \verb.m3. ); la seule diff\'erence portant, entre \verb.m1. et \verb.m2. sur le premier attribut (cf ci-dessus).

Il est int\'eressant d'observer ici que les trois traces obtenues sur cet exemple simple ont la m\^eme suite de ports.

\vspace{3mm}
L'exemple 2 suivant, plus sophistiqu\'e, met en \'evidence de quelle mani\`ere elles diff\`erent en fait sensiblement. Les SO correspondantes sont d\'ecrites dans l'annexe E suivante.
\label{bigexa}

\begin{verbatim}
goal:-q(_).
q(X):-p1(X),p2(X),eq(X,b).
p1(X) :- p(X).
p(a).
p(b).
p(c).
p2(_).
eq(X,X).
\end{verbatim}

Trace obtenue avec le \md \verb,m1, (Byrd simplifi\'e):
\begin{verbatim}
1     1    1    Call    goal
2     2    2    Call    q(_86)
3     3    3    Call    p1(_86)
4     4    4    Call    p(_86)
5     4    4    Exit    p(a)
6     3    3    Exit    p1(a)
7     5    3    Call    p2(a)
8     5    3    Exit    p2(a)
9     6    3    Call    eq(a,b)
10    6    3    Fail    eq(a,b)
11    4    4    Redo    p(a)
12    4    4    Exit    p(b)
13    3    3    Exit    p1(b)
14    7    3    Call    p2(b)
15    7    3    Exit    p2(b)
16    8    3    Call    eq(b,b)
17    8    3    Exit    eq(b,b)
18    2    2    Exit    q(b)
19    1    1    Exit    goal
20    4    4    Redo    p(b)
21    4    4    Exit    p(c)
22    3    3    Exit    p1(c)
23    9    3    Call    p2(c)
24    9    3    Exit    p2(c)
25    10   3    Call    eq(c,b)
26    10   3    Fail    eq(c,b)
27    2    2    Fail    q(_86)
27    1    1    Fail    goal
yes
\end{verbatim}

\begin{figure}[h]
\begin{center}
\includegraphics[width=0.8\linewidth]{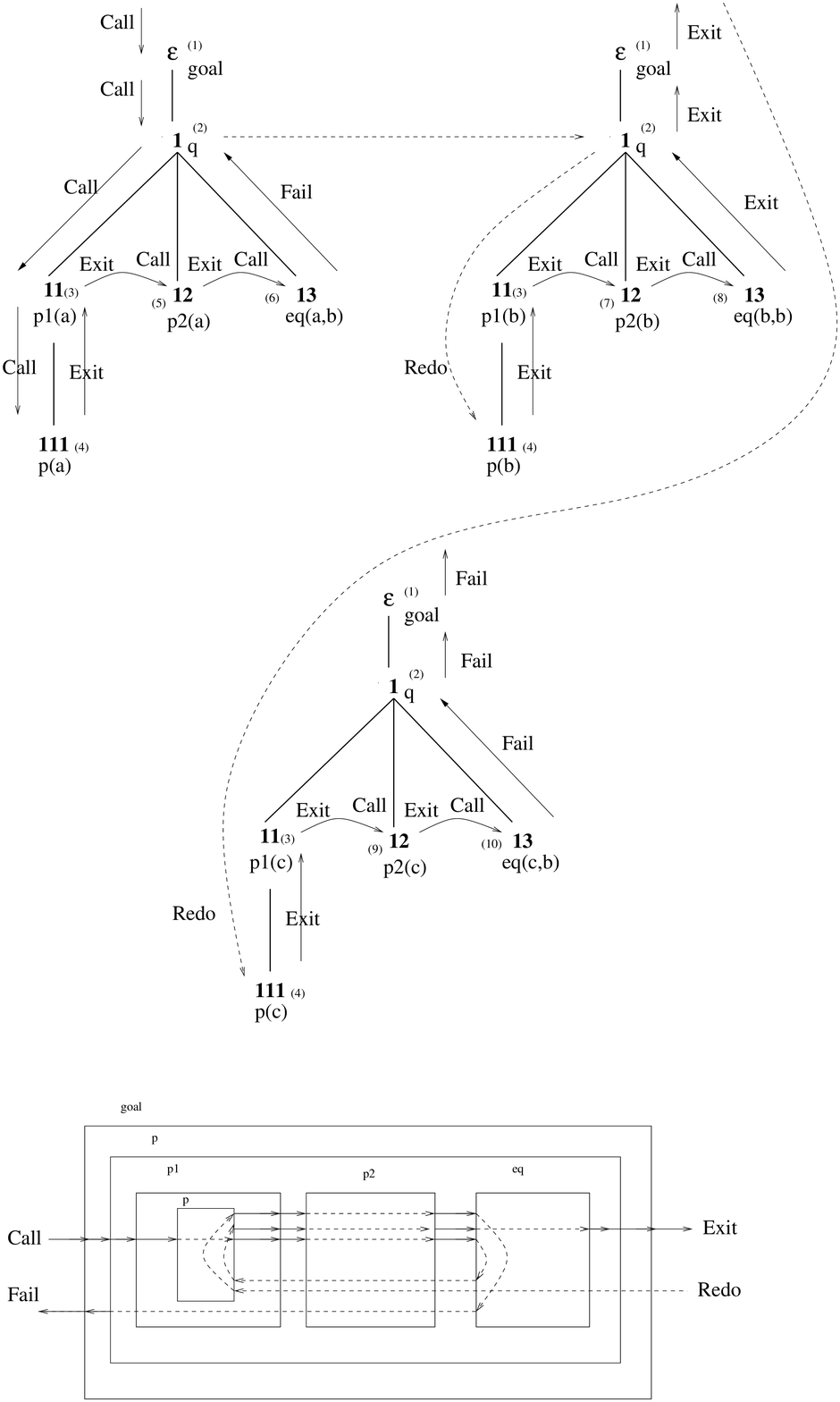}
\end{center}
\caption[Illustration de l'exemple 2 avec arbres et \bts (mod\`ele m1)]{Illustration de l'exemple~\ref{bigexa} avec arbres et \bts (mod\`ele m1, 28 \'ev\'enements)}
\label{fig:exapierre}
\end{figure}

Trace obtenue avec le \md \verb,m2, (traceur de GNU-Prolog).

\begin{verbatim}
1     1    1  Call: goal ?
2     2    2  Call: q(_38) ?
3     3    3  Call: p1(_38) ?
4     4    4  Call: p(_38) ?
5     4    4  Exit: p(a) ?
6     3    3  Exit: p1(a) ?
7     5    3  Call: p2(a) ?
8     5    3  Exit: p2(a) ?
9     6    3  Call: eq(a,b) ?
10    6    3  Fail: eq(a,b) ?
11    3    3  Redo: p1(a) ?
12    4    4  Redo: p(a) ?
13    4    4  Exit: p(b) ?
14    3    3  Exit: p1(b) ?
15    5    3  Call: p2(b) ?
16    5    3  Exit: p2(b) ?
17    6    3  Call: eq(b,b) ?
17    6    3  Exit: eq(b,b) ?
19    2    2  Exit: q(b) ?
20    1    1  Exit: goal ?
21    1    1  Redo: goal ?
22    2    2  Redo: q(b) ?
23    3    3  Redo: p1(b) ?
24    4    4  Redo: p(b) ?
25    4    4  Exit: p(c) ?
26    3    3  Exit: p1(c) ?
27    5    3  Call: p2(c) ?
28    5    3  Exit: p2(c) ?
29    6    3  Call: eq(c,b) ?
30    6    3  Fail: eq(c,b) ?
31    2    2  Fail: q(_38) ?
32    1    1  Fail: goal ?
\end{verbatim}

\begin{figure}[h]
\begin{center}
\includegraphics[width=0.8\linewidth]{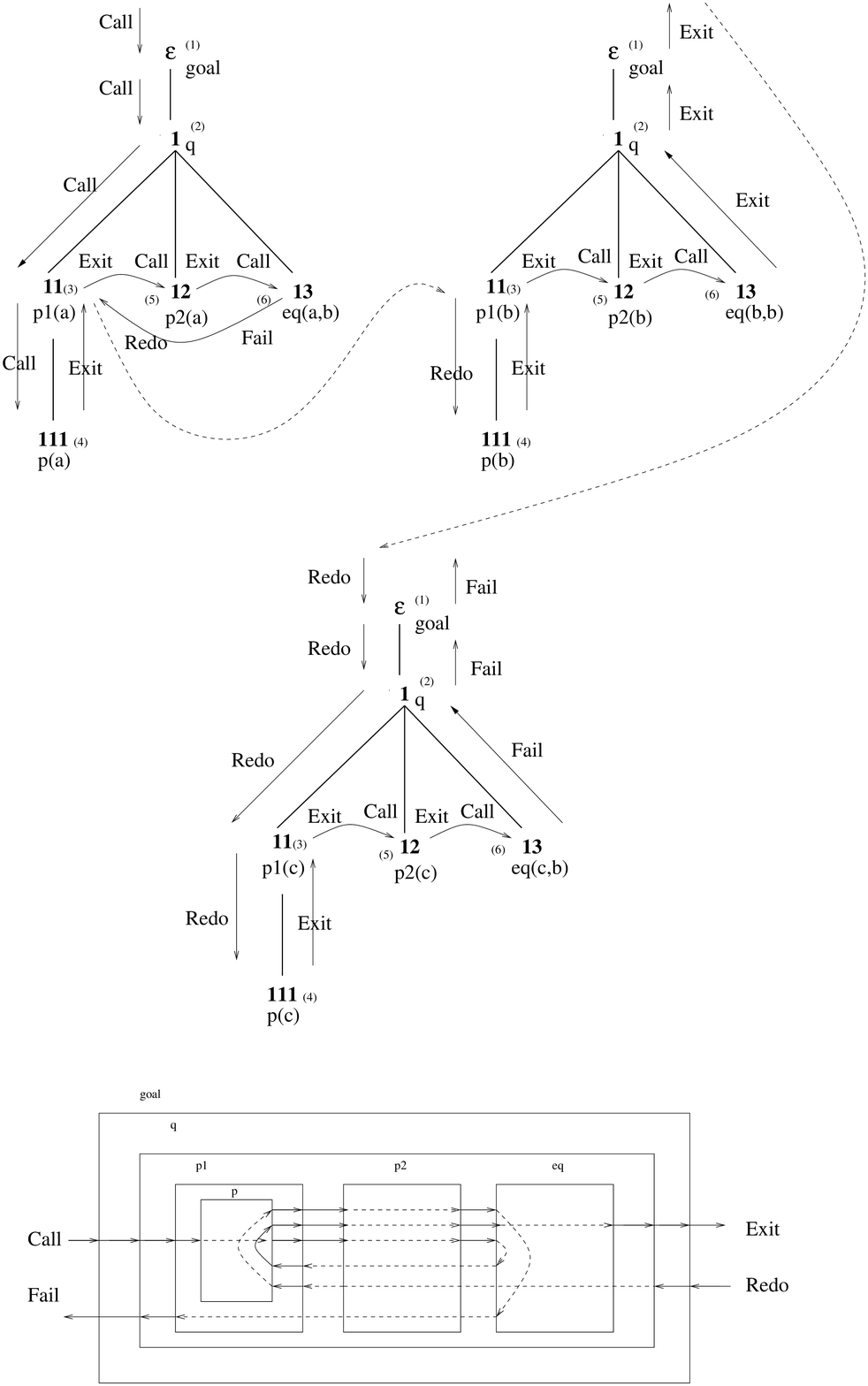}
\end{center}
\caption[Illustration de l'exemple 2 avec arbres et \bts (mod\`ele m2)]{Illustration de l'exemple 2 avec arbres et \bts (mod\`ele m2, 32 \'ev\'enements)}
\label{fig:exagnu}
\end{figure}

Trace obtenue avec le \md \verb,m3, (Byrd, m\'eta-interpr\`ete de l'annexe A et instrumentation du code).

\begin{verbatim}
1     1    1    Call    goal
2     2    2    Call    q(_86)
3     3    3    Call    p1(_86)
4     4    4    Call    p(_86)
5     4    4    Exit    p(a)
6     3    3    Exit    p1(a)
7     5    3    Call    p2(a)
8     5    3    Exit    p2(a)
9     6    3    Call    eq(a,b)
10    6    3    Fail    eq(a,b)
11    5    3    Redo    p2(a)
12    5    3    Fail    p2(a)
13    3    3    Redo    p1(a)
14    4    4    Redo    p(a)
15    4    4    Exit    p(b)
16    3    3    Exit    p1(b)
17    7    3    Call    p2(b)
18    7    3    Exit    p2(b)
19    8    3    Call    eq(b,b)
20    8    3    Exit    eq(b,b)
21    2    2    Exit    q(b)
22    1    1    Exit    goal
23    1    1    Redo    goal
24    2    2    Redo    q(b)
25    8    3    Redo    eq(b,b)
26    8    3    Fail    eq(b,b)
27    7    3    Redo    p2(b)
28    7    3    Fail    p2(b)
29    3    3    Redo    p1(b)
30    4    4    Redo    p(b)
31    4    4    Exit    p(c)
32    3    3    Exit    p1(c)
33    9    3    Call    p2(c)
34    9    3    Exit    p2(c)
35    10   3    Call    eq(c,b)
36    10   3    Fail    eq(c,b)
37    9    3    Redo    p2(c)
38    9    3    Fail    p2(c)
39    3    3    Redo    p1(c)
40    4    4    Redo    p(c)
41    4    4    Fail    p(_86)
42    3    3    Fail    p1(_86)
43    2    2    Fail    q(_86)
44    1    1    Fail    goal
yes
\end{verbatim}

\begin{figure}[h]
\begin{center}
\includegraphics[width=0.8\linewidth]{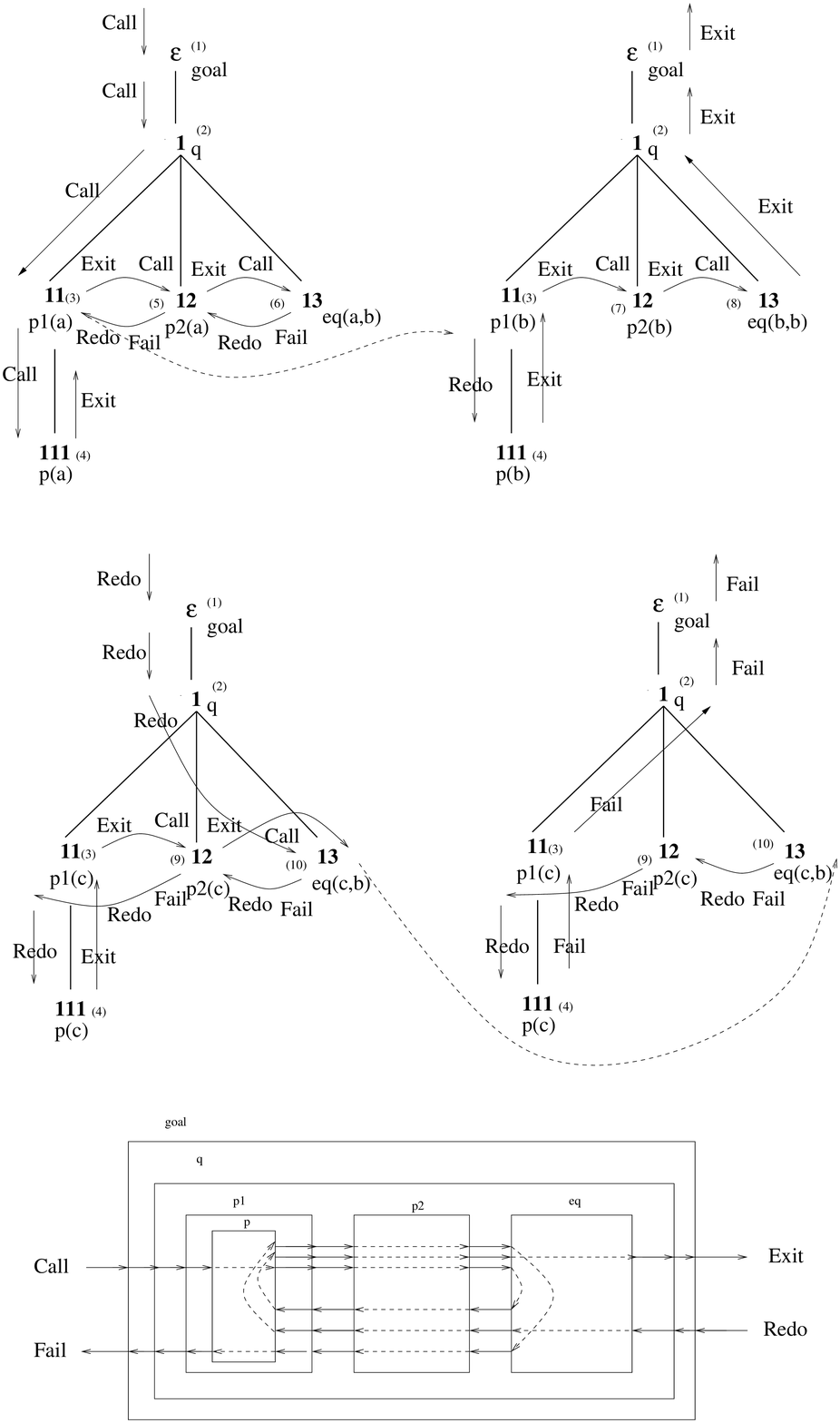}
\end{center}
\caption{Illustration de l'exemple 2 avec arbres et \bts  (mod\`ele m3, 44 \'ev\'enements)}
\label{fig:exabyrd}
\end{figure}

\clearpage

\section{ANNEXE: S\'emantique observationnelle pour les trois mod\`eles (m1, m2, m3)}
\label{appendixE}


On pr\'esente ici une SO qui correspond \ag une forme de recherche d'arbres de preuve complets, param\'etrable par des fonctions de choix (clauses et feuilles). L'objectif est de pouvoir rendre compte avec la m\^eme SO de plusieurs formes de traces.
Cette SO, plus d\'etaill\'ee que la pr\'ec\'edente illustre mieux la distinction entre trace virtuelle et trace actuelle int\'egrale ou non. En effet pour obtenir une trace correspondant au \md \verb.m1. par exemple, il y aura lieu de filtrer la trace virtuelle pour en \'eliminer certains \evs\footnote{Le \md pr\'esent\'e ici suppose que le choix de la trace soit fait de mani\`ere externe. Il s'agit donc, dans cette premi\`ere approche, de donner en fait une forme unifi\'ee \`a trois SO diff\'erentes (une par mod\`ele). La possibilit\'e de pr\'esenter les trois mod\`eles dans une SO unique, bien que possible du fait de l'inclusion partielle des traces, est encore \`a l'\'etude.}.

\vspace{1mm}
La s\'emantique op\'erationnelle de r\'esolution Prolog sous-jacente correspond \ag la SLDT-r\'esolution telle que pr\'esent\'ee dans \cite{Deransart93} p. 57.
C'est une strat\'egie descendante de construction d'arbre de preuve partiels qui repose sur deux fonctions de choix: choix d'une feuille incompl\`ete \ag d\'evelopper, choix d'une clause parmi celles associables \ag la feuille. Un \nd auquel il reste des clauses associ\'ees constitue un point de choix. La SLDT-r\'esolution construit des ``squelettes'' d'arbre et les ``d\'ecore'' de mani\`ere non d\'eterministe. Le d\'eterminisme est introduit de mani\`ere \ag garder la compl\'etude du sch\'ema de r\'esolution et \`a pouvoir reproduire les traces souhait\'ees.

On se limitera donc ici au parcours standard descendant gauche droite; mais d'autres strat\'egies peuvent \^etre consid\'er\'ees \ag condition de garder \`a celles-ci leur propri\'et\'e d'\^etre compl\`etes (``full'' dans \cite{Deransart93}). Le choix des feuilles consistera donc \`a choisir le premi\`ere feuille non visit\'ee (dans l'ordre total des \nd d'un arbre) dont on cherchera \ag d\'evelopper syst\'ematiquement le sous-arbre, et en cas d'\'echec local ou de succ\`es avec un arbre de preuve complet, on reviendra toujours au dernier point de choix cr\'e\'e.

\vspace{2mm}
On pourrait adopter des strat\'egies plus g\'en\'erales, d\'eterministes et compl\`etes, en particulier en autorisant un ordonnancement dynamique des descendants d'un n\oe ud, ou m\^eme en changeant compl\`etement de strat\'egie (en largeur d'abord par exemple), mais ces strat\'egies n'ont plus aucun rapport avec le mod\`ele des \bts; en particulier l'ordonnancement des ports {\bf Call} ou {\bf Redo} n'a plus grand chose \`a voir avec la structure de \bts puisque l'on peut alors ``sauter'' d'un \nd dans un sous-arbre \`a n'importe quel \nd d'un autre sous-arbre, donc d'une \bt \ag n'importe quelle autre.

Les pr\'edicats associ\'es aux \nds, par la fonction $pred$ sont ceux produits par le squelette.

Ce mod\`ele supprime en partie les fonctions externes dans la mesure o\`u succ\`es et \'echecs peuvent \^etre trait\'es comme des variables d'\'etat global (\'etat du sous-arbre courant).

\vspace{2mm}
{\bf Param\`etres de la trace virtuelle}

\noindent
L'\'etat courant a 13 param\`etres: 

 $\{ T , u , n , num , pred , chosen\_claus, claus\_list , first, \sigma, ct , flr , scs, bk3\}$. 


\begin{enumerate}
\item {\bf $T$}: $T$ est un arbre \'etiquet\'e avec un num\'ero de cr\'eation ou re-cr\'eation, une pr\'edication et un sous-ensemble de clauses du programme $P$. Il est d\'ecrit ici par ses fonctions de construction/reconstruction, parcours(cf plus bas) et \'etiquetage. 
Aucune repr\'esentation particuli\`ere n'est requise. Nous utiliserons cependant dans les exemples une notation ``\`a la Dewey''. Chaque n\oe ud est repr\'esent\'e 
par une suite de nombres entiers 
(mais ce pourrait \^etre n'importe quel alphabet ordonn\'e) et d\'enot\'es 
$\epsilon$, $1$, $11$, $12$, $112$, $\dots$. L'ordre lexicographique est le suivant: $u,v,w$ sont des mots, $ui < uiv (v \not = \epsilon$), and $uiv < ujw \ $si$ \ i < j$, $\epsilon$ est le mot vide.

\item {\bf $u \in T$}:  $u$ est le n\oe ud courant dans $T$ (\bt visit\'ee).

\item {\bf $n \in {\cal N}$}:  $n$ est un entier positif associ\'e \`a chaque n\oe ud dans $T$ par la fonction $num$ (ci-dessous). Il correspond \`a son ordre de cr\'eation, re-cr\'eation ou re-visite; c'est aussi le num\'ero de la bo\^{\i}te associ\'ee \ag un n\oe ud. C'est le num\'ero du dernier n\oe ud cr\'e\'e. 

\item {\bf $num: T \rightarrow {\cal N}$}: (abbrev. {\bf $nu$ })  $nu(v)$ est le num\'ero (entier positif) associ\'e au n\oe ud $v$ dans $T$. .

\item {\bf $pred: T \rightarrow {\cal H}$}: (abbrev. {\bf $pd$ }) $pd(v)$ est la pr\'edication associ\'ee au \nd $v$ dans $T$. C'est un \el de l'ensemble d'atomes non clos ${\cal H}$ (base de Herbrand non close). Cette pr\'edication est invariante et correspond \ag un \'el\'ement du dorps de la clause choisie au \nd parent. 

\item {\bf $chosen\_claus: T \rightarrow P$}: (abbrev. {\bf $cc$}) $cc(v)$ est la clause choisie pour r\'esoudre la pr\'edication courante. $cc(v)$ est 
vide si aucune clause ne peut \^etre choisie.

\item {\bf $claus\_list: T \rightarrow 2^P$}:  (abbrev. {\bf $cl$}) $cl(v)$ est une liste de clauses de $P$ (m\^eme ordre que dans $P$) susceptible de contribuer \ag la d\'efinition du pr\'edicat de $pd(v)$ associ\'ee au \nd $v$ dans $T$. Selon les clauses s\'electionn\'ees dans $cl(v)$, on peut obtenir diff\'erentes s\'emantiques op\'erationnelles (mais il y a toujours une seule SO). 
Si la \bt est vide, la predication $pd(v)$ ne peut \^etre r\'esolue et le \nd sera en \'echec ({\em failure}). Cette liste de clauses est d\'efinie par un ordre externe lorsque la predication est appel\'ee (voir $claus\_init$ dans les fonctions externes).

\item {\bf $\sigma: T \rightarrow Subst$}:  $\sigma(v)$ est la substitution courante, celle qui, appliqu\'ee \ag $pd(v)$, donne la pr\'edication appel\'ee au \nd $v$. La substitution vide est not\'ee $\emptyset$. La substitution \'echec est not\'ee $\perp$. La composition de substitutions est not\'ee par justaposition des substitutions, ex $\mu \sigma$.

\item {\bf $first: T \rightarrow Bool$}: (abbrev. $fst$) $fst(v)$ est vrai ssi $v$ est un \nd de $T$ qui n'a pas encore \'et\'e  visit\'e.

\item {\bf $ct \in Bool$}:  $ct$ est l'indicateur de construction achev\'ee (arbre compl\`etement construit et visit\'e, retour \ag la racine) de $T$: $true$ ssi le n\oe ud courant est redevenu $\epsilon$ lors d'une remont\'e dans l'arbre (en succ\`es ou \'echec).

\item {\bf $flr \in Bool$}:  $flr$ est l'indicateur d'\'etat du sous-arbre ($true$ si en \'echec, $false$ sinon, ce qui n'est pas synonyme de succ\`es).

\item {\bf $scs \in Bool$}:  $scs$ est l'indicateur d'\'etat du sous-arbre ($true$ si succ\`es, $false$ sinon, ce qui n'est pas synonyme d'\'echec).

\item {\bf $bk3 \in Bool$}:  $bk3$ est l'indicateur de retour arri\`ere actif de la trace produite pour le \md \verb.m3. seulement ($true$ ssi un parcours inverse de l'arbre courant est en cours).
\end{enumerate}

Noter que la SO utilise comme donn\'ee externe la m\'ethode de trace (\verb.m1., \verb.m2. ou \verb.m3.), qui sont alors des bool\'eens exlcusifs, mais ces \'el\'ements n'ont pas \'et\'e introduits dans l'\'etat afin de ne pas le surcharger.

\vspace{2mm}
{\underline{\bf Fonctions utilitaires}} (manipulation sur les objets d\'ecrits): les fonctions sont pr\'esent\'ees dans l'ordre des objets qu'elles concernent.

\begin{itemize}
\item {\bf $parent: T \rightarrow T$ }: (abbrev. {\bf $pt$}) $pt(v)$ est l'anc\^etre direct de $v$ dans $T$. 
Pour simplifier le \md, on suppose que $pt(\epsilon) = \epsilon$.

\item {\bf $leaf: T \rightarrow Bool$ }: (abbrev. {\bf $lf$}) $lf(v)$ est vraie ssi $v$ est une feuille dans $T$.

\item {\bf $create\_children: T \rightarrow T^*$ }: (abbrev. {\bf $crc$}) $v = crc(v)$ est la liste des nouveaux enfants de $v$ in $T$. 

\item {\bf $first\_elem: liste(E) \rightarrow E$ }: (abbrev. {\bf $fe$}) $v = fe(l)$ est le premier \'el\'ement de la liste $l$.

\item {\bf $has\_a\_next\_node: T \rightarrow Bool$ }: (abbrev. {\bf $hnn$}) $hnn(v)$ est vrai si le \nd $v$ a un \nd suivant dans l'arbre $T$. Faux si  $v$ est $\epsilon$.

\item {\bf $next\_right\_node: T \rightarrow T$ }: (abbrev. {\bf $nrn$}) $w = nrn(v)$ est le \nd suivant $v$ dans $T$. La racine $\epsilon$ ne peut avoir de suivant.

\item {\bf $has\_a\_choice\_point: T \rightarrow Bool$ }: (abbrev. {\bf $hcp$})  $hcp(v)$ est vrai ssi il existe un point de choix $w$ dans le sous-arbre de racine $v$ dans $T$ ($cl(w)$ contient au moins une clause).

\item {\bf $youngest\_child\_with\_choice\_point: T \rightarrow T$}:  (abbrev. {\bf $ycwcp$}) $w = ycwcp(v)$ est le plus r\'ecent point de choix dans le sous-arbre de racine $v$ dans $T$ ($cl(w)$ contient au moins une clause) selon l'ordre lexicographique des \nds dans $T$.

\item {\bf $child\_with\_greates\_choice\_point: T \rightarrow T$}: (abbrev. {\bf $cwgcp$}) $cwgcp(v)$ est le \nd enfant $v$  dans $T$ dont le sous-arbre contient le plus  r\'ecent point de choix.

\item {\bf $greatest\_choice\_point: T \rightarrow T$ }:  (abbrev. {\bf $gcp$}) $w = gcp(v)$ est le plus grand point de choix dans le sous-arbre de racine $v$ dans $T$ ($cl(w)$ contient au moins une clause) selon l'ordre lexicographique des \nds dans $T$. 

\item {\bf $fact: Claus \rightarrow Bool$ }: (abbrev. {\bf $ft$ }) $ft(c)$ est vrai ssi $c$ est un fait.

\item {\bf $update\_number: F_{X},T \rightarrow F_{X} $}: (abbrev. {\bf $upn$}) $upn(nu,v)$ met \ag jour la fonction $nu$ en supprimant toutes les r\'ef\'erences aux \nds d\'econstruits de $T$ jusqu'au \nd $v$ (conserv\'e).

\item {\bf $initialize\_tree\_pred\_and\_first: 2^T, F_{pred}, F_{first}, 2^T \rightarrow 2^T, F_{pred}, F_{first}, $}:  (abbrev. {\bf $itpf$}) $(T',pd',fst')= itpf((T,pd,fst),U)$ (o\`u $U=cc(v)$ est un ensemble de nouveaux enfants du \nd $v$) met \ag jour l'arbre $T$ et les fonctions $pd$ et $fst$ en ajoutant \ag l'arbre $T$ les \nds $U$ ($T'$), associant \ag chaque \nd suppl\'ementaire la pr\'edication correspondante r\'esultant de la clause choisie au \nd parent ($pd'$) et en indiquant que les \nds $U$ n'ont pas encore \'et\'e visit\'es ($fst'$).

\item {\bf $update: 2^T, F_{first}, T \rightarrow 2^T, F_{first}$} (abbrev. $updt$) $(T',fst') = updt((T,fst),v)$ o\`u $T'$ est l'arbre $T$ dans lequel on enl\`eve les sous-arbres posterieurs \ag $v$ sauf leur racine dont l'indicateur de premi\`ere visite est remis \ag $true$. Formellement, on supprime les \nds $y$ tels que $\{y > v \wedge anc(y) > v\}$ et on modifie $fst$ pour les \nds $y$ tels que $\{y > v \wedge anc(y) < v \}$.

\end{itemize}

\vspace{2mm}
{\underline{\bf Fonctions externes}}:

Elles correspondent aux actions non d\'ecrites dans la trace virtuelle mais qui l'influencent effectivement, en particulier tous les aspects de la r\'esolution li\'es \`a l'unification et qui sont omis dans cette description.

\begin{itemize}
\item {\bf $unify: T \rightarrow Substitution$}: (abbrev. $unif$)  $unif(v,\mu): \mu$ est la substitution obtenue par unification de $\sigma(v) pred(v)$ avec la t\^ete de $cc(v)$, la clause choisie. L'\'echec de l'unification se traduit par un unificateur ``\'echec'' $\perp$. 


\item {\bf $claus\_init: T \rightarrow list\_of\_clauses$}: (abbrev. {\bf $cini$}) $lc = cini(v)$ met \`a jour la fonction $claus$ avec la paire $(v, lc)$ o\`u  $lc$ est une liste des clauses renomm\'ees, utiles pour r\'esoudre  $\sigma pred(v)$ et qui sont donc utilisables pour essayer diff\'erentes alternatives pour la r\'esolution  (si la liste est vide il n'y a pas de solution). Selon les clauses mises dans la liste, on obtient diff\'erents mod\`eles d'ex\'ecution.

\item {\bf $choice: list\_of\_clauses \rightarrow clause$}: $c = choice(lc)$ choisit une clause dans la liste (en Prolog ISO la premi\`ere de la liste), mais, selon le choix op\'er\'e, on obtient diff\'erents mod\`eles d'ex\'ecution.


\item {\bf $m1, m2, m3: Bool$}: bool\'eens exclusifs utilis\'es pour distinguer les \md de trace souhait\'e.

\end{itemize}

\vspace{2mm}
{\bf Etats initiaux consid\'er\'es}: 
\begin{quote}
\noindent $S_1: \ \ \{ \{ \epsilon \} , \epsilon , 1, \{ (\epsilon, 1)\}, \{(\epsilon, goal\}, \{\}, \{(\epsilon, cl_{goal})\} , \{ (\epsilon, \emptyset)\}, \{ (\epsilon, true)\},$

$\ \ \ \ \ \ \ \ \ \ \ \ false, false, false, false/true\}$
\end{quote}


\begin{figure*}[ht]\small
\noindent
\reglecontroledeux{\callun{}}
{\neg flr \wedge fst(u) \wedge lf(u) \wedge \neg ct}
{cl'(u) \gets cl \et cc(u) \gets \emptyset \et n' \gets n+1 \et num'(u) \gets n' \et }
{fst'(u) \gets F \et scs' \gets F \et flr' \gets F}
{\{ cl = cini(u) \}}
\saut
\reglecontrole{\choice{}}
{\neg fst(u) \wedge lf(u) \wedge \neg ct \wedge  cc(u) = \emptyset \wedge \neg cl(u) = [] } 
{cc'(u) \gets c \et cl'(u) \gets cl-c } 
{\{ c = choice(cl(u)) \}} 
\saut
\reglecontrole{\factsucceeds{}}
{\neg fst(u) \wedge lf(u) \wedge \neg ct \wedge \mu \not = \perp \wedge  cc(u) \not = \emptyset \wedge ft(cc(u))}
{\sigma'(u) \gets \mu \sigma(u) \et scs' \gets T \et flr' \gets F} 
{\{ unif(u,\mu) \}} 
 \saut
\reglecontroleshort{\claussucceedsshort{}} 
{\neg fst(u) \wedge lf(u) \wedge \neg ct \wedge \mu \not = \perp \wedge cc(u) \not = \emptyset \wedge \neg ft(cc(u)) \wedge U \gets crc(u)}
{ itpf((T,pd,fst),U) \et u' \gets fi(U) \et \sigma'(u) \gets \mu \sigma(u)}
{\{ 
unif(u,\mu) \}}
 \saut
\reglecontrole{\exitun{}}
{\neg fst(u) \wedge scs \wedge \neg hnn(u) \wedge \neg ct \et v \gets pt(u)}
{u' \gets v \et (u = \epsilon) \Rightarrow (ct' \gets T)}
{\{ \}}
 \saut
\reglecontrole{\exitdeux{}}
{ \neg fst(u) \wedge scs \wedge hnn(u) \wedge \neg ct \et v \gets nrn(u)}
{u' \gets v } 
{\{ \}}
 \saut
\reglecontrole{\leafunfail{}}
{\neg fst(u) \wedge lf(u) \wedge \neg ct \wedge cc(u) = \emptyset \et cl(u) = [] \et v \gets pt(u)}
{u' \gets v \et (u = \epsilon) \Rightarrow (ct' \gets T) \et flr' \gets T \et scs' \gets F \et bk3 \gets T}
{\{ \}}
\saut
\reglecontrole{\leafdeuxfailud{}}
{\neg fst(u) \wedge lf(u) \wedge \neg ct \wedge cc(u) \not = \emptyset \wedge \mu = \perp \et v \gets pt(u)}
{u' \gets v \et (u = \epsilon) \Rightarrow (ct' \gets T) \et flr' \gets T \et scs' \gets F}
{\{m1 \vee m2 \et \\ \phantom{xxxxxxxxxxxxxxxxxxxxxxxxxxxxxxxxxxxxxxxxxxxxxxxxxxxxxxxx} unif(u,\mu) \}}
\saut
\reglecontrole{\leafdeuxfailt{}}
{\neg fst(u) \wedge lf(u) \wedge \neg ct \wedge cc(u) \not = \emptyset \wedge \mu = \perp}
{u' \gets v \et (u = \epsilon) \Rightarrow (ct' \gets T) \et flr' \gets T \et scs' \gets F \et bk3 \gets T}
{\{m3 \et \\ \phantom{xxxxxxxxxxxxxxxxxxxxxxxxxxxxxxxxxxxxxxxxxxxxxxxxxxxxxxxxx} unif(u,\mu) \}}
\saut
\reglecontrole{\treefailuda{}}
{\neg fst(u) \wedge \neg lf(u) \wedge flr \wedge \neg ct \wedge \neg hcp(u) \et v \gets pt(u)}
{u' \gets v \et (u = \epsilon) \Rightarrow (ct' \gets T) }
{\{ m1 \vee m2\}}
\saut
\reglecontrole{\treefaildb{}}
{\neg fst(u) \wedge \neg lf(u) \wedge flr \wedge \neg ct \wedge hcp(u) \et v \gets ycwcp(u)}
{u' \gets v }
{\{ m2\}}
\saut
\reglecontroleshort{\redounshortu{}} 
{ \neg fst(u) \wedge hcp(u) \wedge (flr \vee ct) \et v \gets gcp(u) }
{\small updt((T,fst),v) \et u' \gets v  \et cc(v) \gets \emptyset \et ct \Rightarrow (ct' \gets F) \et scs' \gets F \et flr' \gets F} 
{\{m1\}}
 \saut
\reglecontroleshort{\redounshortda{}}
{ \neg fst(u) \wedge hcp(u) \wedge (flr \vee ct) \et gcp(u) \not = u }
{u' \gets cwgcp(u) } 
{\{m2\}}
 \saut
\reglecontroleshort{\redounshortdb{}}
{ \neg fst(u) \wedge hcp(u) \wedge (flr \vee ct)  \et u = gcp(u)}
{updt((T,fst),u) \et cc(u) \gets \emptyset \et scs' \gets F \et flr' \gets F \et ct \Rightarrow (ct' \gets F) } 
{\{m2\}}
 \saut
\reglecontroleshort{\redounshorttu{}}
{ \neg fst(u) \wedge bk3 \wedge cl(u) \not = [] \wedge (flr \vee ct)}
{updt((T,num,fst),u) \et cc(u) \gets \emptyset \et scs' \gets F \et flr' \gets F \et bk3 \gets F \et ct \Rightarrow (ct' \gets F)} 
{\{m3\}}
 \saut
\reglecontroleshort{\redounshorttd{}}
{ \neg fst(u) \wedge bk3 \wedge \neg lf(u) \wedge cl(u) = [] }
{u' \gets rcld(u)} 
{\{m3\}}
 \saut
\reglecontroleshort{\redounshorttt{}}
{ \neg fst(u) \wedge bk3 \wedge lf(u) \wedge hpn(u) \wedge cl(u) = []}
{u' \gets pn(u)} 
{\{m3\}}
 \saut
\reglecontroleshort{\redounshorttq{}}
{ \neg fst(u) \wedge bk3 \wedge lf(u) \wedge \neg hpn(u) \wedge cl(u) = [] \wedge u \not = \epsilon}
{u' \gets pt(u)} 
{\{m3\}}
 \saut
\caption{S\'emantique observationnelle d'une r\'esolution SLDT d\'eterministe avec 3 m\'ethodes de retour arri\`ere}
\label{sovirttracefig2}
\end{figure*}

\begin{figure}[h]
\begin{center}
\includegraphics[width=0.60\linewidth]{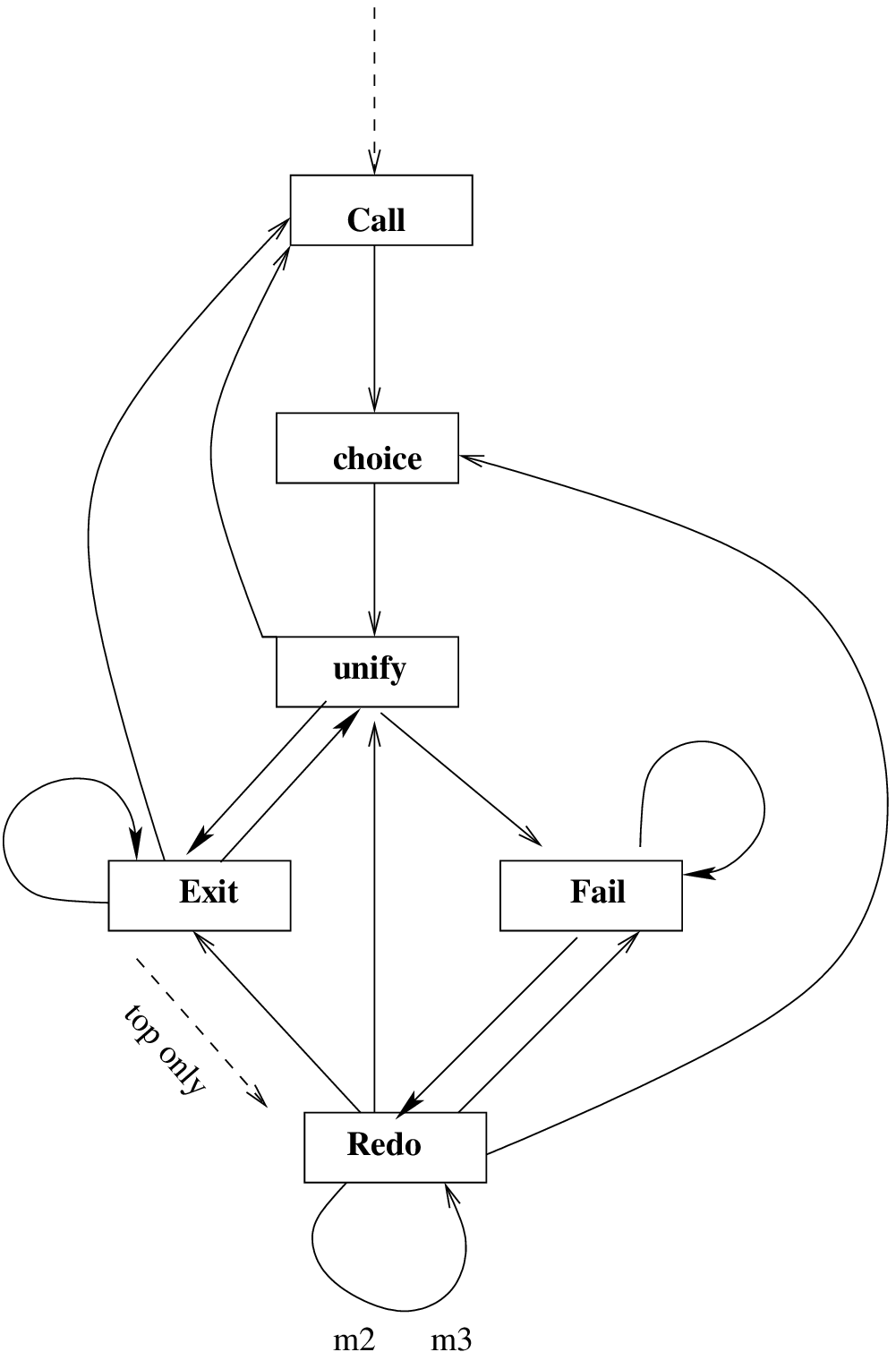}
\end{center}
\caption[Alg\`ebre des ports avec la SO des 3 \mds )(sauf indication explicite, une transition est possible dans tous les mod\`eles]{Alg\`ebre des ports avec la SO des 3 \mds (sauf indication explicite, une transition est possible dans tous les mod\`eles)}
\label{figadequat3}
\end{figure}

La figure~\ref{figadequat3} montre l'alg\`ebre des ports avec la SO des 3 \mds.

\clearpage

\section{ANNEXE: Preuve de correction de la condition d'ad\'equation (Section~\ref{introtrace}, Proposition~\ref{theoadeq} }
\label{appendixF}

On prouve que si la propri\'et\'e suivante est v\'erifi\'ee:

{\bf [Condition d'ad\'equation]}

Etant donn\'es une SO d\'efinie avec un ensemble de \rgs $R$, un sch\'ema de trace ${\cal E}$ et un sch\'ema de reconstruction  ${\cal C}$ pour un sous-ensemble de param\`etres $Q$. Si les deux propri\'et\'es suivantes sont satisfaites pour chaque \rg $r \in R$:
\begin{quote}
$\forall\, e,\,e',\,r', S,\,S',\,S'',$

${\cal E}_r(S,S') = e\  \wedge \ {\cal E}_{r'}(S',S'') = e'$

(1) seule  $Cond_r(e,e')$ est vraie, i.e. $Cond_r(e,e') \bigwedge_{s \not = r} \neg Cond_{s}(e,e')$.

(2) $\ {\cal C}_r(e,e',S/Q) = S'/Q$.
\end{quote}

\vspace{2mm}
alors toute trace actuelle $T_w = <Q_0, w^+_t>$ , d\'efinie par le sch\'ema de trace ${\cal E}$ et telle que $Q_0 = {S_0}{/Q}$, est {\em ad\'equate} pour $Q$ par rapport \ag la trace int\'egrale virtuelle $T_v = <S_0, v^+_t>$;

\vspace{1mm}
c'est \`a dire qu'il existe une fonctions ${\cal F}$ telle que
\begin{quote}
$\forall t\geq 0, \ {\cal F}(w^+_t,Q_0) = Q_t\ $ et

$\forall i \in [0 .. t], \ Q_i = S_i/Q \ \wedge 
\exists r_i \in R, \ \ \, $ tel que $\, w_i = {\cal E}_{r_i}(S_i,S_{i+1})$.
\end{quote}

\vspace{2mm}
Cet ennonc\'e appelle un commentaire afin de bien comprendre ce que signifie l'ad\'equation. L'existence d'une fonction  ${\cal F}$ assure que la lecture de la trace permet de suivre l'\'evolution d'un \'etat restreint aux param\`etres de $Q$, un sous-ensemble de l'\'etat int\'egral virtuel $S$. Ceci n'a de sens que si on part d'un \'etat initial $Q_0$ connu qui ne peut \^etre que l'\'etat int\'egral virtuel retreint initial $S_0/Q$ (donc $Q_0 = S_0/Q$). 

Par ailleurs on notera que la fonction ${\cal F}$ ne pr\'ejuge pas de l'\'etat $S_{t+1}$ atteint apr\`es l'\ev $w_t$. Si la trace est infinie, cela ne pose pas probl\`eme. Si par contre la trace est finie et si aucun \'etat d\'efini comme final n'est atteint ou si la trace est brusquement interrompue (ceci suppose des interactions externes au processus observ\'e non \'etudi\'ees ici), on ne conna\^it rien en g\'en\'eral sur l'\'etat atteint. Ceci n'affaiblit en rien l'id\'ee d'ad\'equation mais montre seulement que tout ``red\'emarrage'' de la trace n\'ecessite de fournir un nouvel \'etat de d\'epart, c'est \ag dire de fournir \ag nouveau un \'etat initial.

L'ad\'equation assure donc deux choses: la premi\`ere est que chaque \ev de trace am\`ene \ag un \'etat $Q$ toujours \'egal \ag l'\'etat int\'egral virtuel restreint courant $S/Q$ (ce qui est assur\'e par le fait que $S_i/Q = Q_i$). 

L'ad\'equation assure \'egalement, en liaison avec la propri\'et\'e pr\'ec\'edente, qu'\ag chaque \ev $w_i$ de la trace actuelle une \rg $<r,S,S'>$ de la SO s'applique de telle mani\`ere que $<r_i,S_i,S_{i+1}>$ en est une instance. Ceci assure que la trace peut ``faire tourner'' la machine d\'efinie par la SO de telle mani\`ere que la trace actuelle est bien susceptible d'avoir \'et\'e extraite \ag partir de celle-ci. Noter que l'ad\'equation n'exige pas que la SO soit d\'ecrite par un automate d\'eterministe; la m\^eme trace actuelle peut donc \^etre produite de diff\'erentes mani\`eres (l'unicit\'e de la \rg $r_i$ qui s'applique n'est pas requise).
La condition propos\'ee ici impose cette unicit\'e, c'est \ag dire qu'\ag une trace actuelle correspond un seul fonctionnement possible de l'automate.
Il s'agit donc seulement d'une condition suffisante d'ad\'equation. Son int\'er\^et cependant est d'assurer qu'\`a toute trace actuelle correspond une lecture unique dans la SO (ceci est en effet trivialement le cas si dans la trace le nom de la \rg de transition fait partie des attributs ou si $Cond_r(e,e')$ ne d\'epend que de $e$, ce qui se v\'erifie si les ports sont isomorphes aux noms des \rgs).

\vspace{2mm}
\begin{proof} 

Rappel: $w^+_t$ repr\'esente la suite finie non vide $w_t \dots w_1 w_0$ et $w^*_t$ est  $w^+_t$ ou la suite vide $\epsilon$ (la suite vide n'est possible que pour $t \leq 0$).
Cette  formulation de l'ad\'equation suppose que les traces actuelles aient au moins deux \evs ($w^+_t, t \geq 1$ d\'ebute toujours avec les \evs $w_0$ et $w_1$).

On d\'efinit alors la valeur de ${\cal F}(w^*_t,Q_0)$ r\'ecursivement, \`a partir de la fonction de reconstruction locale, de la mani\`ere suivante:

\begin{quote}
${\cal F}(\epsilon, Q_0) = Q_0$;

${\cal F}(w_0, Q_0) = Q_0$;  (l'\'etat actuel courant obtenu ne peut \^etre connu qu'avec au moins deux \'ev\'enements de trace).

et pour $t \geq 1 $, \ ${\cal F}(w_t w_{t-1} w^*_{t-2}, Q_0) = $

$\ \ \ \ \ \ \ \ \ \ \ \ \ \ $\ si\ $Cond_r(w_{t-1}, w_t) $ \ alors\  $ {\cal C}_r(w_{t-1}, w_t, {\cal F}(w^*_{t-1}, Q_0)) $.
\end{quote}

La d\'efinition de ${\cal F}$ est donc la suivante:

\begin{quote}
${\cal F}(\epsilon, S) = S$;

${\cal F}(e, S) = S$;

${\cal F}(e' e E, S) = $\ si\ $Cond_r(e, e'))$ \ alors\  ${\cal C}_r(e, e', {\cal F}(e E, S))$ 

avec $E$ suite quelconque d'\'ev\'enements de trace.
\end{quote}

On v\'erifie bien que $\forall t\geq 0, \ {\cal F}(w^+_t,Q_0) = Q_t\ $ car 

${\cal F}(w_t w_{t-1} w^*_{t-2}, Q_0) = {\cal C}_r(w_{t-1}, w_t, {\cal F}(w^*_{t-1}, Q_0)) = {\cal C}_r(w_{t-1}, w_t, Q_{t-1})) = Q_t$.

Noter que selon la condition~1 cette fonction est d\'eterministe.

\vspace{1mm}
Si on consid\`ere alors un \'etat initial $S_0$, un \'etat initial restreint $Q_0 = S_0/Q_0$, et une trace actuelle $T_w = <Q_0, w^+_t>$, d\'efinie par le sch\'ema de trace ${\cal E}$,
la condition d'ad\'equation assure que cette trace est ad\'equate pour $Q$ par rapport \ag la trace virtuelle int\'egrale $T_v = <S_0, v^+_t>$ (rappel: $T_v$ est obtenue par application de fonctions d'extraction ${\cal E}_r(S, S')$ telles que les valeurs des attributs de l'\ev de trace correspondant sont $A_t = S_{t+1}$).

En effet, par hypoth\`ese, l'\'etat initial restreint est $Q_0 = S_0/Q_0$, et du fait de la condition~2 et de la d\'efinition de ${\cal F}$, les \'etats restreints successifs calcul\'es avec la fonction ${\cal F}$ ainsi d\'efinie v\'erifient bien $\forall i \in [1 .. t], \ Q_i = S_i/Q$. 

A chaque \'etape \'egalement, par la condition~1, une seule condition $Cond_{r_{i-1}}(w_{i-1}, w_i)$ est v\'erifi\'ee et une seule \rg $r_{i-1}$ de reconstruction s'applique aux deux \'ev\'enements $w_{i-1}$ et $w_i$: ${\cal C}_{r_{i-1}}(w_{i-1}, w_i, S_{i-1}/Q) = Q_i$. Le premier \ev $w_{i-1}$ a \'et\'e produit par application de la fonction d'extraction aux \'etats virtuels $S_{i-1}$ et $S_i$, ${\cal E}_{r_{i-1}}(S_{i-1}, S_i) = w_{i-1}$, et le second \ev $w_i$, par application de la fonction d'extraction aux \'etats virtuels $S_{i}$ et $S_{i+1}$ pour une certaine \rg $r_i$ avec  ${\cal E}_{r_i}(S_{i}, S_{i+1}) = w_i$.

Comme les \evs successifs de la trace actuelle sont produits par application du sch\'ema de trace ${\cal E}$ et que chaque couple d'\'ev\`enements permet de reconstruire un \'etat virtuel restreint \ag $Q$ par la fonction ${\cal F}$ d\'eduite du sch\'ema de reconstruction, l'ensemble des propri\'et\'es caract\'erisant l'ad\'equation est bien v\'erifi\'e. 

De plus, \ag chaque \ev de trace actuelle $w_i$ ($i \in [0 .. {t-1}]$) correspond une transition $<r_i, S_i, S_{i+1}>$ qui produit l'\ev de trace int\'egrale virtuelle $v_i$ dont les attributs v\'erifient  $A_i/Q = Q_{i+1}$.
On observe ainsi que la trace actuelle, bien que ne donnant qu'une vue partielle du processus observ\'e, permet d'en reconstituer le d\'eroulement complet, \ag condition toutefois d'en conna\ii tre la SO.


\end{proof}


\newpage

\tableofcontents

\newpage
\ 
\newpage
\ 

\end{document}